\documentclass[%
 reprint,
 superscriptaddress,
nofootinbib,
 amsmath,amssymb,
 aps,
 prx
]{revtex4-2}
\usepackage{graphicx}
\usepackage{dcolumn}
\usepackage{bm}
\usepackage[colorlinks=true, allcolors=blue]{hyperref}
\usepackage[capitalise]{cleveref}
\usepackage{braket}
\usepackage[table]{xcolor}
\usepackage{xfrac}
\usepackage{comment}
\usepackage[normalem]{ulem}
\usepackage{mathtools}
\usepackage{layouts}
\usepackage{stmaryrd}
\usepackage{float}
\usepackage{esint}
\usepackage{soul}
\usepackage[T1]{fontenc}

\newcommand{\vect}[1]{\boldsymbol{\mathbf{#1}}}

\begin{document}
\title{Stabilizer subsystem decompositions for single- and multi-mode Gottesman-Kitaev-Preskill codes}
\author{Mackenzie H. Shaw}
\email{msha2420@uni.sydney.edu.au}
\affiliation{ARC Centre of Excellence for Engineered Quantum Systems, School of Physics, The University of Sydney, Sydney, NSW 2006, Australia.}
\author{Andrew C. Doherty}
\affiliation{ARC Centre of Excellence for Engineered Quantum Systems, School of Physics, The University of Sydney, Sydney, NSW 2006, Australia.}
\author{Arne L. Grimsmo}
\affiliation{ARC Centre of Excellence for Engineered Quantum Systems, School of Physics, The University of Sydney, Sydney, NSW 2006, Australia.}
\affiliation{AWS Center for Quantum Computing, Pasadena, CA 91125, USA}
\affiliation{California Institute of Technology, Pasadena, CA 91125, USA}
	
\date{October 26, 2022}
\begin{abstract}
The Gottesman-Kitaev-Preskill (GKP) error correcting code encodes a finite dimensional logical space in one or more bosonic modes, and has recently been demonstrated in trapped ions and superconducting microwave cavities. In this work we introduce a new subsystem decomposition for GKP codes that we call the stabilizer subsystem decomposition, analogous to the usual approach to quantum stabilizer codes. The decomposition has the defining property that a partial trace over the non-logical stabilizer subsystem is equivalent to an ideal decoding of the logical state, distinguishing it from previous GKP subsystem decompositions.
We describe how to decompose arbitrary states across the subsystem decomposition using a set of transformations that move between the decompositions of different GKP codes.
Besides providing a convenient theoretical view on GKP codes, such a decomposition is also of practical use. We use the stabilizer subsystem decomposition to efficiently simulate noise acting on single-mode GKP codes,
and in contrast to more conventional Fock basis simulations, we are able to consider essentially arbitrarily large photon numbers for realistic noise channels such as loss and dephasing.
\end{abstract}
\maketitle
\section{Introduction}\label{sec:intro}

Bosonic codes encode digital quantum information in continuous variable (CV) quantum systems
and have received both theoretical~\cite{Terhal20,Joshi21,Grimsmo21} and experimental~\cite{Ofek16,Heeres17,Hu19,Grimm20,Lescanne20,Gertler21} attention.
The Gottesman-Kitaev-Preskill (GKP) codes~\cite{Gottesman01}
are one of the most intensively studied encodings of this type, and the single-mode square GKP qubit code has recently been realized in both trapped ions~\cite{Fluehmann19,DeNeeve22} and superconducting microwave resonators~\cite{Campagne19,Eickbusch22,Sivak22}.


From a theoretical perspective, bosonic codes can be understood as defining a logical subspace $\mathcal{L}$ of the CV Hilbert space $\mathcal{H}=\mathcal{L}\oplus\mathcal{L}^{*}$, with the infinite dimensional Hilbert space $\mathcal{L}^{*}$ providing the redundancy required for error correction. However, in the case of GKP codes, the non-normalizability of the codewords~\cite{Gottesman01} means that the GKP logical \textquotedblleft subspace\textquotedblright\ is formally not in the CV Hilbert space.

An alternative formulation, which can be applied to any error correcting code, is to consider a decomposition of the Hilbert space such that the logical information in the error correcting code forms a \textit{subsystem} $\mathcal{H}=\mathcal{L}\otimes\mathcal{S}$~\cite{Knill97,Knill00}. In such a decomposition, the partial trace over the non-logical subsystem corresponds to a decoding map $\mathcal{H}\rightarrow\mathcal{L}$. In Ref.~\cite{Pantaleoni20}, Pantaleoni \textit{et al.}\ introduced the concept of a bosonic subsystem decomposition, and defined a subsystem decomposition for single-mode GKP codes based on a modular quadrature.
This subsystem decomposition has been used in numerical studies of GKP codes~\cite{Tzitrin20,Hastrup21,Pantaleoni21-1,Pantaleoni21-2}. 

The subsystem decomposition is, however, not unique and there are good reasons to investigate alternatives. Specifically, the subsystem decomposition of Ref.~\cite{Pantaleoni20} has lower symmetry than the GKP code itself: the logical subsystem differs if one chooses position or momentum as the ``modular quadrature.'' More recent work~\cite{Pantaleoni22} has also linked the modular position subsystem decomposition to the Zak basis~\cite{Zak67}. In all of these cases the decomposition does not represent the logical information one would retrieve by performing noiseless decoding of the GKP code~\cite{Pantaleoni22}.

In this work, we introduce
a subsystem decomposition that resolves these issues. In particular, this new decomposition has the desirable property that tracing over the non-logical subsystem $\mathcal S$ corresponds to a noiseless decoding map for the GKP code.
We refer to this decomposition as the GKP stabilizer subsystem decomposition, as different stabilizer eigenstates correspond to orthogonal basis states of the subsystem $\mathcal{S}$. The stabilizer subsystem decomposition for GKP codes is entirely analogous to the stabilizer/destabilizer formalism of qubit codes~\cite{Aaronson04}.

The stabilizer subsystem decomposition can be applied to all multi-mode qubit or qudit GKP codes (including the concatenation of GKP and qubit stabilizer codes), and is closely related to the Zak basis~\cite{Zak67}.
For any GKP encoding, we show how to write an arbitrary CV state in the corresponding stabilizer subsystem decomposition from the position wavefunction of the state. We use the subsystem decomposition to provide a description of logical Clifford gates on the subsystem decomposition, and show that an ideal implementation of a logical Clifford gate can propagate errors unless a modified round of decoding is performed immediately after the gate.

One practical challenge with GKP codes is the difficulty of numerically simulating GKP codes using a truncated Fock basis, since both the mean and variance of the photon number distribution of physically realizable GKP codestates increases as the codestates approach the infinitely squeezed ``ideal'' codewords. Logical gates can also increase the photon number of the codestates, providing a further need to find new numerical methods to efficiently store and manipulate GKP states~\cite{Bourassa21}.

Using the stabilizer subsystem decomposition we are able to study realistic noise channels such as loss and white-noise dephasing for essentially arbitrary photon numbers. In the case of the single-mode square GKP qubit code our treatment is analytical. We find that GKP codes are far more resilient against pure loss than against dephasing: a square single-mode
code state with ten decibels of GKP squeezing achieves an average gate infidelity below $10^{-3}$ for a loss rate up to ${\sim}4\%$, while it can only tolerate a dephasing rate of ${\sim}0.2\%$ to achieve the same fidelity.
In the case of pure-dephasing, i.e. with white-noise dephasing as the only noise channel, there is a threshold value for the GKP squeezing value and dephasing rate for the GKP code to ``break even'', as the GKP code only performs better than a qubit defined using Fock states $\ket 0$ and $\ket 1$ given the GKP squeezing is above $10$ dB \emph{and simultaneously} the dephasing rate is below $0.1\%$.
We also find that for both pure loss and pure dephasing, there is an optimal finite photon number that minimizes the logical error rate, which is much larger for loss than for dephasing at the same rate, qualitatively consistent with the results of \cite{Hastrup22,Campagne19}

 
Our results are organized as follows. Beginning in \cref{sec:sq}, we present the stabilizer subsystem decomposition for the single-mode square GKP qubit code. Readers wishing to quickly learn the key concepts in the paper can safely begin by reading only \cref{sec:sq}, since it provides a simple explanation of most of the results in the rest of the paper. Then in \cref{sec:notation}, we provide an overview of the established formalism of multi-mode GKP lattices and set up the notation we will use in the remainder of the manuscript. In~\cref{sec:definition}, we define the stabilizer subsystem decomposition in the general case and show that the partial trace over the stabilizer subsystem corresponds to noiseless decoding. In \cref{sec:Zak}, we show how to transform the states of the stabilizer subsystem decomposition of one GKP code to any other code, and describe the method to write the subsystem ``wavefunction'' of a state in terms of its position wavefunction. Finally, we show how to write many practical components of GKP codes conveniently in the stabilizer subsystem decomposition, namely logical Clifford gates (\cref{sec:gates}), approximate GKP codewords, and noise channels such as pure loss, Gaussian displacements and white-noise dephasing (\cref{sec:errors}). Readers focused on applying the stabilizer subsystem decomposition to model noise can safely skip \cref{sec:notation,sec:definition,sec:Zak,sec:gates} and go straight to \cref{sec:errors}. We provide concluding remarks
in \cref{sec:conc}.

\section{Stabilizer Subsystem Decomposition for the Square GKP Qubit Code}\label{sec:sq}

We begin by constructing the stabilizer subsystem decomposition in the simplest non-trivial case: the single-mode square GKP qubit code. To define the subsystem decomposition in \cref{eq:Zak_states,eq:1_mode_defn,eq:1_mode_SSD}, we will make use of the Zak states~\cite{Zak67}, and provide the intuition for why the stabilizer subsystem decomposition accurately describes the GKP logical information stored in an arbitrary state. Then we will outline the key properties of the decomposition in \cref{subsec:properties}, including examples of states and operators decomposed in the subsystem decomposition. In doing so, we foreshadow the numerical techniques for simulating GKP codes that we develop in more detail in \cref{sec:errors}.

\subsection{Preliminaries}\label{subsec:prelims}

The square GKP qubit code encodes a qubit into a single-mode continuous-variable (CV) Hilbert space $\mathcal{H}$, which is described by position and momentum operators that satisfy $[\hat{q},\hat{p}]=i$. We define the displacement operators
\begin{equation}\label{eq:1_mode_W}
    \hat{W}(v_{1},v_{2})=\mathrm{exp}\big(\sqrt{2\pi}i(v_{2}\hat{q}-v_{1}\hat{p})\big)
\end{equation}
for $v_{1},v_{2}\in\mathbb{R}$, which form an operator basis of $\mathcal{L}(\mathcal{H})$, the space of all linear operators acting on $\mathcal H$. The displacement operators obey the commutation relation
\begin{align}\label{eq:1_mode_W_commutation}
    \big\llbracket \hat{W}(u_{1},u_{2}),\hat{W}(v_{1},v_{2})\big\rrbracket&=e^{{-}2i\pi (u_{1}v_{2}-u_{2}v_{1})},
\end{align}
where $\llbracket A,B\rrbracket=ABA^{-1}B^{-1}$ is the group commutator, and the composition rule
\begin{multline}\label{eq:1_mode_W_composition}
    \hat{W}(u_{1},u_{2})\hat{W}(v_{1},v_{2})= \\
    e^{{-}i\pi(u_{1}v_{2}-u_{2}v_{1})}\hat{W}(u_{1}{+}v_{1},u_{2}{+}v_{2}).
\end{multline}
$\hat{W}(v_{1},v_{2})$ ``displaces'' the position and momentum operators such that
\begin{subequations}\label{eq:1_mode_W_qp}
\begin{align}
    \hat{W}(v_{1},v_{2})^{\dag}\hat{q}\hat{W}(v_{1},v_{2})&=\hat{q}+\sqrt{2\pi}v_{1},\\ \hat{W}(v_{1},v_{2})^{\dag}\hat{p}\hat{W}(v_{1},v_{2})&=\hat{p}+\sqrt{2\pi}v_{2}.
\end{align}
\end{subequations}
Note that \cref{eq:1_mode_W,eq:1_mode_W_qp}~differ by a factor of $\sqrt \pi$ from the more standard definition $\hat{D}(\alpha) = \exp\big(\alpha \hat{a}^{\dag}-\alpha^{*}\hat{a}\big)$.

The square GKP qubit code is a stabilizer code with stabilizer group generated by the commuting displacement operators
\begin{subequations}\label{eq:1_mode_stabs}
\begin{align}
    \hat S_1 &= \hat{W}\big(\sqrt{2},0\big)=e^{-2i\sqrt{\pi}\hat{p}}, \\
    \hat S_2 &= \hat{W}\big(0,\sqrt{2}\big)=e^{2i\sqrt{\pi}\hat{q}},
\end{align}
\end{subequations}
along with their inverses. The logical Pauli group is generated by
\begin{subequations}\label{eq:1_mode_Paulis}
\begin{align}
    \bar{X} &= \hat{W}\big(1/\sqrt{2},0\big)=e^{-i\sqrt{\pi}\hat{p}}, \\
    \bar{Z} &= \hat{W}\big(0,1/\sqrt{2}\big)=e^{i\sqrt{\pi}\hat{q}},
\end{align}
\end{subequations}
which anticommute with each other but commute with the stabilizer generators. The ideal codespace is the simultaneous $+1$-eigenspace of both stabilizer generators, and is spanned by the ideal codestates
\begin{align}
    \ket{\bar{0}}&\propto\sum_{s\in\mathbb{Z}}\ket{2s\sqrt{\pi}}_{q},&\ket{\bar{1}}&\propto\sum_{s\in\mathbb{Z}}\ket{(2s+1)\sqrt{\pi}}_{q},
\end{align}
where $\ket{x}_{q}$ is the $x$-eigenstate of the position operator $\hat{q}$.

A particularly useful set of states for describing GKP codes is the Zak basis~\cite{Zak67}, and was first applied to GKP codes in Ref.~\cite{Glancy06}. The Zak states are parameterized by two real numbers, $k_1$ and $k_2$, and are given in the position basis by
\begin{equation}\label{eq:Zak_states}
    \ket{k_1,k_2}_{a}=\sqrt[4]{2\pi a^{2}}\,e^{i\pi k_{1}k_{2}}\!\sum_{s\in\mathbb{Z}}e^{2i\pi ak_{2}s}\big|\sqrt{2\pi}(k_{1}+as)\big\rangle_{q},
\end{equation}
where $a>0$ is a constant. Note that we have rescaled some of the constants in our definition compared to Ref.~\cite{Zak67}. We can interpret the (rescaled) parameter $\sqrt{2\pi}k_{1}$ as the \textit{quasi-position} of the Zak state in the following sense: since a given Zak state has support on position eigenvalues spaced by $\sqrt{2\pi}a$, each Zak state is an eigenstate of the modular-position operator $\hat{q}$ (mod $\sqrt{2\pi}a$), with eigenvalue $\sqrt{2\pi}k_{1}$. Likewise, it can be shown that $\sqrt{2\pi}k_{2}$ represents the quasi-momentum of the Zak state corresponding to the modular-momentum operator $\hat{p}$ (mod $\sqrt{2\pi}/a$).

The full set of Zak states with $k_{1},k_{2}\in\mathbb{R}$ span $\mathcal{H}$ but are not linearly independent, obeying the quasi-periodic boundary conditions
\begin{subequations}\label{eq:1_mode_Zak_BCs}
\begin{align}
    \ket{k_{1}+a,k_{2}}_{a}&=e^{-i\pi a k_{2}}\ket{k_{1},k_{2}}_{a},\\
    \ket{k_{1},k_{2}+1/a}_{a}&=e^{i\pi k_{1}/a}\ket{k_{1},k_{2}}_{a}.
\end{align}
\end{subequations}
As a result, the Zak states only form a non-overcomplete basis of $\mathcal{H}$ when $k_{1}$ is restricted to an interval of length $a$ and $k_{2}$ to an interval of length $1/a$. Moreover, the Zak basis is orthonormal, satisfying
\begin{equation}
    {\vphantom{\ket{x}}}_{a}\!\braket{k^{}_{1},k^{}_{2}|k_{1}',k_{2}'}_{a}=\delta(k^{}_{1}-k_{1}')\delta(k^{}_{2}-k_{2}')
\end{equation}
as long as $k_{1}$ and $k_{2}$ are restricted as above.

An alternative formulation of the Zak states is to define $\ket{0,0}_{a}$ as the unique simultaneous $+1$-eigenstate of the displacements $\hat{W}(a,0)$ and $\hat{W}(0,1/a)$. The remaining Zak states are then given by the property
\begin{equation}\label{eq:Zak_translation_property}
    \ket{k_{1},k_{2}}_{a}=\hat{W}(k_{1},k_{2})\ket{0,0}_{a}.
\end{equation}
Setting $a=\sqrt{2}$, we observe that the $\ket{0,0}_{\!\sqrt{2}}$ Zak state is a simultaneous $+1$-eigenstate of the GKP operators $\hat{S}_{1}$ and $\bar{Z}$, so we can write
\begin{align}\label{eq:1_mode_codestates_Zak}
    \ket{\bar{0}}&=\ket{0,0}_{\!\sqrt{2}},&\ket{\bar{1}}&=\bar{X}\ket{\bar{0}}=\ket{1/\sqrt{2},0}_{\!\sqrt{2}}.
\end{align}
The remaining $a=\sqrt{2}$ Zak states can be viewed as displaced GKP codestates.

\subsection{Stabilizer Subsystem Decomposition for the GKP Code}

\begin{figure}
\includegraphics{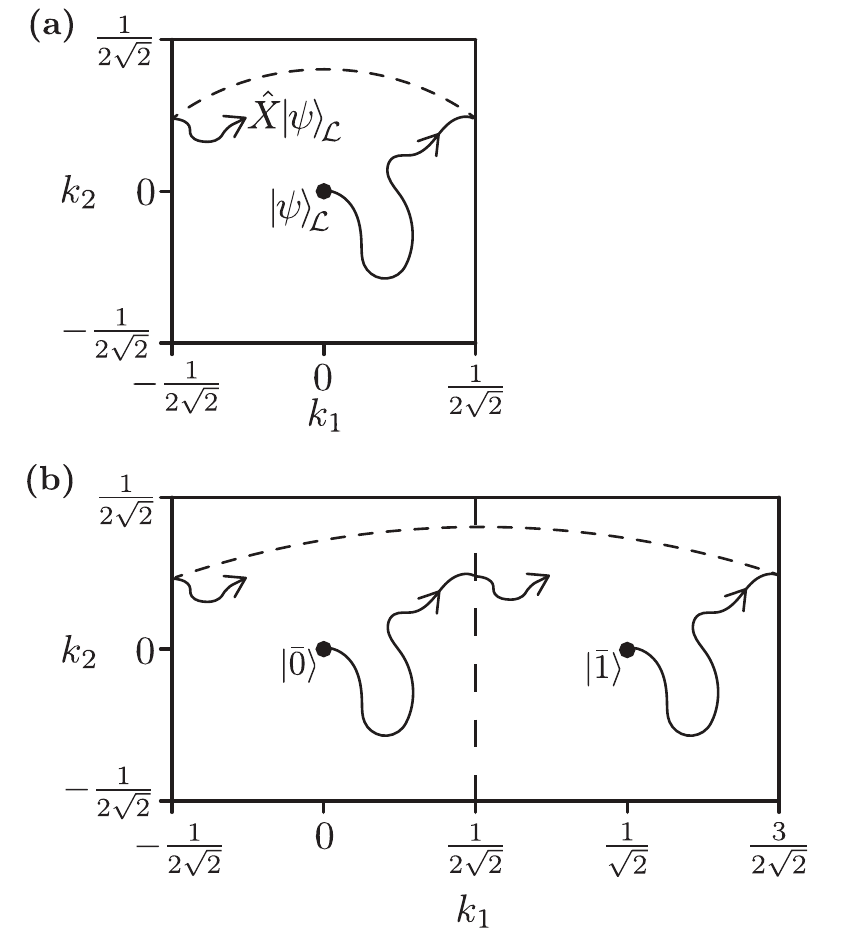}
\caption{Diagrams representing (a) the stabilizer subsystem for the single-mode square GKP qubit code, (b) the Zak basis with $a=\sqrt{2}$. In subplot (a), each point represents the two-dimensional stabilizer subspace $V_{k_{1},k_{2}}$; while in (b) each point represents a single Zak state $\ket{k_{1},k_{2}}_{\sqrt{2}}$. Applying a random walk of displacement operators to an ideal GKP codestate $\ket{\psi}_{\mathcal{L}}\otimes\ket{0,0}$ does not affect the logical subsystem until the state reaches one of the quasi-periodic boundaries of the cell; for example causing an $\hat{X}$ error as shown in (a). The corresponding path is traced out twice in (b) since each basis state $\ket{\psi}\otimes\ket{k_{1},k_{2}}$ of the square GKP code consists of superpositions of states $\ket{k_{1},k_{2}}_{\sqrt{2}}$ and $\ket{k_{1}+1/\sqrt{2},k_{2}}_{\sqrt{2}}$ in the Zak basis.}
\label{fig:square_Zak}
\end{figure}

To define the stabilizer subsystem decomposition we first define the \textit{stabilizer subspaces} $V_{k_{1},k_{2}}$, each of which is a simultaneous eigenspace of the stabilizer generators $\hat{S}_{1},\hat{S}_{2}$. In particular, we define $V_{k_{1},k_{2}}$ as the set of states $\ket{\phi}\in\mathcal{H}$ satisfying
\begin{subequations}\label{eq:1_mode_stabilizer_eigenvalues}
\begin{align}
    \hat{S}_{1}\ket{\phi}&=e^{-2i\sqrt{\pi}\hat{p}}\ket{\phi}=e^{-2\sqrt{2}i\pi k_{2}}\ket{\phi},\\
    \hat{S}_{2}\ket{\phi}&=e^{2i\sqrt{\pi}\hat{q}}\ket{\phi}=e^{2\sqrt{2}i\pi k_{1}}\ket{\phi},
\end{align}
\end{subequations}
for $k_{1},k_{2}\in\big({-}2^{-3/2},2^{-3/2}\big]$. Here, $\sqrt{2\pi}k_{1}$ and $\sqrt{2\pi}k_{2}$ represent the quasi-position $\hat{q}$ (mod $\sqrt{\pi}$) and quasi-momentum $\hat{p}$ (mod $\sqrt{\pi}$) of $\ket{\phi}$ (respectively). It is straightforward to show that each subspace $V_{k_{1},k_{2}}$ is two-dimensional and spanned by the $a=\sqrt{2}$ Zak states $\ket{k_{1},k_{2}}_{\sqrt{2}}$ and $\ket{k_{1}+1/\sqrt{2},k_{2}}_{\sqrt{2}}$. With this connection to Zak states we can see that the union of subspaces $V_{k_{1},k_{2}}$ for $k_{1},k_{2}\in\big({-}2^{-3/2},2^{-3/2}\big]$ spans the full Hilbert space $\mathcal{H}$.

Since each stabilizer subspace $V_{k_{1},k_{2}}$ is two-dimensional, we can define a qubit within each subspace labelled by the orthonormal \textit{stabilizer states} $\ket{\mu,k_{1},k_{2}}$, where $\mu=0,1$. The na\"ive way to do so would be to define the $\ket{0,k_{1},k_{2}}$ stabilizer state as $\ket{k_{1},k_{2}}_{\sqrt{2}}$ and $\ket{1,k_{1},k_{2}}$ as $\ket{k_{1}+1/\sqrt{2},k_{2}}_{\sqrt{2}}$. This is justified since $\ket{k_{1},k_{2}}_{\sqrt{2}}$ is ``closest'' to the ideal codestate $\ket{\bar{0}}=\ket{0,0}_{\sqrt{2}}$, while $\ket{k_{1}+1/\sqrt{2},k_{2}}_{\sqrt{2}}$ is ``closest'' to $\ket{\bar{1}}=\ket{1/\sqrt{2},0}_{\sqrt{2}}$, see~\cref{eq:1_mode_codestates_Zak,fig:square_Zak}.

However, we want to ensure that the qubit state represents the GKP logical information stored in the state. In particular, we impose the defining property of the stabilizer states that
\begin{equation}\label{eq:displacement_property}
    \ket{\psi,k_{1},k_{2}}=\hat{W}(k_{1},k_{2})\ket{\bar{\psi}}
\end{equation}
is a displaced ideal codestate for all qubit states $\ket{\psi}=\alpha\ket{0}+\beta\ket{1}$ with ideal GKP encoding $\ket{\bar{\psi}}=\alpha\ket{\bar{0}}+\beta\ket{\bar{1}}$.

To see the importance of \cref{eq:displacement_property}, consider performing a round of ideal GKP error correction on the state $\ket{\psi,k_{1},k_{2}}$ as follows. First, we measure the stabilizer generators, which reveals the values of $k_{1}$ and $k_{2}$ via \cref{eq:1_mode_stabilizer_eigenvalues}. Then, we apply the displacement $\hat{W}(k_{1},k_{2})^{\dag}$ that returns the state to the ideal codespace. With this definition, we ensure that the qubit information $\ket{\psi}$ in the state $\ket{\psi,k_{1},k_{2}}$ is the same as the logical information one would obtain by performing an ideal round of error-correction and reading out the resultant ideal codestate. Equivalently, this enforces that the partial trace over the stabilizer subsystem correspond to an ideal GKP decoding map, as we show in \cref{subsec:partial_trace}. Strictly enforcing \cref{eq:displacement_property} is, in fact, the key difference between our subsystem decomposition and previous definitions~\cite{Pantaleoni20}.

Enforcing \cref{eq:displacement_property} gives the stabilizer states in terms of Zak states
\begin{subequations}\label{eq:1_mode_defn}
\begin{align}
    \ket{0,k_{1},k_{2}}&=\ket{k_{1},k_{2}}_{\sqrt{2}},\\
    \ket{1,k_{1},k_{2}}&=e^{i\pi k_{2}/\sqrt{2}}\ket{k_{1}+1/\sqrt{2},k_{2}}_{\sqrt{2}}.
\end{align}
\end{subequations}
The additional $e^{i\pi k_{2}/\sqrt{2}}$ phase on the definition of $\ket{1,k_{1},k_{2}}$ arises from the differing geometric phases in the definition of the Zak states \cref{eq:Zak_translation_property} and the stabilizer states \cref{eq:displacement_property}:
\begin{subequations}
    \begin{align}
        \ket{1,k_{1},k_{2}}&=\hat{W}(k_{1},k_{2})\ket{\bar{1}}\\
        &=\hat{W}(k_{1},k_{2})\hat{W}(1/\sqrt{2},0)\ket{0,0}_{\sqrt{2}}\\
        &=e^{i\pi k_{2}/\sqrt{2}}\ket{k_{1}+1/\sqrt{2},k_{2}}_{\sqrt{2}}.
    \end{align}
\end{subequations}

This phase has two additional consequences. First, it ensures that all the logical Pauli operators act as a tensor product between the logical and stabilizer subsystems, as we will see in \cref{subsubsec:operators}. A similar result is described in Ref.~\cite{Ketterer16}. Second, the phase ensures that the full symmetry of the square GKP code is preserved in the subsystem decomposition.

It is also interesting to compare \cref{eq:1_mode_defn} with the Zak-basis representation of the modular-position subsystem decomposition \cite{Pantaleoni20,Pantaleoni22}. Once a rescaling of $k_{1},k_{2}$ is taken into account, the only difference between the two decompositions is the $k_{2}$-dependent phase (see \cref{sec:comparison_to_giacomo}). In this sense the stabilizer subsystem decomposition for the single-mode square GKP qubit code can be thought of as a \textquotedblleft rephasing\textquotedblright\ of the modular-position subsystem decomposition that symmetrizes the treatment of position and momentum.

The states $\ket{\mu,k_{1},k_{2}}$ form a basis for $\mu=0,1$ and $k_{1},k_{2}\in\big({-}2^{-3/2},2^{-3/2}\big]$, so we can define a subsystem decomposition
\begin{align}\label{eq:1_mode_SSD}
    \mathcal{H}&=\mathcal{L}\otimes\mathcal{S},&
    \ket{\mu}&\otimes\ket{k_{1},k_{2}}=\ket{\mu,k_{1},k_{2}},
\end{align}
where $\mathcal{L}$ is the logical subsystem and $\mathcal{S}$ is the stabilizer subsystem. Similar to results obtained in \cite{Pantaleoni20}, $\mathcal{L}$ is a two-dimensional subsystem while $\mathcal{S}$ is isomorphic to the full Hilbert space $\mathcal{H}$ by associating the stabilizer subsystem basis states $\ket{k_{1},k_{2}}\in\mathcal{S}$ with $a=1$ Zak states $\ket{\sqrt{2}k_{1},\sqrt{2}k_{2}}_{1}\in\mathcal{H}$ of the full Hilbert space. For this reason we call the basis of the stabilizer subsystem $\ket{k_{1},k_{2}}$ the Zak basis of $\mathcal{S}$. We note here for clarity that the stabilizer subsystem decomposition~\cref{eq:1_mode_SSD} applies to the square GKP code, which is a stabilizer code and \textit{not} a ``subsystem code'' in the sense of Ref.~\cite{Poulin05}.

It is worth briefly reiterating why the subsystem decomposition \cref{eq:1_mode_SSD} is non-trivial. The key feature of the stabilizer subsystem decomposition is that the state in the logical subsystem is the information one would obtain if one performed a round of ideal quantum error correction and logical read-out. This feature is enforced by \cref{eq:displacement_property} and appears as a state-dependent $e^{i\pi k_{2}/\sqrt{2}}$ phase when defining the subsystem basis states in terms of Zak states, see \cref{eq:1_mode_defn}. We can use the connection to error correction to justify the use of the stabilizer subsystem decomposition in the analysis of GKP codes, as we will now do in the rest of the manuscript.

\subsection{Properties}\label{subsec:properties}

Now that we have defined the stabilizer subsystem decomposition for the square GKP code, we outline its key properties. We begin by presenting the quasi-periodic boundary conditions of $\mathcal{S}$ in \cref{subsubsec:BCs}, which provide an intuitive picture of how uncorrectable errors on the oscillator cause logical errors in the logical subsystem $\mathcal{L}$. In \cref{subsubsec:states}, we provide examples of states decomposed into the subsystem decomposition. In particular, the decomposition of approximate GKP codestates follows a simplified version of the general method developed in \cref{sec:errors} for numerical simulations of GKP codestates. Finally in \cref{subsubsec:operators}, we decompose examples of operators, including logical Clifford operators, into the subsystem decomposition, and discuss how operators that do not decompose into tensor products can spread errors in the GKP code.

\subsubsection{Boundary Conditions}\label{subsubsec:BCs}
We begin by noting that the stabilizer states $\ket{\mu,k_{1},k_{2}}$ obey quasi-periodic boundary conditions given by
\begin{subequations}\label{eq:sq_BCs}
\begin{align}
\ket{\mu,k_{1}+1/\sqrt{2},k_{2}}&=e^{-i\pi k_{2}/\sqrt{2}}\ket{\mu\oplus 1,k_{1},k_{2}},\label{eq:sq_XBC}\\
\ket{\mu,k_{1},k_{2}+1/\sqrt{2}}&=e^{i\pi k_{1}/\sqrt{2}}(-1)^{\mu}\ket{\mu,k_{1},k_{2}},\label{eq:sq_ZBC}
\end{align}
\end{subequations}
where here $\oplus$ denotes addition mod 2. These are analogous to the Zak state boundary conditions \cref{eq:1_mode_Zak_BCs}, except that the boundary conditions also affect the logical information. In particular, \cref{eq:sq_XBC} applies a Pauli $\hat{X}$ operator to the logical information while \cref{eq:sq_ZBC} applies a Pauli $\hat{Z}$.

For illustrative purposes, consider a toy error model consisting of a random walk of displacement errors applied to an ideal square GKP codestate $\ket{\bar{\psi}}$, as depicted in \cref{fig:square_Zak}. The logical information in the state remains unchanged as long as the random walk does not cross a boundary, i.e.~while the error remains correctable. Once it crosses a boundary, applying \Cref{eq:sq_XBC} or (\ref{eq:sq_ZBC}) causes a logical Pauli operator to be applied to the logical subsystem, corresponding to a logical error on the state, reflecting the fact that the correctable error has now become uncorrectable. The applied logical Pauli operator is identical to the logical error that would be applied if an ideal decoder acted on the displaced codestate.

\subsubsection{States}\label{subsubsec:states}

Arbitrary single-mode CV states can be decomposed into the square subsystem decomposition using \Cref{eq:Zak_states,eq:1_mode_codestates_Zak}. For example, ideal GKP codestates $\ket{\bar{\mu}}_{\text{sq}} = \ket{\mu}\otimes\ket{0,0}$ are tensor product states by definition. Position eigenstates $\ket{x}_{q}$ and momentum eigenstates $\ket{x}_{p}$ are also tensor product states given by
\begin{subequations}\label{eq:sq_qp_eigenstates}
\begin{align}
\ket{x}_{q} &=
\frac{1}{\sqrt[4]{\pi}}\ket{\mu_{x}}\otimes
\int_{-\frac{1}{2\sqrt{2}}}^{\frac{1}{2\sqrt{2}}}\!\! dk_{2}\,e^{-i\pi(k_{x}+\sqrt{2}n_{x})k_{2}}\ket{k_{x},k_{2}}\!,\\
\ket{x}_{p} &=
\frac{1}{\sqrt[4]{\pi}}\ket{\pm_{x}}\otimes
\int_{-\frac{1}{2\sqrt{2}}}^{\frac{1}{2\sqrt{2}}}\!\! dk_{1}\,
e^{i\pi(k_{x}+\sqrt{2}n_{x})k_{1}}\ket{k_{1},k_{x}}\!,
\end{align}
\end{subequations}
where we decompose $x=\sqrt{2\pi}k_{x}+\sqrt{\pi}n_{x}$ such that $k_{x}\in\big({-}2^{-3/2},2^{-3/2}\big]$ and $n_{x}\in\mathbb{Z}$, $\mu_{x}=n_{x}$ (mod 2), and we write $\ket{\pm_{x}}=\ket{+}$ if $\mu_{x}=0$ and $\ket{\pm_{x}}=\ket{-}$ if $\mu_{x}=1$. Intuitively, the position eigenstate $\ket{x}_{q}$ corresponds to a product state with logical subsystem state $\ket{0}$ ($\ket{1}$, respectively) if $x$ rounds to an even (odd) multiple of $\sqrt{\pi}$. Similarly, the momentum eigenstate $\ket{x}_{p}$ corresponds to a product state with logical subsystem state $\ket{+}$ ($\ket{-}$, respectively) if $x$ rounds to an even (odd) multiple of $\sqrt{\pi}$.

In contrast, approximate codestates are ``entangled’’ across the two subsystems. We define approximate codestates by $\ket{\bar{\psi}_{\Delta}}\propto e^{-\Delta^{2}\hat{a}^{\dag}\hat{a}}\ket{\bar{\psi}}$ with constant of proportionality such that $\ket{\bar{\psi}_{\Delta}}$ is normalized, and where $e^{-\Delta^{2}\hat{a}^{\dag}\hat{a}}$ is the non-unitary \textit{envelope} operator. To find the analytical form of $\ket{\bar{\psi}_{\Delta}}$ in the subsystem decomposition, we first utilize the characteristic function of the envelope operator (see \cref{sec:Gaussian_chi_representation})
\begin{equation}\label{eq:envelope_op}
    e^{-\Delta^{2}\hat{a}^{\dag}\hat{a}}\propto\int_{\mathrlap{\mathbb{R}}}\;dv_{1}dv_{2}\,e^{-\frac{\pi}{2}\mathrm{coth}\big(\!\frac{\Delta^{2}}{2}\!\big)\left(v_{1}^{2}+v_{2}^{2}\right)}\hat{W}(v_{1},v_{2}).
\end{equation}
With the envelope operator written in this form it is straight-forward to apply it to an ideal codestate $\ket{\psi}\otimes\ket{0,0}$ using \cref{eq:displacement_property}. However, since the integral in \cref{eq:envelope_op} is over $v_{1},v_{2}\in\mathbb{R}$, we must apply the boundary conditions \cref{eq:sq_BCs} to obtain a valid subsystem decomposition, giving
\begin{multline}\label{eq:sq_approx}
\ket{\bar{\psi}_{\Delta}}\propto\sum_{\vect{s}\in\mathbb{Z}^{2}}\hat{P}(\vect{s})\ket{\psi}\otimes\int \!d^{2}\vect{v}\bigg(e^{-\frac{\pi}{2}\mathrm{coth}\big(\!\frac{\Delta^{2}}{2}\!\big)|\vect{v}+\vect{s}/\!\sqrt{2}|^{2}}\\
\times e^{i\pi(v_{1}s_{2}-v_{2}s_{1})/\!\sqrt{2}}\ket{v_{1},v_{2}}\!\bigg),
\end{multline}
where $\hat{P}(\vect{s})=e^{i\pi s_{1}s_{2}/2}\hat{X}^{s_{1}}\hat{Z}^{s_{2}}$, the region of integration is $v_{1},v_{2}\in\big({-}2^{-3/2},2^{-3/2}\big]$, and we have written $\vect{v}=(v_{1},v_{2})$. Note that the boundary conditions introduce logical Pauli operators acting on the logical subsystem, reflecting the fact that the envelope operator introduces errors on the ideal codestate.

\begin{figure}
    \centering
    \includegraphics{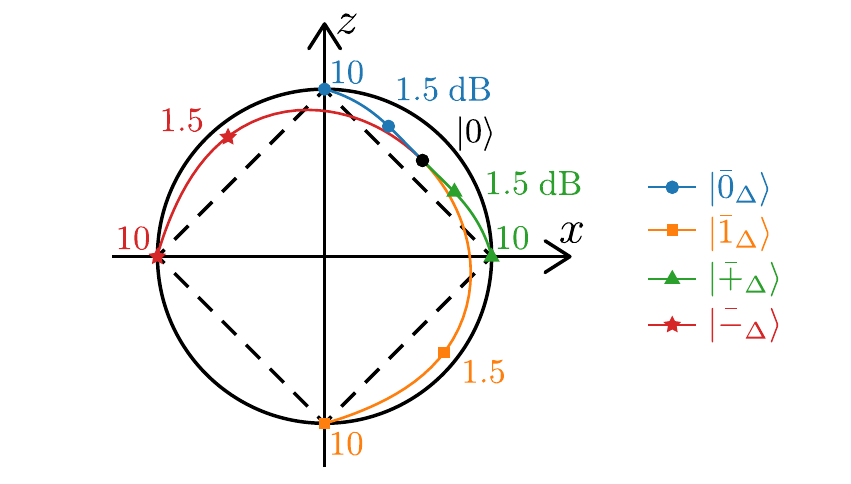}
    \caption{(Color) Decoded states $\mathrm{tr}_{\mathcal{S}}\big(\!\ket{\bar{\psi}_{\Delta}}\!\bra{\bar{\psi}_{\Delta}}\!\big)$ for $\ket{\psi}=\ket{0},\ket{1},\ket{+},\ket{-}$, where $\ket{\bar{\psi}_{\Delta}}$ is an approximate single-mode square GKP codestate. The decoded states are plotted on the $xz$-plane of the Bloch sphere (solid outline) as a function of $\Delta$ (labelled on the plot). As $\Delta_{\mathrm{dB}}\rightarrow+\infty$ ($\Delta\rightarrow0$), each state approaches the ideal logical $\ket{\psi}_{\mathcal{L}}$ state respectively; while as $\Delta_{\mathrm{dB}}\rightarrow-\infty$ ($\Delta\rightarrow+\infty$), each state approaches the vacuum state $\ket{0}$, which is outside the stabilizer octahedron (dotted line) and is }distillable to a magic state~\cite{Baragiola19}.
    \label{fig:bloch_sphere}
\end{figure}

To quantify the logical information stored in a state, we can apply the partial trace over $\mathcal{S}$, which gives an expression of the form
\begin{equation}\label{eq:sq_approx_partial_trace}
    \mathrm{tr}_{\mathcal{S}}\big(\ket{\bar{\psi}_{\Delta}}\!\bra{\bar{\psi}_{\Delta}}\big)\propto\sum_{\vect{s},\vect{t}\in\mathbb{Z}^{2}}I^{\Delta}_{\vect{s},\vect{t}}\hat{P}(\vect{s})\ket{\psi}\!\bra{\psi}\hat{P}(\vect{t}).
\end{equation}
We derive \cref{eq:sq_approx_partial_trace} and provide the analytical form of $I_{\vect{s},\vect{t}}^{\Delta}$ in \cref{sec:analytic_partial_trace} due to the length of the equations. To numerically evaluate \cref{eq:sq_approx_partial_trace} we can truncate the infinite sums over $\vect{s},\vect{t}\in\mathbb{Z}^{2}$, which is justified as long as $|I_{\vect{s},\vect{t}}^{\Delta}|\rightarrow0$ sufficiently fast as $|\vect{s}|,|\vect{t}|\rightarrow\infty$. Importantly, numerically evaluating \cref{eq:sq_approx_partial_trace} also becomes \textit{easier} as $\Delta\rightarrow0$ since $|I_{\vect{s},\vect{t}}^{\Delta}|$ converges to zero faster as $\Delta$ becomes small, requiring fewer terms in the sum to be included. Intuitively, this is because the characteristic function of the envelope operator \cref{eq:envelope_op} decays exponentially away from the origin, and the rate of decay increases as $\Delta\rightarrow0$.

In \cref{fig:bloch_sphere}, we plot the logical state given by \cref{eq:sq_approx_partial_trace}, where we have quoted $\Delta$ in decibels using the formula $\Delta_{\mathrm{dB}}=-10\mathrm{log}_{10}(\Delta^{2})$. In the limit $\Delta\rightarrow0$, the approximate codestate $\ket{\bar{\psi}_{\Delta}}$ approaches the ideal GKP codestate $\ket{\bar{\psi}}$ and its partial trace approaches the pure state $\ket{\psi}$. In the limit $\Delta\rightarrow\infty$, $e^{-\Delta^{2}\hat{a}^{\dag}\hat{a}}\rightarrow\ket{0}\!\bra{0}$, the projector onto the vacuum state, and as such $\ket{\bar{\psi}_{\Delta}}\rightarrow\ket{0}$. In the square code, the logical information stored in the vacuum state is a mixed state which lies outside the stabilizer octahedron, and has been shown to be distillable to a magic state \cite{Baragiola19}.

In fact, as we show in \cref{sec:errors}, the procedure we have just followed to obtain the partial trace of $\ket{\bar{\psi}_{\Delta}}$ can be generalized to apply quantum channels $\mathcal{E}$ to ideal GKP codestates. In particular, one needs to obtain the characteristic function of the map, apply the boundary conditions of the code, and then take the partial trace. The resulting object is an effective \textit{logical} channel that reflects the change in logical information stored in the state under the action of $\mathcal{E}$. Moreover, the logical channel can be obtained numerically by truncating each of the infinite series that arise from the boundary conditions so long as the characteristic function of the channel tends to zero sufficiently fast. The details and results of this method are discussed in more detail in \cref{sec:errors}.

\subsubsection{Operators}\label{subsubsec:operators}

We can also decompose arbitrary CV operators into the subsystem decomposition. In order to do so, we first define the stabilizer subsystem operators $\hat{k}_{1},\hat{k}_{2}$, which act on the Zak basis states $\ket{k_{1},k_{2}}\in\mathcal{S}$ via
\begin{align}
    \hat{k}_{1}\ket{k_{1},k_{2}}&=k_{1}\ket{k_{1},k_{2}},&\hat{k}_{2}\ket{k_{1},k_{2}}&=k_{2}\ket{k_{1},k_{2}}.
\end{align}
Since $\hat{k}_{1}$ and $\hat{k}_{2}$ are simultaneously diagonalizable we also have $[\hat{k}_{1},\hat{k}_{2}]=0$. Moreover, the eigenvalues of $\hat{k}_{1},\hat{k}_{2}$ lie in the range $({-}2^{-3/2},2^{-3/2}]$.

From \cref{eq:1_mode_stabilizer_eigenvalues}, we see that the stabilizer generators are product operators that act trivially on the logical subsystem:
\begin{subequations}
\begin{align}
\hat{S}_{1}&=e^{-2i\sqrt{\pi}\hat{p}}=\hat{I}\otimes e^{-2\sqrt{2}i\pi \hat{k}_{2}},\\
\hat{S}_{2}&=e^{2i\sqrt{\pi}\hat{q}}=\hat{I}\otimes e^{2\sqrt{2}i\pi \hat{k}_{1}}.
\end{align}
\end{subequations}
The operators $\sqrt{2\pi}\hat{k}_{1}=\hat{q}$ (mod $\sqrt{\pi}$) and $\sqrt{2\pi}\hat{k}_{2}=\hat{p}$ (mod $\sqrt{\pi}$) can also be interpreted as modular quadrature operators.

It is straightforward to show directly that logical Pauli operators decompose to tensor products given by
\begin{subequations}
\begin{align}
\bar{X}&=e^{-i\sqrt{\pi}\hat{p}}=\hat{X}\otimes e^{-\sqrt{2}i\pi \hat{k}_{2}},\\
\bar{Z}&=e^{i\sqrt{\pi}\hat{q}}=\hat{Z}\otimes e^{\sqrt{2}i\pi \hat{k}_{1}},
\end{align}
\end{subequations}
where $\bar{X}$, $\bar{Z}$ are logical Pauli operators acting on $\mathcal{H}$ while $\hat{X}$, $\hat{Z}$ are Pauli operators acting on $\mathcal{L}\cong\mathbb{C}^{2}$. Note that the non-trivial action of $\bar{X}$, $\bar{Z}$ on the stabilizer subsystem is necessary for them to satisfy the identities $\bar{X}^{2}=\hat{S}_{1}$, $\bar{Z}^{2}=\hat{S}_{2}$. When a logical Pauli operator is applied to a state, the logical subsystem is transformed exactly by the corresponding Pauli operator, and the distribution of the state in the stabilizer subsystem is unchanged since $e^{-\sqrt{2}i\pi\hat{k}_{2}}$ only multiplies each Zak basis state by a phase. As such, $\bar{X}$ and $\bar{Z}$ can be considered as ideal Pauli operators when acting as a gate.

When considered as a measurement operator however, the phase on the stabilizer subsystem can affect the measurement outcome. In particular, the phase ensures that the spectrum of $\bar{X},\bar{Z}$ is indeed the set of modulus 1 complex numbers, consistent with the fact that displacements are unitary but not Hermitian operators. Alternatively, one can define the Hermitian and unitary ideal Pauli measurement operators $\hat{X}_{\text{m}}=\hat{X}\otimes\hat{I}$ and $\hat{Z}_{\text{m}}=\hat{Z}\otimes\hat{I}$. These operators can in theory be measured by performing an ideal round of error correction and then performing a measurement of the original Pauli operators $\bar{X},\bar{Z}$, since $k_{1},k_{2}=0$ in the ideal codespace. We note that such a measurement is impossible in practice as the ideal round of error correction requires the preparation of an ideal GKP codeword. It is interesting to note that the measurement operators $\hat{X}_{\text{m}}$, $\hat{Z}_{\text{m}}$ do \textit{not}, in general, coincide with ``binned quadrature operator'' measurements, which we discuss in more detail in \cref{sec:logical_state_tomography}.

Unlike logical Pauli operators, displacements are not tensor product operators in the subsystem decomposition in general. As an example, consider the displacement $\hat{W}(2^{-3/2},0)=e^{-i\sqrt{\pi}\hat{p}/2}=\sqrt{\bar{X}}$, which acts on states in the subsystem decomposition as
\begin{multline}
    \sqrt{\bar{X}}\ket{\psi}\otimes \ket{k_{1},k_{2}}=\\
    \left\{
    \begin{aligned}
    \ket{\psi}&\otimes e^{-i\pi k_{2}/\!\sqrt{8}}\big|k_{1}+2^{-3/2},k_{2}\big\rangle,&k_{1}&\leq0,\\
    \hat{X}\!\ket{\psi}&\otimes e^{-3i\pi k_{2}/\!\sqrt{8}}\big|k_{1}-2^{-3/2},k_{2}\big\rangle,&k_{1}&>0.
    \end{aligned}
    \right.
\end{multline}
Note that the states with $k_{1}>0$ have been mapped over the $\hat{X}$-boundary [\cref{eq:sq_XBC}], while the remaining $k_{1}\leq0$ states have not. $\sqrt{\bar{X}}$ is thus an entangling operator across the subsystem decomposition.

For the square code, the Fourier transform operator $e^{i\pi\hat{a}^{\dag}\hat{a}/2}$ is a tensor product operator given by
\begin{equation}\label{eq:sq_Had}
    e^{i\pi\hat{a}^{\dag}\hat{a}/2}=\hat{H}\otimes\hat{R}(\pi/2)
\end{equation}
where $\hat{H}=(\hat{X}+\hat{Z})/\sqrt{2}$ is the Hadamard operator, and $\hat{R}(\pi/2)\ket{k_{1},k_{2}}=\ket{-k_{2},k_{1}}$ rotates the vector $(k_{1},k_{2})$ anticlockwise by $\pi/2$. This reflects the fact that $e^{i\pi\hat{a}^{\dag}\hat{a}/2}=\bar{H}$ is a logical Hadamard operator on the square GKP code, and demonstrates that the stabilizer subsystem decomposition is symmetric in $\hat{q}$ and $\hat{p}$.

However, logical Clifford gates are \textit{not} tensor product operators in general. For example, consider the logical phase gate $\bar{S}=e^{i\hat{q}^{2}\!/2}$, which implements the gate $\hat{S}=\mathrm{diag}(1,i)$. It is straightforward to show from \cref{eq:Zak_translation_property,eq:1_mode_Zak_BCs,eq:1_mode_defn} that
\begin{equation}\label{eq:sq_phase_stabilizer_states}
    \bar{S}\ket{\mu,k_{1},k_{2}}=i^{\mu}\ket{\mu,k_{1},k_{1}+k_{2}},
\end{equation}
where $\mu=0,1$. Therefore, $\bar{S}$ acts as an ideal phase gate only on Zak basis states $\ket{\mu,k_{1},k_{2}}$ with $k_{1}+k_{2}\in({-}2^{-3/2},2^{-3/2}]$. States lying outside this range incur an additional $\bar{Z}$ logical error due to the boundary condition \cref{eq:sq_ZBC}. In particular, we have
\begin{equation}\label{eq:sq_phase}
\begin{multlined}
    \bar{S}\ket{\psi}\otimes\ket{k_{1},k_{2}}=\\
    \!\left\{\!\!
    \begin{aligned}
    \hat{S}^{\dag}\!\ket{\psi}&\otimes e^{i\pi k_{1}/\!\sqrt{2}}\big|k_{1},k_{1}{+}k_{2}{-}2^{-1/2}\big\rangle,&k_{1}{+}k_{2}&{>}2^{-3/2},\\
    \hat{S}^{\dag}\!\ket{\psi}&\otimes e^{-i\pi k_{1}/\!\sqrt{2}}\big|k_{1},k_{1}{+}k_{2}{+}2^{-1/2}\big\rangle,&k_{1}{+}k_{2}&{\leq}{-}2^{-3/2},\\
    \hat{S}\!\ket{\psi}&\otimes \big|k_{1},k_{1}+k_{2}\big\rangle,&\text{e}&\text{lse}.
    \end{aligned}
    \right.
\end{multlined}
\end{equation}
In \cref{sec:gates} we show that logical Clifford gates by default do not act as tensor product acting on general GKP codes, including two-qubit gates such as the logical controlled-NOT and controlled-$Z$ gates acting on the square GKP code. Indeed, only a few special Clifford gates do in fact decompose as tensor products, such as the Hadamard gate in the square GKP code and the permutation gate $\hat{H}\hat{S}^{\dag}$ in the hexagonal GKP code. Intuitively, a Clifford gate will decompose as a tensor product operator only if it does not deform the decoding ``primitive cell'' of the code, an idea that we discuss more formally in \cref{sec:Zak,sec:gates}.

In the remainder of the paper, we generalize the stabilizer subsystem decomposition to general multi-mode GKP codes (\cref{sec:notation,sec:definition}), and describe in more detail the applications of the subsystem decomposition to Clifford gates (\cref{sec:gates}) and the modelling of noise (\cref{sec:errors}). Readers that are only interested in modelling noise can skip \cref{sec:notation,sec:definition,sec:Zak,sec:gates} and resume reading at \cref{sec:errors}.

\section{General GKP Codes}\label{sec:notation}

Now that we have introduced the stabilizer subsystem decomposition for the square GKP qubit code, we discuss the generalization of the stabilizer subsystem decomposition to general multi-mode GKP qudit codes in the following three sections. In this section, we introduce the notation we will use throughout the rest of the paper, and review the properties of multi-mode GKP codes. Then in \cref{sec:definition} we define the stabilizer subsystem decomposition in general. We provide a method of decomposing arbitrary CV states into a general stabilizer subsystem decomposition in \cref{sec:Zak}, before moving on to applications of the stabilizer subsystem decomposition in the remainder of the paper.

\subsection{Preliminaries}

We start with a discrete-variable system consisting of the tensor product of $n$ finite-dimensional Hilbert spaces each with dimension $d_{j}$, $j=1,\dots,n$, which we write as a single vector $\vect{d}=(d_{1},\dots,d_{n})$.
We will allow for $d_{j}=1$ for any of the dimensions, which we refer to as a \textquotedblleft qunaught\textquotedblright\, since no logical information can be stored in the system. The Hilbert space of the system, $\mathcal{H}_{\vect{d}}=\bigotimes_{j=1}^{n}\mathbb{C}^{d_{j}}$, is spanned by an orthonormal computational basis $\ket{\mu}$, where $\mu=(\mu_{1},\dots,\mu_{n})\in \bigoplus_{j=1}^{n}\mathbb{Z}_{d_{j}}$ is a dit-string. On each qudit we define the Pauli $\hat{X}_{(d)}$ and $\hat{Z}_{(d)}$ operators that have action 
\begin{subequations}\label{eq:PauliXZ}
\begin{align}
\hat{X}_{(d)}\ket{a}={}&\ket{a+1~(\mathrm{mod}\,d)},\\
\hat{Z}_{(d)}\ket{a}={}&e^{2i\pi  a/d}\ket{a}.
\end{align}
\end{subequations}
From now on we will no longer explicitly write $(d)$ to indicate the dimension of the Pauli operators. For a general Pauli operator acting on the whole system $\mathcal{H}_{\vect{d}}$ we use the notation
\begin{equation}\label{eq:general_Pauli}
\hat{P}_{\vect{d}}(\vect{s})=\bigotimes_{j=1}^{n}\,\exp\!\bigg(\frac{i\pi}{d_{j}} s_{j}s_{j+n}\!\bigg)\hat{X}^{s_{j}}\hat{Z}^{s_{j+n}},
\end{equation}
where $\vect{s}\in\mathbb{Z}^{2n}$, c.f.~$\hat{P}(\vect{s})$ as in \cref{eq:sq_approx}. We allow every component of the vector $\vect{s}=(s_{1},\dots,s_{2n})$ to be an integer with unrestricted range for later convenience. Because of this, the Pauli operators $\hat{P}_{\vect{d}}(\vect{s})$ and $\hat{P}_{\vect{d}}(\vect{s}')$ may be identical even if $\vect{s}\neq\vect{s}'$.

Next, consider a CV system consisting of $n$ modes with Hilbert space $\mathcal{H}$. Such a system can be described by $n$ position and momentum operators, which we write in a column vector $\vect{\hat{\xi}}=[\hat{q}_{1}~\cdots~\hat{q}_{n}~\hat{p}_{1}~\cdots~\hat{p}_{n}]^{T}$. These obey the canonical commutation relations $[\hat q_j,\hat p_{j'}] = i\delta_{jj'}$.

We define the $n$-mode displacement operators
\begin{equation}\label{eq:Wv}
\hat{W}(\vect{v}) = \mathrm{exp}\big(\sqrt{2\pi}i\,\vect{\hat{\xi}}^{T}\!\Omega\,\vect{v}\big),
\end{equation}
for $\vect{v}\in\mathbb{R}^{2n}$, where
\begin{equation}
\Omega=\begin{bmatrix}0_{n}&I_{n}\\-I_{n}&0_{n}\end{bmatrix}
\end{equation}
defines the standard symplectic bilinear form $\vect{u}^{T}\Omega\vect{v}$
in $\mathbb{R}^{2n}$. The displacement operators obey the commutation relation
\begin{equation}
    \big\llbracket \hat{W}(\vect{u}),\hat{W}(\vect{v})\big\rrbracket=\exp\big({-}2i\pi \vect{u}^{T}\Omega\vect{v}\big),
\end{equation}
c.f.~\cref{eq:1_mode_W,eq:1_mode_W_commutation}. Moreover, they ``displace'' the position and momentum operators such that
\begin{equation}
    \hat{W}(\vect{v})^{\dag}\vect{\hat{\xi}}\hat{W}(\vect{v})=\vect{\hat{\xi}}+\sqrt{2\pi}\vect{v},
\end{equation}
c.f.~\cref{eq:1_mode_W_qp}, where on the left-hand side the displacement operators are acting component-wise on the vector $\vect{\hat{\xi}}$.

We define the $n$-mode Zak states
\begin{equation}\label{eq:Zak_multi_mode}
\ket{\vect{k}}_{\vect{a}}=
\ket{k_1,\dots,k_{2n}}_{\vect{a}} =
\bigotimes_{j=1}^{n}\ket{k_{j},k_{j+n}}_{a_{j}},
\end{equation}
where $\vect{a}=(a_{1},\dots,a_{n})$ is a list of constants $a_{j}>0$. Each Zak state is a simultaneous eigenstate of the displacement operators $\hat{W}(a_{j}\vect{e}_{j})$ and $\hat{W}(\vect{e}_{n+j}/a_{j})$ for $j=1,\dots,n$, where $\vect{e}_{J}\in\mathbb{R}^{2n}$ is the vector with a 1 in the $J$-th component and 0's elsewhere.
Restricting each component $k_{j}$ to an interval of length $a_{j}$ and each component $k_{j+n}$ to an interval of length $1/a_{j}$ forms a basis of Zak states.

We also consider Gaussian unitary operators $\hat{U}_{S}$ that are parameterized by a $2n\times2n$ real symplectic matrix $S$ with $S^{T}\Omega S=\Omega$. The action of a Gaussian unitary on the position and momentum operators is defined as the linear transformation
\begin{equation}\label{eq:Gaussian_unitary}
    \hat{U}_{S}^{\dag}\vect{\hat{\xi}}\hat{U}_{S}^{}=S\vect{\hat{\xi}}.
\end{equation}
Given a symplectic matrix $S$, the operator $\hat{U}_{S}$ can be found from the unitary metaplectic representation of $S$~\cite{Simon93,Dutta95}. Every Gaussian operator $\hat{U}_{S}$ can be written as the product of unitaries generated by Hamiltonians quadratic in $\vect{\hat{\xi}}$. Alternatively, we can interpret $S$ as defining a canonically transformed set of modes $\vect{\hat{\Xi}}=S^{-1}\vect{\hat{\xi}}$ such that we can write $\hat{U}_{S}^{}\hat{W}(\vect{v})\hat{U}_{S}^{\dag}=\hat{W}(S\vect{v})=\mathrm{exp}(\sqrt{2\pi}i\,\vect{\hat{\Xi}}{}^{T}\Omega\vect{v})$.

\subsection{Multi-mode GKP encodings}

We can now introduce multi-mode qudit GKP codes, which encode a discrete variables system $\mathcal{H}_{\vect{d}}$ consisting of $k$ qudits and $k-n$ qunaught states into an $n$-mode continuous variables system $\mathcal{H}$. To specify the GKP encoding we provide two pieces of information $(\Sigma,\vect d)$, where $\Sigma$ is a $2n\times 2n$ real symplectic matrix, and $\vect d$ is a list of dimensions of length $n$ consisting of $k$ elements that are integers greater than or equal to 2 and $n-k$ elements equal to 1. Together, they specify a set of $2n$ mutually commuting stabilizer generators
\begin{align}\label{eq:SJ}
\hat S_J &= \hat{W}(\vect{m}_{J}),& \vect{m}_{J}&=d_{J\;(\mathrm{mod}\,n)}^{1/2}\,(\Sigma)_{J},
\end{align}
where $(\Sigma)_{J}$ is the $J$-th column of $\Sigma$, such that the set of operators $\hat{S}^{}_{J}$ and $\hat{S}_{J}^{\dag}$ generate the stabilizer group. Note our index convention of using $j$ when the index runs from $1$ to $n$, and $J$ when the index runs from $1$ to $2n$. Also note that the (mod $n$) in \cref{eq:SJ} is needed only due to our definition of $\vect{d}$ being length $n$ (and not $2n$). The codespace is defined by the simultaneous $+1$-eigenspace of the stabilizer generators. We define the logical Pauli operators
\begin{align}\label{eq:logical_Pauli}
\bar{X}_{j}&=\hat{W}(\vect{\bar{m}}_{j}),&
\bar{Z}_{j}&=\hat{W}(\vect{\bar{m}}_{j+n}),
\end{align}
where $\vect{\bar{m}}_{J}=\vect{m}_{J}/d_{J\,(\mathrm{mod}\,n)}$. We use the bar in the notation $\vect{\bar{m}}_{J}$ both to denote that the vector $\vect{\bar{m}}_{J}$ is ``dual'' to the vector $\vect{m}_{J}$ (in the lattice sense, as will be explained below), and to remind the reader that $\hat{W}(\vect{\bar{m}}_{J})$ corresponds to a logical Pauli operator. Note that on the qunaught modes where $d_{j}=1$ the logical Pauli operators $\bar{Z}_{j}$ and $\bar{X}_{j}$ coincide with the stabilizers $\hat{S}_{j}$ and $\hat{S}_{j+n}$.

As an illustrative example, consider the case of a single qudit with dimension $d$ encoded in a single-mode square GKP code. The parameters for this encoding are $\big(\Sigma_\text{sq},(d)\big)$, with $\Sigma_\text{sq} = I_2$, such that the stabilizers and logical Paulis are simply
\begin{subequations}\label{eq:sq_qudit_example}
\begin{align}
    \hat S_1 &= \hat{W}(\sqrt{d}\,{\vect{e}}_{1})=e^{-i\sqrt{2\pi d}\,\hat{p}}, \\
    \hat S_2 &= \hat{W}(\sqrt{d}\,{\vect{e}}_{2})=e^{i\sqrt{2\pi d}\,\hat{q}},\\
    \bar X &=  \hat{W}({\vect{e}}_{1}/\sqrt{d})=e^{-i\sqrt{2\pi/d}\,\hat{p}}, \\
    \bar Z &= \hat{W}({\vect{e}}_{2}/\sqrt{d})=e^{i\sqrt{2\pi/d}\,\hat{q}},
\end{align}
\end{subequations}
where $\vect{e}_{1}=[1,0]^{T}$ and $\vect{e}_{2}=[0,1]^{T}$. The encoding is called square because of the square lattice generated by $\vect{m}_1=\sqrt{d}\,{\vect{e}}_{1}$ and $\vect{m}_2=\sqrt{d}\,{\vect{e}}_{2}$.

\begin{figure}
    \centering
    \includegraphics{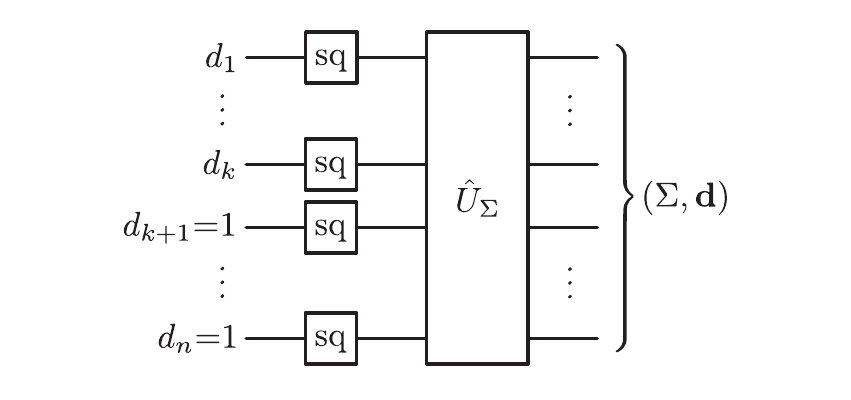}
    \caption{A circuit representing the encoding of $k$ qudits into an ideal GKP code described by $(\Sigma,\vect{d})$. On the left-hand side, a state is described in the Hilbert space $\mathcal{H}_{\vect{d}}$ consisting of a tensor product of $k$ qudits and $k-n$ qunaught states. The qudits are encoded first into the ideal codespace of $n$ independent square GKP codes (each labelled sq), followed by a Gaussian unitary operator $\hat{U}_{\Sigma}$.
    }
    \label{fig:encoding_circuit}
\end{figure}

One way to understand a general multi-mode GKP encoding defined by arbitrary $(\Sigma,\vect{d})$ is as follows. Since $\Sigma$ is a symplectic matrix, it defines a new set of canonically transformed modes
\begin{equation}\label{eq:logical_modes}
\vect{\bar{\xi}} = \Sigma^{-1} \vect{\hat \xi},
\end{equation}
which we refer to as the \emph{logical} modes. Written in terms of the logical modes, the stabilizer generators and logical Paulis take the form
\begin{subequations}\label{eq:logical_mode_ops}
\begin{align}
\hat S_j &= e^{-i \sqrt{2\pi d_{j}} \bar{p}_{j}}, & \hat{S}_{j+n}&=e^{i\sqrt{2\pi d_{j}}\bar{q}_{j}},\\
\bar{X}_{j}&= e^{-i\sqrt{2\pi/d_{j}} \bar{p}_{j}}, & \bar{Z}_{j} &= e^{i\sqrt{2\pi/d_{j}}\bar{q}_{j}},
\end{align}
\end{subequations}
which we recognize as $n$ independent square-GKP encodings into the logical modes $\bar{q}_j, \bar{p}_j$.
This picture
is illustrated in~\cref{fig:encoding_circuit}, where $k$ qudits and $n-k$ qunaught states are first encoded into $n$ square-lattice logical GKP modes, before applying a unitary Gaussian transformation $\hat U_{\Sigma}$.

To clarify the role of the qunaught states, note that the corresponding logical mode is constrained to a unique state that could equivalently be defined as the $+1$ eigenstate of $\bar Z$ for a square lattice GKP code with arbitrary $d$. The qunaught modes thus play precisely the same role as stabilizers in conventional qudit stabilizer codes~\cite{Poulin05}, and can be used to provide additional protection against errors. A special case is when $\hat U_\Sigma$ can be decomposed into a tensor product of single-mode Gaussian unitaries, followed by a logical Clifford circuit acting on the encoded GKP qudits. This case can be understood as a concatenation of $n$ single-mode GKP codes with a conventional discrete qudit stabilizer code.
For example, the GKP-surface code~\cite{Vuillot19,Noh20} belongs to this class, where there are $n$ physical modes and $n-1$ logical qunaught modes (corresponding to the surface code stabilizer generators), leaving a single logical GKP mode encoding a qubit with $d=2$.


\subsection{GKP lattices and primitive cell decoding}\label{subsec:latticeandcell}

A useful way of understanding multi-mode GKP codes is with lattice theory, which was first applied to GKP codes in Refs.~\cite{Gottesman01,Harrington04} and revisited recently in Refs.~\cite{Conrad22,Royer22}.
We define the GKP lattice
\begin{equation}\Lambda=\bigg\{\vect{\ell}=\sum_{J=1}^{2n}s_{J}\vect{m}_{J}\;\bigg|\,\;\vect{s}\in\mathbb{Z}^{2n}\bigg\},
\end{equation}
which we say is generated by the vectors $\vect{m}_{J}$. $\Lambda$ has the property that $\hat{W}(\vect{\ell})$ is a stabilizer (up to a $\pm1$ phase) for every $\vect{\ell}\in\Lambda$. The symplectic dual lattice of $\Lambda$ is defined by 
\begin{equation}
    \bar{\Lambda}=\big\{\vect{\bar{\ell}}~\big|~\vect{\bar{\ell}}^{\,T}\Omega\vect{\ell}\in\mathbb{Z},\;\forall\,\vect{\ell}\in\Lambda\big\}.
\end{equation}
$\bar{\Lambda}$ is generated by the dual vectors $\vect{\bar{m}}_{J}$ and has the property that $\hat{W}\big(\vect{\bar{\ell}}\,\big)$ is a (possibly trivial) logical Pauli operator for every $\vect{\bar{\ell}}\in\bar{\Lambda}$. For example, the single-mode square qudit code defined in \cref{eq:sq_qudit_example} has GKP stabilizer lattice $\Lambda=\sqrt{d}\,\mathbb{Z}^{2}$ and dual lattice $\bar{\Lambda}=\mathbb{Z}^{2}/\sqrt{d}$.

Conversely, if one starts with a lattice $\Lambda$ which defines a GKP code, one can always find the corresponding pair $(\Sigma,\vect{d})$. For the lattice $\Lambda$ to define a GKP code, it must be full-rank and symplectic, i.e.\ $\vect{\ell}^{T}\Omega\vect{\ell}'\in\mathbb{Z},\;\forall\;\vect{\ell},\vect{\ell}'\in\Lambda$. For consistency with previous literature~\cite{Gottesman01,Harrington04,Conrad22,Royer22}, we define the lattice generator matrix $M$ whose \textit{rows} are $\vect{m}_{J}^{T}$. There is freedom in the choice of matrix $M$ (or, equivalently, the generators $\vect{m}_{J}$) for the lattice $\Lambda$; in particular, left-multiplication of any integral unimodular matrix $N$ transforms $M\mapsto NM$ via row operations that preserve the lattice. Each choice of $M$ corresponds to a different choice of stabilizer generators, but not all of these choices of stabilizer generator can be expressed in terms of logical modes $\vect{\bar{\xi}}$ as in \cref{eq:logical_modes}. To solve this issue, one can always use Gaussian elimination to find a generator $M$ written in \textit{standard form} such that $M^{T}=\Sigma D^{1/2}$ for a symplectic matrix $\Sigma$ and diagonal matrix $D=\mathrm{diag}(\vect{d},\vect{d})$, which provide the parameters in our original description $(\Sigma,\vect{d})$. We note that the standard form of $M$, and equivalently the choice of $\Sigma$, is not unique for a given lattice $\Lambda$. However, specifying $\Sigma$ fixes the choice of stabilizer generators $\hat S_J$ and the labelling of each logical Pauli operator.

The lattice perspective of GKP codes also provides a convenient description of GKP decoding in terms of primitive cells of the dual lattice. We define a primitive cell $\mathcal{P}$ of the dual lattice $\bar{\Lambda}$ as any subset $\mathcal{P}\subset\mathbb{R}^{2n}$ satisfying the property that any vector $\vect{v}\in\mathbb{R}^{2n}$ can be uniquely written as
\begin{equation}\label{eq:primitive_cell}
    \vect{v}=\vect{\bar \ell} + \{\vect{v}\}_{\mathcal{P}},
\end{equation}
where $\vect{\bar \ell}\in \bar{\Lambda}$ and $\{\vect{v}\}_{\mathcal{P}}\in\mathcal{P}$.

To perform an ideal round of GKP error correction, one first measures the eigenvalues of each stabilizer generator $\hat{W}(\vect{m}_{J})$, giving a complex measurement outcome. By introducing a vector $\vect{v}\in\mathbb{R}^{2n}$, we can write the set of measurement outcomes as $\exp(2i\pi \vect{v}^{T}\Omega\vect{m}_{J})$ for each stabilizer. However, the choice of $\vect{v}$ is unique only up to the addition of vectors in the dual lattice. As such, to uniquely assign a displacement that returns the state to the codespace, one must choose a primitive cell $\mathcal{P}$ of $\bar{\Lambda}$, such that $\vect{v}$ can be written uniquely as $\vect{v}=\vect{\bar{\ell}}+\{\vect{v}\}_{\mathcal{P}}$ for $\vect{\bar{\ell}}\in\bar{\Lambda}$ and $\{\vect{v}\}_{\mathcal{P}}\in\mathcal{P}$.

The choice of $\mathcal{P}$ defines a decoder, in which we pick the unique vector $\vect{v}=\{\vect{v}\}_{\mathcal{P}}\in\mathcal{P}$ that reproduces the syndrome $e^{2i\pi \vect{v}^{T}\Omega\vect{m}_{J}}$ and perform the correction $\hat{W}(\vect{v})^{\dag}$, returning the state to the ideal codespace. The canonical choice of $\mathcal{P}$ is given by the Voronoi cell $\mathcal{V}(\bar{\Lambda})$ of the dual lattice, which contains the set of points closer to the origin than any other point in the lattice. However, other choices of $\mathcal{P}$ can be made, for example, to account for the effects of logical Clifford gates (\cref{sec:gates}) and maximum likelihood decoding~\cite{Conrad22}.

\section{Stabilizer Subsystem Decomposition}\label{sec:definition}

In this section, we describe how to construct the stabilizer subsystem decomposition for arbitrary multi-mode GKP codes.
We do this by first defining the stabilizer states $\ket{\mu,\vect{k}}_{(\Sigma,\vect{d})}$, which are labelled by two variables $\mu\in\bigoplus_{j=1}^{n}\mathbb{Z}_{d_{j}}$ (a dit-string label) and $\vect{k}\in\mathbb{R}^{2n}$, where the former represents logical information and the latter the stabilizer eigenvalues. The stabilizer states span the Hilbert space $\mathcal{H}$, but are not linearly independent, so we limit the range of $\vect{k}$ to a subset of $\mathbb{R}^{2n}$ such that the resulting states form a linearly independent basis. We can then define a subsystem decomposition between the logical $\mathcal{L}$ and stabilizer $\mathcal{S}$ labels. Finally, we outline two key features of the decomposition: the decomposition of stabilizers and logical Paulis as tensor product operators, and the correspondence of the partial trace to a noiseless primitive cell decoding of the GKP code. A practical example of the general framework is provided in \cref{sec:repetition_code}.

\subsection{Stabilizer States}\label{subsec:stabiliser_states}

We begin by stepping through the definition of the stabilizer states. For any pair of parameters $(\Sigma,\vect{d})$, we define the state $\ket{0_{n},\vect{0}_{2n}}_{(\Sigma,\vect{d})}$ as the simultaneous $+1$-eigenstate of the stabilizer group and each logical Pauli $\hat{Z}$ operator $\bar{Z}_{j}$ for $j=1,\dots,\,n$, which corresponds to the ideal GKP state encoding the codeword $\ket{0}^{\otimes n}$.

Next, we define the remaining ideal GKP codewords
\begin{equation}\label{eq:ideal_codewords}
    \ket{\mu,\vect{0}_{2n}}_{(\Sigma,\vect{d})}=\ket{\bar{\mu}}=\bigotimes_{j=1}^{n}\bar{X}_{j}^{\mu_{j}}\ket{0_{n},\vect{0}_{2n}}_{(\Sigma,\vect{d})}\,,
\end{equation}
where $\mu=(\mu_{1},\dots,\mu_{n})\in \bigoplus_{j=1}^{n}\mathbb{Z}_{d_{j}}$ is a dit-string throughout this section. Here, $\bigotimes_{j=1}^{n}\bar{X}_{j}^{\mu_{j}}$ is a product of Pauli $X$ operators such that $\ket{\mu,\vect{0}_{2n}}_{(\Sigma,\vect{d})}$ is the ideal codeword $\ket{\bar{\mu}}$.

Finally, we define the remaining stabilizer states as:
\begin{equation}\label{eq:stabiliser_states}
    \ket{\mu,\vect{k}}_{(\Sigma,\vect{d})}=\hat{W}(\vect{k})\ket{\mu,\vect{0}_{2n}}_{(\Sigma,\vect{d})}
\end{equation}
where $\vect{k}\in\mathbb{R}^{2n}$. The state $\ket{\mu,\vect{k}}_{(\Sigma,\vect{d})}$ can be viewed as an ideal GKP codeword $\ket{\bar{\mu}}$ that has incurred a displacement error $\hat{W}(\vect{k})$. We will also notate
\begin{equation}
    \ket{\psi,\vect{k}}_{(\Sigma,\vect{d})}=\sum_{\mu}c_{\mu}\ket{\mu,\vect{k}}_{(\Sigma,\vect{d})}
\end{equation} where $\ket{\psi}=\sum_{\mu}c_{\mu}\ket{\mu}$.

We call the states $\ket{\mu,\vect{k}}_{(\Sigma,\vect{d})}$ \textit{stabilizer} states due to the property:
\begin{equation}\label{eq:stabiliser_eigenvalues}
    \hat S_{J}\ket{\mu,\vect{k}}_{(\Sigma,\vect{d})}=e^{2i\pi \vect{k}^{T}\Omega\vect{m}_{J}}\ket{\mu,\vect{k}}_{(\Sigma,\vect{d})},
\end{equation}
where we recall that $\hat S_J = \hat{W}(\vect{m}_{J})$ is a stabilizer generator of the code [\cref{eq:SJ}].
In other words, the subspace $V_{\vect{k}}$ spanned by
\begin{equation}
    \bigg\{\ket{\mu,\vect{k}}_{(\Sigma,\vect{d})}\bigg|\,\mu\in\bigoplus_{j=1}^{n}\mathbb{Z}_{d_{j}}\bigg\}
\end{equation}
is the simultaneous eigenspace of the stabilizers $\hat{S}_{J}$ with eigenvalues $\exp\big(2i\pi \vect{k}^{T}\Omega\vect{m}_{J}\big)$ respectively. The stabilizer states span the Hilbert space but are not linearly independent, obeying the quasi-periodic boundary conditions:
\begin{subequations}\label{eq:BCs}
\begin{align}
    \ket{\mu,\vect{k}+\vect{\bar{m}}_{j}}_{(\Sigma,\vect{d})}&=e^{i\pi\, \vect{k}^{T}\Omega\vect{\bar{m}}_{j}}\ket{\mu+\vect{e}_{j},\vect{k}}_{(\Sigma,\vect{d})}\label{eq:XBC}\\
    \ket{\mu,\vect{k}+\vect{\bar{m}}_{j+n}}_{(\Sigma,\vect{d})}&=e^{i\pi\, \vect{k}^{T}\Omega\vect{\bar{m}}_{j+n}}\,e^{2i\pi \mu_{j}/d_{j}}\ket{\mu,\vect{k}}_{(\Sigma,\vect{d})}\label{eq:ZBC}
\end{align}
\end{subequations}
where the addition $\mu+\vect{e}_{j}$ in \cref{eq:XBC} is taken mod $d_{j}$ on each of its components. Comparing to \cref{eq:PauliXZ}, one can interpret \cref{eq:XBC} as applying a logical $\hat{X}$ to the qudit label $\mu$ and a $\vect{k}$-dependent phase to the stabilizer label upon application of the boundary condition, while \cref{eq:ZBC} applies a logical $\hat{Z}$ to $\mu$ along with a $\vect{k}$-dependent phase.

\subsection{The Subsystem Decomposition}\label{subsec:subsystem}

To obtain a non-overcomplete basis of $\mathcal{H}$ from the stabilizer states, we must restrict $\vect{k}$ to a primitive cell $\mathcal{P}$ of $\bar{\Lambda}$ [defined in \cref{eq:primitive_cell}], such that the set of states $\ket{\mu,\vect k}_{(\Sigma,\vect d)}$ for $\vect{k}\in\mathcal{P}$
forms a (linearly independent) basis of $\mathcal{H}$. We shall justify this claim in \cref{subsec:Zak} by making a direct connection to Zak states, which have already been shown to form a basis over a given primitive cell~\cite{Zak67}. From this we construct a subsystem decomposition
\begin{equation}
    \mathcal{H}=\mathcal{L}\otimes_{\mathcal{G}}\mathcal{S},
\end{equation}
where $\mathcal{G}=(\Sigma,\vect{d},\mathcal{P})$ represents the three parameters required to specify the decomposition. We define the tensor product such that
\begin{equation}\label{eq:subsystem_decomposition}
    \ket{\psi}\otimes\!{\vphantom{\ket{\psi}}}_{\mathcal{G}}\ket{\vect{k}}=\ket{\psi,\vect{k}}_{(\Sigma,\vect{d})}
\end{equation}
for $\vect{k}\in\mathcal{P}$. Note that this definition implicitly defines a Zak basis $\{\ket{\vect{k}}_{\mathcal{S}}\mid\vect{k}\in\mathcal{P}\}$ of $\mathcal{S}$.
We may also define the stabilizer states \cref{eq:stabiliser_states} to be normalized such that
\begin{equation}\label{eq:subsystem_normalisation}
    \bigg(\sum_{\mu}\ket{\mu}\!\bra{\mu}\!\bigg)\otimes_{\mathcal{G}}\bigg(\int_{\mathcal{P}}d^{2n}\vect{k}\ket{\vect{k}}\!\bra{\vect{k}}\!\bigg)=I.
\end{equation}

We call $\mathcal{L}$ the logical subsystem, which is isomorphic to the discrete-variable Hilbert space $\mathcal{H}_{\vect{d}}$. We call $\mathcal{S}$ the stabilizer subsystem, which is isomorphic to the full Hilbert space $\mathcal{H}$ via the correspondence
\begin{equation}\label{eq:subsystem_mapping}
    \ket{\vect{k}}_{\mathcal{S}}\leftrightarrow\ket{D^{1/2}\Sigma^{-1}\vect{k}}_{\vect{1}_{n}}
\end{equation}
where $D=\mathrm{diag}(\vect{d},\vect{d})$. The right-hand side of \cref{eq:subsystem_mapping} is a Zak state with $\vect{a}=(1,\dots,1)$, and the set $\{\ket{D^{1/2}\Sigma^{-1}\vect{k}}_{\vect{1}_{n}}\mid\vect{k}\in\mathcal{P}\}$ forms a basis of $\mathcal{H}$.

We can make a connection here to the stabilizer-destabilizer formalism of discrete-variable codes~\cite{Aaronson04}. The set of \textit{destabilizers} of the decomposition $\mathcal{G}$ is given by $\mathcal{D}_{\mathcal{G}}=\{\hat{W}(\vect{v})\mid\vect{v}\in\mathcal{P}\}$, which represent the ``lowest weight'' displacements that give rise to a given error syndrome. The full set of displacement operators $\{\hat{W}(\vect{v})\mid\vect{v}\in\mathbb{R}^{2n}\}$ is generated by the destabilizers $\mathcal{D}_{\mathcal{G}}$ and logical operators $\hat{W}(\vect{\bar{m}}_{J})$ via the equation $\vect{v}=\vect{\bar{\ell}}+\{\vect{v}\}_{\mathcal{P}}$. The set of displacement operators in turn forms an operator basis for $\mathcal{L}(\mathcal{H})$, the space of linear operators on $\mathcal{H}$. This is analogous to how qubit stabilizers, destabilizers and logical operators generate the Pauli group, which in turn forms a basis for the space of linear operators acting on the $n$-qubit Hilbert space.

The square single-mode GKP qubit code (discussed extensively in \cref{sec:sq} has parameters
\begin{align}\label{eq:sq_G}
    \mathcal{G}_{\text{sq}}&=\big(I_{2},(2),\mathcal{V}_{\text{sq}}\big),&\mathcal{V}_{\text{sq}}&=\big({-}2^{-3/2},2^{-3/2}\big]^{\times2}.
\end{align}
As a second example, the hexagonal single-mode GKP qubit code has parameters
\begin{align}\label{eq:hex_G}
    \mathcal{G}_{\text{hex}}&=\big(\Sigma_{\text{hex}},(2),\mathcal{V}_{\text{hex}}\big), & \Sigma_{\text{hex}} &= \begin{bmatrix}\sqrt[4]{4/3} & -1/\sqrt[4]{12} \\ 0 & \sqrt[4]{3/4}\end{bmatrix},
\end{align}
where $\mathcal{V}_{\text{hex}}$ is illustrated in \cref{fig:patch_figures}(d). The hexagonal code has the property that both $\Lambda$ and $\bar{\Lambda}$ are hexagonal lattices in $\mathbb{R}^{2}$. The Voronoi cell $\mathcal{V}_{\text{hex}}$ is hexagonal in shape and has boundaries corresponding to all three Pauli operators. As we will discuss in \cref{sec:Zak}, there are two differences between the hexagonal and square GKP codes: the generators of the lattice $\Sigma$ and the Voronoi cells $\mathcal{V}$, both of which need to be taken into account when comparing the properties of the codes.


\subsection{Stabilizers and Logical Paulis}\label{subsec:logical_Paulis}

Next, we wish to discuss the decomposition of stabilizers and logical Pauli operators in the stabilizer subsystem decomposition. Before doing so, we must introduce operators that act on each of the subsystems. For the logical subsystem $\mathcal{L}$ this is straightforward since it is isomorphic to a discrete-variable Hilbert space. For the stabilizer subsystem $\mathcal{S}$, one \textit{could} use the isomorphism with the corresponding CV Hilbert space $\mathcal{H}$ to define, for example, ladder operators $\hat{a}^{\vphantom{\dag}}_{j,\mathcal{S}},\hat{a}_{j,\mathcal{S}}^{\dag}$ for each mode of $\mathcal{S}$. Instead, we will find it more useful to instead consider $\mathcal{S}$ as a subsystem of $\mathcal{H}$, and use operators on $\mathcal{H}$ that act trivially on $\mathcal{L}$ to define the desired operators, analogous to our definition of $\hat{k}_{1}$ and $\hat{k}_{2}$ in the single-mode square case (see \cref{subsubsec:operators}).

The first such operator that we wish to define is a vector-valued operator $\vect{\hat{k}}$, that has action on the full Hilbert space:
\begin{equation}\label{eq:definition_k_hat}
    \hat{\vect{k}}\big(\!\ket{\psi}\otimes_{\mathcal{G}}\ket{\vect{k}}\!\big)=\vect{k}\big(\!\ket{\psi}\otimes_{\mathcal{G}}\ket{\vect{k}}\!\big),
\end{equation}
where $\vect k=(k_1,\dots,k_{2n})$, and the right-hand side may be viewed as a vector with components $k_{J}(\ket{\psi}\otimes_{\mathcal{G}}\ket{\vect{k}})$. By definition, $\vect{\hat{k}}$ acts trivially on $\mathcal{L}$ and thus can be viewed as an operator acting solely on $\mathcal{S}$. $\vect{\hat{k}}$ can be interpreted as a modular quadrature operator $\vect{\hat{k}}=\big\{\vect{\hat{\xi}}/\sqrt{2\pi}\big\}_{\mathcal{P}}$, where we recall that $\{\vect{v}\}_{\mathcal{P}}$ is the remainder of $\vect{v}$ in the primitive cell $\mathcal{P}$ such that $\vect{v}=\vect{\bar{\ell}}+\{\vect{v}\}_{\mathcal{P}}$ for some $\vect{\bar{\ell}}\in\bar{\Lambda}$.

From the stabilizer eigenvalue equation [\cref{eq:stabiliser_eigenvalues}], we can decompose the stabilizer generators as
\begin{equation}\label{eq:stabilisers_in_subsystem}
    \hat S_J = \hat{W}(\vect{m}_{J})=\hat{I}\otimes_{\mathcal{G}} e^{2i \pi\,\vect{\hat{k}}^{T}\Omega\vect{m}_{J}}.
\end{equation}
Physically, the eigenvalues of $\vect{\hat{k}}$, which lie within the primitive cell $\mathcal{P}$, can be measured simply by measuring the set of stabilizer generators~\cite{Gottesman01,Weigand20}.

For the logical Pauli operators we have
\begin{subequations}\label{eq:Paulis_in_subsystem}
\begin{align}
    \bar{X}_{j}&=\hat{W}(\vect{\bar{m}}_{j})=\hat{X}_{j}\otimes_{\mathcal{G}}e^{2i\pi\,\vect{\hat{k}}^{T}\Omega\vect{\bar{m}}_{j}},\\
    \bar{Z}_{j}&=\hat{W}(\vect{\bar{m}}_{j+n})=\hat{Z}_{j}\otimes_{\mathcal{G}}e^{2i\pi\,\vect{\hat{k}}^{T}\Omega\vect{\bar{m}}_{j+n}},
\end{align}
\end{subequations}
where $\bar{X}_{j},\bar{Z}_{j}$ represent the logical Pauli operators acting on $\mathcal{H}$, while $\hat{X}_{j},\hat{Z}_{j}$ represent the finite-dimensional Pauli operators acting on $\mathcal{L}$. The fact that each logical Pauli operator can be written as a tensor product of operators acting on $\mathcal{L}$ and $\mathcal{S}$ ensures that it perfectly applies the corresponding gate to the logical information of any given state.

Alternatively, it is also possible to write the stabilizers and logical operators in terms of the logical modes $\vect{\bar{\xi}}=(\bar{q}_{1},\dots,\bar{q}_{n},\bar{p}_{1},\dots,\bar{p}_{n})$ that we introduced in \Cref{eq:logical_modes}. We define the modular quadrature operators of the logical modes $\vect{\bar{k}}=(\bar{k}_{1},\dots,\bar{k}_{2n})$ as
\begin{align}
    \bar{k}_{j}&=\big\{\bar{q}_{j}/\sqrt{2\pi}\big\}_{d_{j}^{-1/2}}\,, & \bar{k}_{j+n}&=\big\{\bar{p}_{j}/\sqrt{2\pi}\big\}_{d_{j}^{-1/2}}\,,
\end{align}
where we have written $x=a\lfloor x\rceil_{a}+\{x\}_{a}$ for $x\in\mathbb{R}$ such that $\lfloor x\rceil_{a}\in\mathbb{Z}$ and $\{x\}_{a}\in(-a/2,a/2]$. From \Cref{eq:logical_mode_ops} we see that the eigenstates of each stabilizer generator $\hat{S}_{j}$ (respectively, $\hat{S}_{j+n}$) are eigenstates of $\bar{p}_{j}$ ($\bar{q}_{j}$) mod $\sqrt{2\pi/d_{j}}$, and are thus also the eigenstates of $\bar{k}_{j+n}$ ($\bar{k}_{j}$). The operator $\vect{\bar{k}}$ acts trivially on $\mathcal{L}$ because every basis state $\ket{\psi}\otimes_{\mathcal{G}}\ket{\vect{k}}$ is an eigenstate of the stabilizers regardless of its logical state. Since the stabilizer generators are simultaneously diagonalizable, we also have $\big[\bar{k}_{J},\bar{k}_{J'}\big]=0$. Finally, we can relate the operators $\vect{\bar{k}}$ and $\vect{\hat{k}}$ to each other via the equation
\begin{equation}
    \hat{\vect{k}}=\big\{\Sigma\vect{\bar{k}}\big\}_{\mathcal{P}}.
\end{equation}
This equation can be used to rewrite \cref{eq:stabilisers_in_subsystem,eq:Paulis_in_subsystem} in terms of $\vect{\bar{k}}$, giving
\begin{subequations}
\begin{align}
    \hat{S}_{j}&=\hat{I}\otimes_{\mathcal{G}}e^{-2i\pi\sqrt{d_{j}}\bar{k}_{j+n}},\\ \hat{S}_{j+n}&=\hat{I}\otimes_{\mathcal{G}}e^{2i\pi\sqrt{d_{j}}\bar{k}_{j}},\\
    \bar{X}_{j}&=\hat{X}_{j}\otimes_{\mathcal{G}}e^{-2i\pi\bar{k}_{j+n}/\sqrt{d_{j}}},\\
    \bar{Z}_{j}&=\hat{Z}_{j}\otimes_{\mathcal{G}}e^{2i\pi\bar{k}_{j}/\sqrt{d_{j}}}.
\end{align}
\end{subequations}

\subsection{The Partial Trace}\label{subsec:partial_trace}

The central feature of using the stabilizer subsystem decomposition is the property that the partial trace over the stabilizer subsystem $\mathrm{tr}_{\mathcal{S}}$ corresponds to an ideal decoding map over a primitive cell $\mathcal{P}$, which we now show.

We consider an ideal decoding map composed of a round of ideal error correction over $\mathcal{P}$, followed by a read-out of the logical information in the resultant ideal codestate.
This ideal decoding map can be described in the subsystem decomposition in three steps:
\begin{enumerate}
    \item First, there is a measurement of $\vect{\hat{k}}$ with some measurement outcome $\vect{v}\in\mathcal{P}$, which projects the initial state into the $\vect{\hat{k}}=\vect{v}$ eigenspace via the projection operator $\ket{\vect{v}}_{\mathcal{S}}\!\bra{\vect{v}}$ (up to normalization).
    \item To return the state to the $\vect{\hat{k}}=\vect{0}$ eigenspace (the ideal codespace), we apply the displacement $\hat{W}(\vect{v})^{\dag}$.
    \item Finally, ideal read-out is then described by the map $\hat{\rho}\otimes_{\mathcal{G}}\ket{\vect{0}}\!\bra{\vect{0}}\mapsto\hat{\rho}$.
\end{enumerate}
Considering these three steps together and averaging over the possible measurement outcomes results in a quantum channel $\mathcal{H}\mapsto\mathcal{L}$ described by the Kraus operators $\big\{{\vphantom{\ket{\vect{v}}}}_{\mathcal{S}}\!\bra{\vect{v}}\:\big|~\vect{v}\in\mathcal{P}\big\}$, where ${\vphantom{\ket{\vect{v}}}}_{\mathcal{S}}\!\bra{\vect{v}}$ is an operator that maps $\mathcal{H}\rightarrow\mathcal{L}$. Thus, the post-decoding state of an arbitrary density operator $\hat{\rho}$, unconditional on the stabilizer measurement outcomes, can be found by taking the partial trace over the $\mathcal{S}$ subsystem
\begin{equation}\label{eq:partial_trace}
    \int_{\mathcal{P}}\!d^{2n}\vect{v}\,{\vphantom{\ket{0}}}_{\mathcal{S}}\!\bra{\vect{v}}\!\hat{\rho}\!\ket{\vect{v}}_{\mathcal{S}}=\mathrm{tr}_{\mathcal{S}}(\hat{\rho}).
\end{equation}
Equivalently, if one performs a round of ideal error-correction over $\mathcal{P}$ and averages over the outcomes, one obtains the ideal GKP codestate $\mathrm{tr}_{\mathcal{S}}(\hat{\rho})\otimes_{\mathcal{G}}\ket{\vect{0}}\!\bra{\vect{0}}$. The partial trace can in some sense be considered
an ``ideal'' measure of the logical information in a state $\hat{\rho}$. Indeed, the ideal decoding procedure is not implementable physically since the measurement of the stabilizer generators requires the use of a non-normalizable ideal GKP codestate. We leave it to future work to describe approximate error-correction in the stabilizer subsystem decomposition.

In some contexts, it is also useful to consider the state $\hat{\rho}_{\vect{v}}\propto{\vphantom{\ket{0}}}_{\mathcal{S}}\!\bra{\vect{v}}\!\hat{\rho}\!\ket{\vect{v}}_{\mathcal{S}}$, which is the decoded state conditioned on the measurement outcome $\vect{v}$ (not averaged over them), because some stabilizer outcomes can result in decoded states $\hat{\rho}_{\vect{v}}$ with less noise than others. Such considerations have previously been explored in qubit codes~\cite{Chamberland17} and in GKP magic state distillation~\cite{Baragiola19}.

We remark that this ideal GKP decoding map is \textit{not} equivalent to a decoding map defined by logical state tomography with binned measurement operators. Indeed, we show in \cref{sec:logical_state_tomography} that the logical state tomography with binned measurement operators (binned-LST) decoding map is not completely positive, and is more susceptible to logical errors than the ideal decoding map.

\begin{figure*}
\includegraphics[width = 0.875\linewidth]{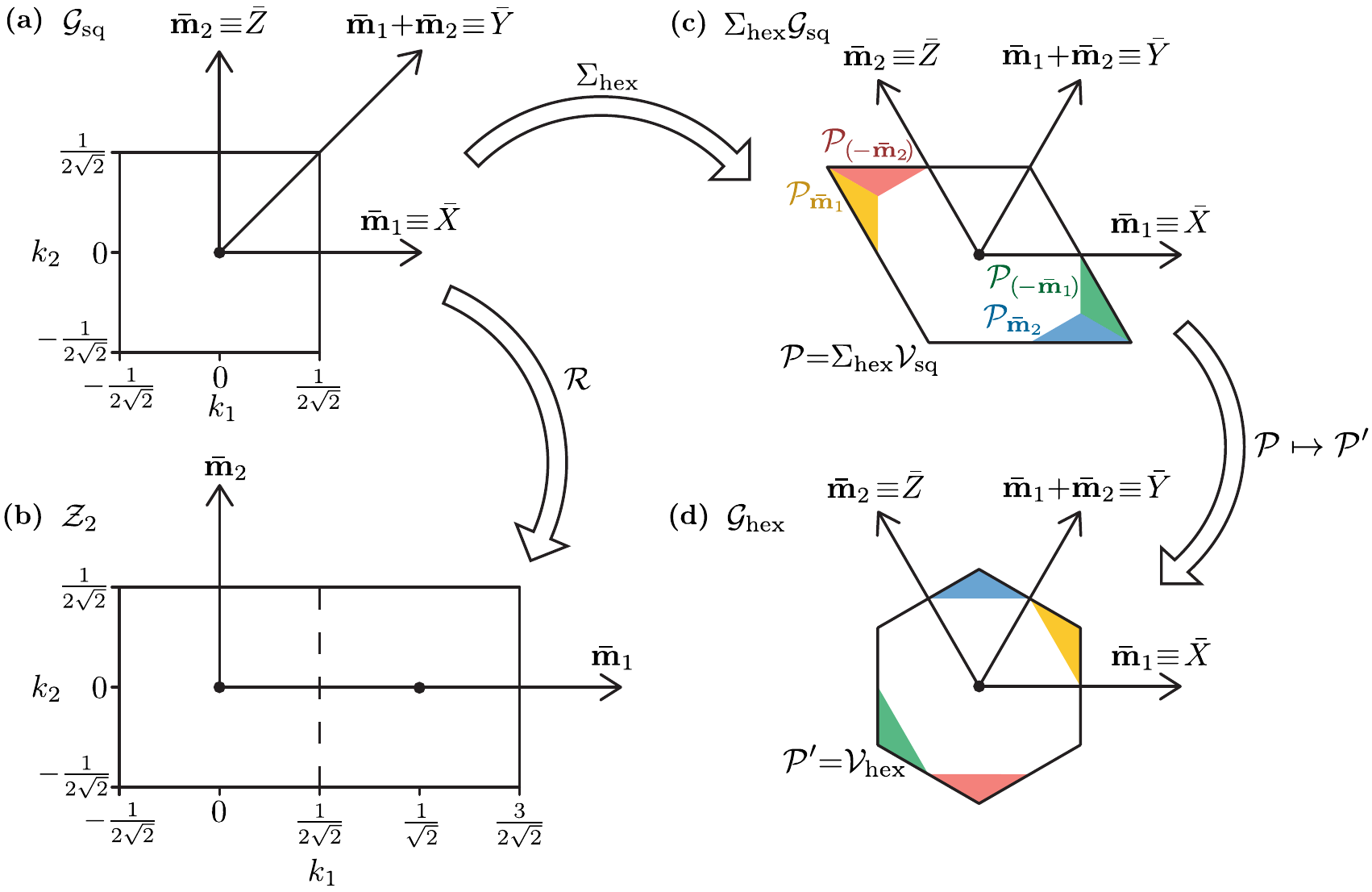}
\caption{(Color) Primitive cell diagrams for (a) the single-mode square GKP qubit code $\mathcal{G}_{\text{sq}}$~[\cref{eq:sq_G}], (b) the Zak basis $\mathcal{Z}_{\sqrt{2}}$~[\cref{eq:Zak_G}], (c) the square code transformed by the hexagonal Gaussian transformation $\Sigma_{\text{hex}}\mathcal{G}_{\text{sq}}$, and (d) the single-mode hexagonal GKP qubit code $\mathcal{G}_{\text{hex}}$~[\cref{eq:hex_G}]. Subplots (a) and (b) are related by an unfolding operation $\mathcal{R}$; (a) and (c) are related by a Gaussian transformation $\Sigma_{\text{hex}}$; and (c) and (d) are related by a cell transformation $\mathcal{P}=\Sigma_{\text{hex}}\mathcal{V}_{\text{sq}}\mapsto\mathcal{P}'=\mathcal{V}_{\text{hex}}$, which is performed by shifting each of the coloured regions $\mathcal{P}_{\vect{\bar{\ell}}}$ by the dual lattice vector $\vect{\bar{\ell}}$ to form $\mathcal{P}'$. In subplots (a), (c) and (d), each point on the plot represents the two-dimensional simultaneous eigenspace of the GKP stabilizers; while in (b) each point represents a single state.}
\label{fig:patch_figures}
\end{figure*}



\section{Transformations of $\otimes_{\mathcal{G}}$}\label{sec:Zak}

Next, we turn our attention to transformations of the subsystem decomposition $\otimes_{\mathcal{G}}$. In particular, we are interested in describing transformations that map basis states in one subsystem decomposition to basis states of another: $\ket{\psi}\otimes_{\mathcal{G}}\ket{\vect{k}}\mapsto\ket{\psi'}\otimes_{\mathcal{G}'}\ket{\vect{k}'}$. We wish to do this for three reasons. First, these transformations provide a recipe to decompose arbitrary CV states in the $\otimes_{\mathcal{G}}$ subsystem decomposition (\cref{subsec:state_decompositions}). Second, the transformations provide useful tools to analyze logical Clifford gates (\cref{sec:gates}) and the relationship between different GKP codes. Finally, the transformations allow us to make explicit the relationship between the subsystem decompositions and the Zak basis states [\cref{eq:Zak_multi_mode}] of $\mathcal{H}$ (\cref{subsec:Zak}). To do this, we introduce three transformations---cell transformations, Gaussian transformations, and dimension transformations---that allow one to relate any two arbitrary stabilizer subsystem decompositions $\mathcal{G}$ and $\mathcal{G}'$  over $n$ modes to each other. An example of each transformation is depicted in \cref{fig:patch_figures}, which we will refer to throughout the section.

\subsection{Cell transformations}\label{subsec:cell}

The first transformation we consider is a primitive cell transformation $(\Sigma, \vect{d}, \mathcal{P})\mapsto(\Sigma,\vect{d},\mathcal{P}’)$, which can be achieved by applying the boundary conditions \cref{eq:BCs} to each basis state in $\mathcal{P}$ (as described below). Cell transformations are important to consider because different primitive cells correspond to different decoders with different error-correction properties. As such, the same state decomposed across two subsystem decompositions differing only by their primitive cell can encode different information in their logical subsystems.

In order to describe the cell transformation $\mathcal{G}=(\Sigma, \vect{d}, \mathcal{P})\mapsto\mathcal{G}'=(\Sigma,\vect{d},\mathcal{P}’)$, we start by noting from the definition of a primitive cell [\cref{eq:primitive_cell}] that any vector $\vect{k}\in\mathcal{P}$ can be uniquely written
\begin{equation}\label{eq:cell_vector_transformation}
    \vect{k}=\vect{\bar{\ell}}+\{\vect{k}\}_{\mathcal{P}'}\,,
\end{equation}
where $\vect{\bar{\ell}}\in\bar{\Lambda}$ and $\{\vect{k}\}_{\mathcal{P}'}\in\mathcal{P}'$. Therefore, to write any basis state $\ket{\psi}\otimes_{\mathcal{G}}\ket{\vect{k}}$ as a basis state of $\otimes_{\mathcal{G}'}$, we can simply apply the boundary conditions associated with the dual lattice vector $\bar{\vect{\ell}}$, resulting in the equation
\begin{equation}\label{eq:cell_transformation}
    \ket{\psi}\otimes_{\mathcal{G}}\ket{\vect{k}}=\big(\hat{P}_{\vect{d}}(\vect{s})\ket{\psi}\!\big)\otimes_{\mathcal{G}'}\!\big(e^{i\pi\vect{k}^{T}\!\Omega\vect{\bar{\ell}}}\ket{\vect{k}-\vect{\bar{\ell}}\,}\!\big),
\end{equation}
where we have written $\vect{\bar{\ell}}=\sum_{J=1}^{2n}s_{J}\vect{\bar{m}}_{J}$ and $\hat{P}_{\vect{d}}(\vect{s})$ is an $n$-qudit Pauli operator defined in \Cref{eq:general_Pauli}. Importantly, $\vect{k}-\vect{\bar{\ell}}=\{\vect{k}\}_{\mathcal{P}'}$ is in the new primitive cell $\mathcal{P}'$ of the subsystem decomposition $\otimes_{\mathcal{G}'}$. Note that this transformation applies a $\vect{k}$-dependent Pauli operator to the logical subsystem, reflecting the fact that the new decoder defined by $\mathcal{P}'$ results in different logical information stored in a subsystem basis state.

It is also convenient to consider the subsets $\mathcal{P}_{\vect{\bar{\ell}}}\subseteq\mathcal{P}$ for $\vect{\bar{\ell}}\in\bar{\Lambda}$, defined as the set of points $\vect{k}$ that are mapped by \Cref{eq:cell_vector_transformation} to $\mathcal{P}'$ with the dual lattice vector $\vect{\bar{\ell}}$. For example, in \cref{fig:patch_figures}(c) and (d), we show the cell transformation of a rhombus primitive cell $\mathcal{P}=\Sigma_{\text{hex}}\mathcal{V}_{\text{sq}}$ to a hexagonal primitive cell $\mathcal{P}'=\mathcal{V}_{\text{hex}}$, while the remaining parameters $\Sigma=\Sigma_{\text{hex}}$ and $\vect{d}=(2)$ remain fixed. To perform this transformation, the initial rhombus cell $\mathcal{P}$ is split into regions $\mathcal{P}_{\vect{\bar{\ell}}}\subset\mathcal{P}$, each of which will be shifted by the dual lattice vector $\vect{\bar{\ell}}$ to form the new hexagonal cell $\mathcal{P}'$. In particular, the yellow, green, blue and red regions are shifted by $+\vect{\bar{m}}_{1}$, $-\vect{\bar{m}}_{1}$, $+\vect{\bar{m}}_{2}$ and $-\vect{\bar{m}}_{2}$ respectively; while the unshaded region is \textquotedblleft shifted\textquotedblright\ by the zero vector $\vect{0}\in\bar{\Lambda}$. The basis vectors in each state are transformed by the boundary conditions \cref{eq:cell_transformation}, which apply a Pauli operator to the logical subsystem and a phase to the stabilizer subsystem. In general, each region can be found by performing the decomposition \cref{eq:cell_vector_transformation} to each vector $\vect{k}\in\mathcal{P}$.

\subsection{Gaussian transformations}\label{subsec:Gaussian}

Next, we consider Gaussian transformations of the basis states. Applying a Gaussian unitary operator $\hat{U}_{S}$ to each basis state in the subsystem decomposition transforms $\mathcal{G}\mapsto S(\mathcal{G})=(S\Sigma,\vect{d},S\mathcal{P})$ such that
\begin{equation}\label{eq:gaussian_transformation}
    \hat{U}_{S}\big(\!\ket{\psi}\otimes_{\mathcal{G}}\ket{\vect{k}}\!\big)=\ket{\psi}\otimes_{S(\mathcal{G})}\ket{S\vect{k}}.
\end{equation}
Unlike the cell transformation, a Gaussian transformation affects both the generators of the lattice $\Sigma\mapsto S\Sigma$ and the cell $\mathcal{P}\mapsto S\mathcal{P}$. As such, to obtain a transformation that only alters the lattice generators in $\Sigma$, one must combine a Gaussian transformation with a cell transformation that restores the original cell $\mathcal{P}$. This is particularly relevant in the special case where $S=\Sigma N\Sigma^{-1}$ for some symplectic integral matrix $N$, in which case $\hat{U}_{S}$ is the Gaussian unitary operator that implements the logical Clifford gate with action $\hat{U}^{}_{S}\bar{P}^{}_{\vect{d}}(\vect{s})\hat{U}_{S}^{\dag}=\bar P^{}_{\vect d}(N\vect s)$. Here, the transformation not only applies a logical Clifford gate but also alters the primitive cell, which we discuss in more detail in \cref{sec:gates}.

We can now describe the full set of transformations required to transform the single-mode square GKP code $\mathcal{G}_{\text{sq}}$ [\cref{eq:sq_G}] to the single-mode hexagonal GKP code $\mathcal{G}_{\text{hex}}$ [\cref{eq:hex_G}], depicted in \cref{fig:patch_figures}(a), (c) and (d). First, we perform a Gaussian transformation $\hat{U}_{\Sigma_{\text{hex}}}$, resulting in the intermediate subsystem decomposition $\Sigma_{\text{hex}}\mathcal{G}_{\text{sq}}$ [\cref{fig:patch_figures}(c)]. Then, we perform a cell transformation from the rhombus primitive cell $\Sigma_{\text{hex}}\mathcal{V}_{\text{sq}}$ to the hexagonal primitive cell $\mathcal{V}_{\text{hex}}$. When comparing the properties of the square and hexagonal GKP codes it is important to consider both of these transformations, as both of them alter the properties of the decomposition.


\subsection{Dimension transformations}\label{subsec:dimension}

Finally, we describe how to transform the dimension, \vect d, of the subsystem decomposition. Central to this description is the single-mode \textit{unfolding} operation $\mathcal{R}_{j}$ (defined below), which transforms the dimension of the $j$-th mode $d_{j}\mapsto1$, creating a qunaught mode. While there are many possible transformations that map a given mode from a qudit mode to a qunaught mode, we choose to describe the unfolding operation $\mathcal{R}_{j}$ (defined below) due to its simple action on the basis states of $\otimes_{\mathcal{G}}$. The net effect of this operation is to take the logical information in the $j$-th mode and instead label it in the stabilizer mode, thus increasing the size of the primitive cell [see \cref{fig:patch_figures}(b)]. The unfolding operation is therefore a generalization of the relationship between the single-mode square GKP qubit code and the Zak basis as shown in \cref{fig:square_Zak}. The ``inverse'' of the unfolding operation is a $d$-fold \textit{folding} operation $\mathcal{F}_{j}(d)$, which turns a qunaught mode back into a qudit mode with dimension $d$. Arbitrary dimension transformations $\vect{d}\mapsto\vect{d}'$ can then be described as unfolding and folding operations on each of the modes (although in some cases it is necessary to perform a cell transformation following the unfolding operation in order for the folding operation to be well-defined).

Intuitively, the unfolding operation $\mathcal{R}_{j}$ consists of demoting the logical $\bar{Z}_{j}=\hat{W}(\vect{\bar{m}}_{j+n})$ operator on the $j$-th mode to a stabilizer of the new code, and removing the corresponding $\bar{X}_{j}=\hat{W}(\vect{\bar{m}}_{j})$ from the logical Pauli group. Since $\hat{W}(\vect{\bar{m}}_{j+n})$ is now in the stabilizer group, states related by a displacement $\hat{W}(\vect{\bar{m}}_{j})$ which previously had identical support in the stabilizer subsystem are now distinct. As a result, $\mathcal{R}_{j}$ \textquotedblleft unfolds\textquotedblright\ (or copies) the cell $\mathcal{P}$ $d_{j}$ times along the vector $\vect{\bar{m}}_{j}$, with the new patch given by $\mathcal{R}_{j}(\mathcal{P})=\bigcup_{a=0}^{d_{j}-1}(\mathcal{P}+a\vect{\bar{m}}_{j})$.

Concretely, the action of $\mathcal{R}_{1}$ (acting on the first mode for simplicity) on states in the subsystem decomposition is given by
\begin{equation}\label{eq:unfolding_transform}
    \ket{\mu_{1}{\oplus}\mu}\otimes_{\mathcal{G}}\ket{\vect{k}}=e^{i\pi\mu^{}_{1}\vect{\bar{m}}_{1}^{T}\Omega\vect{k}}\ket{0{\oplus}\mu}\otimes_{\mathcal{R}_{1}(\mathcal{G})}\ket{\vect{k}+\mu_{1}\vect{\bar{m}}_{1}},
\end{equation}
where $\oplus$ represents the direct sum of two dit-strings $\mu_{1}\in\mathbb{Z}_{d_{1}}$ and $\mu\in \bigoplus_{j=2}^{n}\mathbb{Z}_{d_{j}}$, $\vect{k}\in\mathcal{P}$, and $\mathcal{R}_{1}(\mathcal{G})$ is determined below. Note that the 0 label in the right-hand side of \cref{eq:unfolding_transform} is redundant since it represents a qunaught degree of freedom. The corresponding folding operation $\mathcal{F}_{1}(d_{1})$ can be found by inverting \Cref{eq:unfolding_transform}.

Finally, we must consider how $\mathcal{R}_{j}$ transforms the parameters $\mathcal{G}$ of the subsystem decomposition. In terms of lattice generators, $\mathcal{R}_{j}$ transforms $\vect{m}_{j+n}\mapsto\vect{m}_{j+n}/d_{j}$ (which adds $\bar{Z}_{j}$ to the stabilizer group), and correspondingly $\vect{\bar{m}}_{j}\mapsto d_{j}\vect{\bar{m}}_{j}$ (which removes $\bar{X}_{j}$ from the logical Pauli group), while all other lattice and dual lattice generators are unchanged. It is straight-forward to show that the equivalent transformation of the parameters is given by
\begin{align}
    \mathcal{R}_{j}(\mathcal{G})&=\bigg(\Sigma A_{j}\big(\!\sqrt{d_{j}}\big),\vect{d}_{d_{j}\rightarrow1},\bigcup_{a=0}^{d_{j}-1}(\mathcal{P}+a\vect{\bar{m}}_{j})\bigg),
\end{align}
where $A_{j}(\lambda)$ is a diagonal matrix with $\lambda$ and $\lambda^{-1}$ on the $j$-th and $(j+n)$-th positions respectively and ones elsewhere, and $\vect{d}_{d_{j}\rightarrow1}=(d_{1},\dots,d_{j-1},1,d_{j+1},\dots,d_{n})$. In \cref{subsec:Zak} we will also make use of the all-mode unfolding operation $\mathcal{R}=\mathcal{R}_{1}\circ\,\cdots\,\circ\mathcal{R}_{n}$, which results in a trivial subsystem decomposition $\mathcal{R}(\mathcal{G})$ with dimension vector $\vect{1}_{n}$.

As an example consider the unfolding of the square GKP code $\mathcal{G}_{\text{sq}}$, shown in \cref{fig:patch_figures}(a) and (b). The unfolded square GKP code $\mathcal{R}(\mathcal{G}_{\text{sq}})$ is a qunaught code with trivial logical subsystem $\mathcal{L}\cong\mathbb{C}$, such that each point in the unfolded primitive cell $\mathcal{R}(\mathcal{V}_{\text{sq}})$ represents a single basis state of the full Hilbert space $\mathcal{H}$. $\mathcal{R}(\mathcal{V}_{\text{sq}})$ has double the area of $\mathcal{V}_{\text{sq}}$. The left half of $\mathcal{R}(\mathcal{V}_{\text{sq}})$ corresponds to states in $\mathcal{G}_{\text{sq}}$ with a $\ket{0}$ logical state, while the right half corresponds to $\ket{1}$. In general, a basis state $\ket{\psi}\otimes_{\text{sq}}\ket{\vect{k}}$ of $\mathcal{G}_{\text{sq}}$ is a superposition of two basis states $\ket{\vect{k}}$ and $\ket{\vect{k}+\vect{e}_{1}/\sqrt{2}}$ of the unfolded decomposition $\mathcal{R}(\mathcal{G}_{\text{sq}})$ (where we have omitted the redundant logical information $\ket{0}$ in the notation). By comparison to \cref{fig:square_Zak}, one can see that the basis states of $\mathcal{R}(\mathcal{G}_{\text{sq}})$ correspond to Zak states as we will now discuss.

\subsection{Zak states}\label{subsec:Zak}

We can now make explicit the connection between states in a stabilizer subsystem decomposition $\mathcal{G}$ and Zak states~\cite{Zak67}, which are known to be closely related to GKP codes~\cite{Ketterer16,Pantaleoni22}.

Since the $n$-mode Zak states $\ket{\vect{k}}_{\vect{a}}$ [\cref{eq:Zak_multi_mode}] are simultaneous eigenstates of the displacement operators $\hat{W}(a_{j}\vect{e}_{j})$ and $\hat{W}(\vect{e}_{j+n}/a_{j})$, we can identify them with the stabilizer states of an $n$-mode qunaught rectangular GKP code
\begin{equation}\label{eq:Zak_stabilizer_states}
    \ket{\vect{k}}_{\vect{a}}=\ket{0_{n},\vect{k}}_{(\Sigma_{\mathcal{Z},\vect{a}},\vect{1}_{n})},
\end{equation}
where $\Sigma_{\mathcal{Z},\vect{a}}=\mathrm{diag}\big(a_{1},\dots,a_{n},1/a_{1},\dots,1/a_{n}\big)$ and holds for all $\vect{k}\in\mathbb{R}^{2n}$. We can obtain a Zak basis by restricting
\begin{align}\label{eq:Zak_patch}
    k_{j}&\in\bigg(\!{-}\frac{1}{2a},a-\frac{1}{2a}\bigg],&k_{j+n}&\in\bigg(\!{-}\frac{1}{2a},\frac{1}{2a}\bigg]
\end{align}
for $j=1,\dots,n$. The Zak basis coincides with the Zak basis states of a trivial stabilizer subsystem decomposition
\begin{equation}\label{eq:Zak_G}
    \mathcal{Z}_{\vect{a}}=(\Sigma_{\mathcal{Z},\vect{a}},\vect{1}_{n},\mathcal{P}_{\mathcal{Z},\vect{a}})
\end{equation}
that we call the Zak subsystem decomposition, where $\mathcal{P}_{\mathcal{Z},\vect{a}}$ is defined by \cref{eq:Zak_patch}. As foreshadowed in \cref{subsec:dimension}, applying an all-mode unfolding operation $\mathcal{R}$ to a square qudit GKP code $\mathcal{G}_{\text{sq},\vect{d}}$ results in a Zak subsystem decomposition $\mathcal{Z}_{\sqrt{\vect{d}}}$, where $\sqrt{\vect{d}}=(\sqrt{d_{1}},\dots,\sqrt{d_{n}})$.

With this connection, we can now apply any of the transformations discussed above in \cref{subsec:cell,subsec:Gaussian,subsec:dimension} to relate arbitrary GKP stabilizer subsystem decompositions to a Zak basis $\mathcal{Z}_{\sqrt{\vect{d}}}$. Starting from an arbitrary $\mathcal{G}=(\Sigma,\vect{d},\mathcal{P})$, we can apply three transformations to result in a decomposition described by $\mathcal{Z}_{\sqrt{\vect{d}}}$:
\begin{enumerate}
    \item First we apply a Gaussian transformation $\Sigma^{-1}$, resulting in the decomposition $\Sigma^{-1}(\mathcal{G})=(I_{2n},\vect{d},\Sigma^{-1}\mathcal{P})$.
    \item Then, we apply a cell transformation $\Sigma^{-1}\mathcal{P}\mapsto \mathcal{V}_{\text{sq},\vect{d}}$, where $\mathcal{V}_{\text{sq},\vect{d}}$ is the Voronoi cell of the square GKP code with dimension vector $\vect{d}$.
    \item Finally, we apply an all-mode unfolding operation $\mathcal{R}$, which results in the Zak subsystem decomposition $\mathcal{Z}_{\sqrt{\vect{d}}}$.
\end{enumerate}
Since the transformations described in \cref{subsec:cell,subsec:Gaussian,subsec:dimension} preserve the linear independence and completeness of the states in each subsystem decomposition, and the Zak states form a basis of $\mathcal{H}$ when restricted to the primitive cell $\mathcal{P}_{\mathcal{Z},\sqrt{\vect{d}}}$, we can confirm that the restriction of $\vect{k}$ to a primitive cell $\mathcal{P}$ of the dual lattice $\bar\Lambda$ does indeed result in a basis $\big\{\!\ket{\mu,\vect{k}}\,\big|\;\mu\in \bigoplus_{j=1}^{n}\mathbb{Z}_{d_{j}},\vect{k}\in\mathcal{P}\big\}$ of the full Hilbert space $\mathcal{H}$.

\subsection{Wavefunctions in the subsystem decomposition}\label{subsec:state_decompositions}

The relationship between basis states of an arbitrary subsystem decomposition and the Zak basis also provides a practical method of decomposing arbitrary states $\ket{\phi}\in\mathcal{H}$ into an arbitrary stabilizer subsystem decomposition $\otimes_{\mathcal{G}}$. In particular, we wish to calculate the overlap
\begin{equation}\label{eq:inner_product}
    \big(\!\bra{\mu}\otimes_{\mathcal{G}}\bra{\vect{k}}\!\big)\ket{\phi}={\vphantom{\ket{}}}_{(\Sigma,\vect{d})}\!\braket{\mu,\vect{k}|\phi}
\end{equation}
for all $\mu\in\bigoplus_{j=1}^{n}\mathbb{Z}_{d_{j}}$ and $\vect{k}\in\mathcal{P}$.

To perform this calculation we work with stabilizer states $\ket{\mu,\vect{k}}_{(\Sigma,\vect{d})}$ as defined in \cref{eq:stabiliser_states}. We can write the stabilizer state $\ket{\mu,\vect{k}}_{(\Sigma,\vect{d})}$ as a Zak state:
\begin{align}
    \ket{\mu,\vect{k}}_{(\Sigma,\vect{d})}&=\hat{U}_{\Sigma}\ket{\mu,\Sigma^{-1}\vect{k}}_{(I_{2n},\vect{d})}\label{eq:stab_Gaussian}\\
    &=\hat{U}_{\Sigma}\,e^{i\pi\vect{\bar{\ell}}(\mu)^{T}\Omega\vect{k}}\big|\Sigma^{-1}\big(\vect{k}+\vect{\bar{\ell}}(\mu)\big)\big\rangle_{\mathcal{Z}_{\sqrt{\vect{d}}}}\label{eq:stab_unfolding},
\end{align}
where $\vect{\bar{\ell}}(\mu)=\sum_{j=1}^{n}\mu_{j}\vect{\bar{m}}_{j}$. Note that $\vect{\bar{m}}_{j}$ for $j\leq n$ represents the logical $\bar{X}_{j}$ operator, such that $\hat{W}\big(\vect{\bar{\ell}}(\mu)\big)=\prod_{j=1}^{n}\bar{X}_{j}^{\mu_{j}}$. In \cref{eq:stab_Gaussian}, we have applied the Gaussian transformation $\hat{U}_{\Sigma}$ [\cref{eq:gaussian_transformation}]; and in \cref{eq:stab_unfolding}, we have applied the all-mode unfolding operation $\mathcal{R}$ defined by \cref{eq:unfolding_transform}.

Since we are working with stabilizer states and not with a subsystem decomposition, we do not need to explicitly apply a cell transformation. Instead, the primitive cell of the original decomposition $\mathcal{G}$ is incorporated implicitly into the calculation by only calculating overlaps where $\vect{k}\in\mathcal{P}$. Indeed, when $\vect{k}$ runs over $\mathcal{P}$ and $\mu$ runs over all possible ditstrings, the ``unfolded'' vector $\Sigma^{-1}\big(\vect{k}+\vect{\bar{\ell}}(\mu)\big)$ [\cref{eq:stab_unfolding}] runs over a primitive cell that is related by a cell transformation to our previous choice of Zak primitive cell $\mathcal{P}_{\mathcal{Z},\sqrt{\vect{d}}}$.

Finally, we can use the position-representation of the $n$-mode Zak states given by \cref{eq:Zak_states,eq:Zak_multi_mode} to write the overlap \cref{eq:inner_product} as a sum of overlaps
\begin{equation}\label{eq:position_state}
    \vphantom{\ket{}}_{q}\!\braket{x|\hat{U}_{\Sigma}^{\dag}|\phi}.
\end{equation}
Therefore, any state for which one can calculate the overlap \cref{eq:position_state} can be decomposed into any subsystem decomposition $\mathcal{G}$. We relegate the full equation relating the overlap \cref{eq:inner_product} with \cref{eq:position_state} to \cref{sec:state_decompositions_eq} due to its length.

\section{Logical Clifford Gates}\label{sec:gates}

We now move on to analyzing logical Clifford gates in the stabilizer subsystem decomposition. To simplify our discussion we only consider GKP codes that encode $k$ qubits in $n$ modes, i.e.\ the dimension vector $\vect{d}$ consists of $k$ 2's followed by $n-k$ 1's. One appealing feature of GKP codes is that Gaussian operators, which are often easy to implement experimentally, are sufficient to generate the logical Clifford group. Given a $k$-qubit Clifford operator $\hat{A}$, we can always write $\hat{A}\hat{P}_{\vect{d}}(\vect{s})\hat{A}^{\dag}=\hat{P}_{\vect{d}}(N_{A}\vect{s})$, where $N_{A}$ is an integral symplectic matrix that acts trivially on the components of $\vect{s}$ corresponding to the $n-k$ qunaught states, and $\hat{P}_{\vect{d}}(\vect{s})$ is a Pauli operator defined in \cref{eq:general_Pauli}. Equivalently, $N_{A}$ is the symplectic representation of the Gaussian operator $\bar{A}_{\text{sq}}$ that applies the logical $\hat{A}$ gate on the square GKP $k$-qubit code. For a general GKP $k$-qubit code, a logical implementation of the Clifford gate $\hat{A}$ is given by $\bar{A}=\hat{U}_{S_A}$ with $S_A = \Sigma N_{A}\Sigma^{-1}$, such that $\bar{A}\bar{P}_{\vect{d}}(\vect{s})\bar{A}^{\dag}=\bar{P}_{\vect{d}}(N_{A}\vect{s})$. We note that the logical implementation $\bar{A}$ (and, equivalently, the symplectic matrices $S_{A}$ and $N_{A}$) is not unique for a given Clifford gate $\hat{A}$.

We begin our analysis by considering the action of $\bar{A}$ on stabilizer states, given by
\begin{equation}\label{eq:Clifford_stabiliser_states}
    \bar{A}\ket{\psi,\vect{k}}_{(\Sigma,\vect{d})}=\ket{A(\psi),S_A\vect{k}}_{(\Sigma,\vect{d})},
\end{equation}
where $A(\psi)$ is the qubit label representing the state $\ket{A(\psi)}=\hat{A}\ket{\psi}$, c.f.~\cref{eq:sq_phase_stabilizer_states}. Since the state $\ket{\psi,\vect{k}}_{(\Sigma,\vect{d})}$ can be interpreted as an ideal GKP codestate that has incurred a displacement error $\hat{W}(\vect{k})\ket{\bar{\psi}}$, we can see that the logical Clifford gate maps the error $\vect{k}$ to $S_{A}\vect{k}$. The Clifford gate can therefore cause a logical error if the initial error is correctable (i.e.\ $\vect{k}\in\mathcal{P}$) but the final error is not ($S_{A}\vect{k}\notin\mathcal{P}$). However, one can exactly counteract this effect if one simply performs a round of error correction over a modified patch $S_A \mathcal{P}$ immediately after the application of the gate, an idea which was first introduced in Ref.~\cite{Noh22} and generalized in Ref.~\cite{Shaw22-1}.

An interesting alternative viewpoint is to consider $\bar{A}$ as a Gaussian transformation of the stabilizer subsystem decomposition 
$\mathcal{G}\mapsto S_{A}(\mathcal{G})=(\Sigma N_{A},\vect{d},S_{A}\mathcal{P})$
via \cref{eq:gaussian_transformation}. Here, the right multiplication of $\Sigma$ by the integral symplectic matrix $N_{A}$ can be viewed as a column operation that relabels the generators of the GKP lattice and dual lattice (equivalent to a row operation $N_{A}^{T}$ acting on the generator matrix $M$) while leaving the lattice invariant. The relabelling occurs in such a way that the Clifford operator $\hat{A}$ is applied to the logical Pauli labels of the dual lattice $\bar{\Lambda}$, i.e.
\begin{equation}\label{eq:clifford_rewriting}
    \ket{\psi}\otimes_{(\Sigma N_{A},\vect{d},S_{A}\mathcal{P})}\ket{S_{A}\vect{k}}=\big(\hat{A}\ket{\psi}\!\big)\otimes_{(\Sigma,\vect{d},S_{A}\mathcal{P})}\ket{S_{A}\vect{k}}\!.
\end{equation}
However, to write the right-hand side of \cref{eq:clifford_rewriting} in terms of the original subsystem decomposition $\mathcal{G}$, one must perform a cell transformation $S_{A}\mathcal{P}\mapsto\mathcal{P}$ via \cref{eq:cell_transformation}, which in general applies a Pauli operator to the logical subsystem.

In practical terms, the logical information in the \textit{original} subsystem decomposition $\mathcal{G}$ \textit{after} applying $\bar{A}$ corresponds to performing a round of error correction over the \textit{original} primitive cell $\mathcal{P}$ after the gate (and thus incurring additional errors as described above). In contrast, the logical information in the \textit{modified} subsystem decomposition $(\Sigma,\vect{d},S_{A}\mathcal{P})$ after applying $\bar{A}$ corresponds to performing error correction over the \textit{modified} primitive cell $S_{A}\mathcal{P}$. In this case, no cell transformation is performed after \cref{eq:clifford_rewriting} and so no logical errors are introduced.

In \cref{subsubsec:operators} we already discussed the examples of the logical Hadamard [\cref{eq:sq_Had}] and logical phase [\cref{eq:sq_phase}] gates for the single-mode square GKP code. In particular, we saw that the logical Hadamard gate $\bar{H}_{\text{sq}}$ \textit{is} a tensor product operator in the subsystem decomposition, while the logical phase gate $\bar{S}_{\text{sq}}$ is not. This is a consequence of how the corresponding symplectic matrices for the Hadamard $S_{H}^{\text{(sq)}}$ and phase gates $S_{S}^{\text{(sq)}}$ transform the square Voronoi cell $\mathcal{V}_{\text{sq}}$. For the Hadamard gate, we have $S^{\text{(sq)}}_{H}\mathcal{V}_{\text{sq}}=\mathcal{V}_{\text{sq}}$, so no cell transformation is required following \cref{eq:clifford_rewriting}, and hence $\bar{H}$ decomposes as a tensor product operator. For the phase gate, we have $S_{S}^{\text{(sq)}}\mathcal{V}_{\text{sq}}\neq\mathcal{V}_{\text{sq}}$, and the required cell transformation $S_{S}^{\text{(sq)}}\mathcal{V}_{\text{sq}}\mapsto\mathcal{V}_{\text{sq}}$ ensures that $\bar{S}_{\text{sq}}$ cannot be written as a tensor product operator.

As a further example, consider the logical controlled-$Z$ gate. Since this is a two-qubit gate, we consider it acting on two copies of the square GKP code, which is described by the parameters
\begin{equation}
    \mathcal{G}_{\text{sq}}^{2}=\big(I_{4},(2,2),({-}2^{-3/2},2^{-3/2}]^{\times 4}\big).
\end{equation}
The controlled-$Z$ gate is described by
\begin{subequations}
\begin{align}
    \hat{C}_{Z}&=\mathrm{diag}(1,1,1,-1),\\
    N_{C_{Z}}&=S_{C_{Z}}^{\text{(sq)}}=\begin{bmatrix}1&0&0&0\\0&1&0&0\\0&1&1&0\\1&0&0&1\end{bmatrix},\\
    \bar{C}_{Z,\text{sq}}&=\mathrm{exp}(i\hat{q}\otimes\hat{q}).
\end{align}
\end{subequations}
Since $S_{C_{Z}}^{\text{(sq)}}\mathcal{V}_{\text{sq}}^{2}\neq\mathcal{V}_{\text{sq}}^{2}$, the controlled-$Z$ gate cannot be written as a tensor product operator.

We can perform a similar analysis for the hexagonal GKP code [\cref{eq:hex_G}]. Consider the logical permutation gate $\bar{R}^{}_{\text{hex}}=\bar{H}^{}_{\text{hex}}\bar{S}_{\text{hex}}^{\dag}$, which has
\begin{subequations}
\begin{align}
    \hat{R}&=\frac{1}{\sqrt{2}}\begin{bmatrix}1&-i\\1&i\end{bmatrix},&N_{R}&=\begin{bmatrix}1&-1\\1&0\end{bmatrix},\\
    S_{R}^{\text{(hex)}}&=\begin{bmatrix}1/2&-\sqrt{3}/2\\\sqrt{3}/2&1/2\end{bmatrix},&\bar{R}_{\text{hex}}&=\exp(i\pi\hat{a}^{\dag}\hat{a}/3).
\end{align}
\end{subequations}
Since $S_{R}^{\text{(hex)}}$ is a $\pi/3$ anticlockwise rotation, we have $S_{R}^{\text{(hex)}}\mathcal{V}_{\text{hex}}=\mathcal{V}_{\text{hex}}$, so $\bar{R}_{\text{hex}}$ is a tensor product operator across $\mathcal{G}_{\text{hex}}$. However, the logical Hadamard, phase, controlled-NOT and controlled-$Z$ gates all do not preserve the hexagonal Voronoi cell and are not tensor product operators in $\mathcal{G}_{\text{sq}}$.

Our formalism also allows for other logical gate proposals such as those using gauge fixing techniques~\cite{Vuillot19-2,Royer22}, which could be used to implement code deformation and lattice surgery schemes. In particular, consider the projection into the $+1$-eigenspace of an arbitrary logical Pauli operator $\hat{P}$. In the stabilizer subsystem decomposition this consists of demoting the logical Pauli operator to a stabilizer via an appropriate unfolding operation, followed by a round of error correction to project into the $+1$-eigenspace. In symbols, this corresponds to the sequence $\mathcal{R}_{1}\circ S_{A}$, where $\hat{A}$ is a Clifford operator such that $\hat{A}\hat{P}\hat{A}^{\dag}=\hat{Z}\otimes \hat{I}^{\otimes (n-1)}$. We leave it to future work to apply this formalism to practical problems involving GKP gauge fixing.

\section{Numerical modeling of noise}\label{sec:errors}

One appealing feature of subsystem decompositions is that the partial trace operation provides a straight-forward method of extracting qudit-level information from CV states. Here, we apply the stabilizer subsystem decomposition partial trace to model noise sources such as loss and dephasing in GKP codes. In particular, we are interested in obtaining the qudit-to-qudit \textit{logical} map corresponding to applying a given noise map $\mathcal{N}$ on approximate GKP codewords. We develop a general method for calculating the logical noise map from $\mathcal{N}$ using the partial trace in \cref{subsec:method_overview}, from which the average gate fidelity can be directly calculated. Excitingly, our method becomes more accurate the closer the state is to an ideal GKP codeword, allowing us to calculate logical noise maps for states with essentially arbitrarily large photon numbers (\cref{sec:very_small_Delta}). We then apply the method to calculate the average gate fidelity of the pure loss (\cref{subsec:loss}), Gaussian random displacement (\cref{subsec:random_walk}) and white-noise dephasing (\cref{subsec:dephasing}) noise channels acting on single-mode square GKP approximate codestates. We find that in the regime of small loss and high-quality approximate codestates, pure loss outperforms the Gaussian random displacement channel obtained by post-composing pure loss with a quantum-limited amplification noise channel, consistent with Ref.~\cite{Hastrup22}. Finally, we show that the square GKP code is far more vulnerable to dephasing than pure loss; indeed, when comparing the average gate infidelity of the GKP codes with a trivial $\{\ket{0},\ket{1}\}$ Fock space encoding, we find that a squeezing of at least 12 dB and a dephasing rate below $\sigma^{2}=10^{-3}$ (defined below) is necessary to break-even.

\subsection{Description of Method}\label{subsec:method_overview}

In this subsection we present the methods we use to determine the logical noise channel $\mathcal{N}_{\mathcal{L}}$ corresponding to a CV noise channel $\mathcal{N}$. We develop the method for a general GKP stabilizer subsystem decomposition $\mathcal{G}=(\Sigma,\vect{d},\mathcal{P})$, although in the remainder of the section (\cref{subsec:envelope,subsec:loss,subsec:random_walk,subsec:dephasing}) we will only consider the square single-mode GKP code $\mathcal{G}_{\text{sq}}=(I_{2},(2),\mathcal{V}_{\text{sq}})$ [\cref{eq:sq_G}].

We start by defining the logical noise map $\mathcal{N}_{\mathcal{L}}$ as
\begin{equation}\label{eq:logical_noise_channel_defn}
    \mathcal{N}_{\mathcal{L}}(\hat{\rho})=\mathrm{tr}_{\mathcal{S}}\big(\mathcal{N}(\hat{\rho}\otimes_{\mathcal{G}}\ket{\vect{0}}\!\bra{\vect{0}})\big),
\end{equation}
where $\mathcal{N}$ is an arbitrary CV noise map. One can understand $\mathcal{N}_{\mathcal{L}}$ as composed of three steps:
\begin{enumerate}
    \item an ideal encoding map $\hat{\rho}\mapsto\bar{\rho}=\hat{\rho}\otimes_{\mathcal{G}}\ket{\vect{0}}\!\bra{\vect{0}}$, followed by
    \item the CV noise map $\mathcal{N}$, and finally
    \item an ideal decoding map over the primitive cell $\mathcal{P}$, which is equivalent to the partial trace over $\mathcal{S}$ (as described in \cref{subsec:partial_trace}).
\end{enumerate}

Importantly, we have defined $\mathcal{N}$ as acting on \textit{ideal} codestates $\bar{\rho}$. In practice though, we are typically interested in noise channels acting on non-ideal codestates $\tilde{\rho}$ since the ideal codestates are non-normalizable. However, we can always define a map $\mathcal{C}$ that maps the ideal codestates $\ket{\bar{\mu}}$ to the non-ideal codestates $\ket{\tilde{\mu}}$. Then, the map $\mathcal{N}'=\mathcal{N}\circ\mathcal{C}$ can be used in place of $\mathcal{N}$ in \cref{eq:logical_noise_channel_defn} to define the corresponding logical noise map acting on the approximate codestates.

To find the logical map $\mathcal{N}_{\mathcal{L}}$ from $\mathcal{N}$, we first need to write $\mathcal{N}$ in terms of displacement operators in its characteristic function~\cite{Conrad21}
\begin{equation}\label{eq:translation_operator_basis}
    \mathcal{N}(\hat{\rho})=\iint_{\mathrlap{\mathbb{R}^{2n}}}\;\;d^{2n}\vect{u}\,d^{2n}\vect{v}\,c(\vect{u},\vect{v})\hat{W}(\vect{u})\hat{\rho}\hat{W}(\vect{v})^{\dag}.
\end{equation}
In general, the characteristic function $c(\vect{u},\vect{v})$ can be obtained either from the Liouville superoperator representation of $\mathcal{N}$ or from a Kraus decomposition, as described in \cref{sec:Gaussian_chi_representation}. Alternatively, we derive the characteristic function of an arbitrary Gaussian CPTP map in \cref{sec:Gaussian_chi_representation}, giving us many of the characteristic functions used below. Finally, the characteristic function $c(\vect{u},\vect{v})$ of a composition of maps $\mathcal{N}=\mathcal{N}_{2}\circ\mathcal{N}_{1}$ can be written in terms of the characteristic functions of the maps $\mathcal{N}_{i}$ with the equation:
\begin{multline}\label{eq:chi_composition}
    c(\vect{u},\vect{v})=\iint_{\mathrlap{\mathbb{R}^{2n}}}\;d^{2n}\vect{\tilde{u}}\,d^{2n}\vect{\tilde{v}}\bigg(e^{i\pi(\vect{u}^{T}\!\Omega\vect{\tilde{u}}-\vect{v}^{T}\!\Omega\vect{\tilde{v}})}\\
    \times c_{1}(\vect{u}-\vect{\tilde{u}},\vect{v}-\vect{\tilde{v}})\,c_{2}(\vect{\tilde{u}},\vect{\tilde{v}})\bigg),
\end{multline}
where $c_{i}(\vect{u},\vect{v})$ is the characteristic function of $\mathcal{N}_{i}$. \Cref{eq:chi_composition} can be used, for instance, to calculate the characteristic function of the map $\mathcal{N}\circ\mathcal{C}$, where $\mathcal{C}$ is the non-ideal codestate map from the previous paragraph.

To calculate the logical noise map, we first apply $\mathcal{N}$ to an ideal codestate via the equation
\begin{equation}\label{eq:noise_stabiliser_states}
    \mathcal{N}\big(\!\ket{\psi,\vect{0}}\!\bra{\psi,\vect{0}}\!\big)=\iint_{\mathrlap{\mathbb{R}^{2n}}}\;d^{2n}\vect{u}\,d^{2n}\vect{v}\;c(\vect{u},\vect{v})\ket{\psi,\vect{u}}\!\bra{\psi,\vect{v}},
\end{equation}
where we have omitted the $(\Sigma,\vect{d})$ label of the stabilizer states to save space. We can see \cref{eq:noise_stabiliser_states} holds from the definition of stabilizer states as displaced ideal codestates [\cref{eq:stabiliser_states}]. Moreover, \cref{eq:noise_stabiliser_states} can be applied to an arbitrary ideal encoded density matrix $\hat{\rho}\otimes_{\mathcal{G}}\ket{\vect{0}}\!\bra{\vect{0}}$ using the linearity of $\mathcal{N}$.

The use of stabilizer states on the right-hand side of \cref{eq:noise_stabiliser_states} is necessary due to the domain of integration $\mathbb{R}^{2n}$. Therefore, to write \cref{eq:noise_stabiliser_states} in terms of the subsystem decomposition $\otimes_{\mathcal{G}}$, we must apply the boundary conditions \cref{eq:BCs}, giving
\begin{multline}\label{eq:noise_subsystem_decomp}
    \mathcal{N}\big(\hat{\rho}\otimes_{\mathcal{G}}\!\ket{\vect{0}}\!\bra{\vect{0}}\!\big)=\;\sum_{\mathclap{\vect{s},\vect{t}\in\mathbb{Z}^{2n}}}\;\;\iint_{\mathrlap{\mathcal{P}}}\;d^{2n}\vect{u}\,d^{2n}\vect{v}\bigg(\!c_{\vect{s},\vect{t}}(\vect{u},\vect{v})\\
    \times\hat{P}_{\vect{d}}(\vect{s})\hat{\rho}\hat{P}_{\vect{d}}(\vect{t})^{\dag}\otimes_{\mathcal{G}}\ket{\vect{u}}\!\bra{\vect{v}}\!\bigg),
\end{multline}
where $\hat{P}_{\vect{d}}(\vect{s})$ is a Pauli operator defined in \cref{eq:general_Pauli}, and
\begin{equation}
    c_{\vect{s},\vect{t}}(\vect{u},\vect{v})=c\big(\vect{u}+\vect{\bar{\ell}}(\vect{s}),\vect{v}+\vect{\bar{\ell}}(\vect{t})\big)e^{i\pi\left(\vect{u}^{T}\!\Omega\vect{\bar{\ell}}(\vect{s})-\vect{v}^{T}\!\Omega\vect{\bar{\ell}}(\vect{t})\right)},
\end{equation}
with $\vect{\bar{\ell}}(\vect{s})=\sum_{J=1}^{2n}s_{J}\vect{\bar{m}}_{J}$. Note that applying the boundary conditions has introduced logical Pauli operators on the logical subsystem, altering the logical information stored in the state.

Finally, we obtain the logical noise map by taking the partial trace over the stabilizer subsystem, giving
\begin{equation}\label{eq:logical_noise_channel}
    \mathcal{N}_{\mathcal{L}}(\hat{\rho})=\;\sum_{\mathclap{\vect{s},\vect{t}\in\mathbb{Z}^{2n}}}\;\bigg(\int_{\mathcal{P}}\!\!d^{2n}\vect{v}\,c_{\vect{s},\vect{t}}(\vect{v},\vect{v})\!\bigg)\,\hat{P}_{\vect{d}}(\vect{s})\hat{\rho} \hat{P}_{\vect{d}}(\vect{t})^{\dag}.
\end{equation}

In order to evaluate the logical noise map \cref{eq:logical_noise_channel} in practice, we need to truncate the infinite series over $\vect{s},\vect{t}$. This truncation is possible if the modulus of the characteristic function $|c(\vect{u},\vect{v})|$ goes to $0$ sufficiently fast as $|\vect{u}|,|\vect{v}|\rightarrow\infty$. Indeed, in some cases this property does not hold, for example in the unphysical scenario of applying pure loss to an ideal codestate, and in these cases our method cannot be applied in practice. However, in all the other examples we consider, $|c(\vect{u},\vect{v})|$ decays to 0 exponentially as $|\vect{u}|,|\vect{v}|\rightarrow\infty$, and we find that a truncation of each component of $\vect{s}$, $\vect{t}$ to $-1,0,1$ is sufficient to achieve accurate results in almost all regimes of interest, as discussed in more detail later in this section. In the case of the square GKP code, the integrals in \cref{eq:logical_noise_channel} can often be analytically evaluated due to the Cartesian product structure of the square Voronoi cell $\mathcal{V}_{\text{sq}}$ [\cref{eq:sq_G}]. However, for subsystem decompositions with more complicated primitive cells $\mathcal{P}$, the integrals must be evaluated numerically. In this work we only consider square GKP codes for simplicity.

With the logical noise map, one can then directly extract quantities such as the average gate fidelity, which is defined as~\cite{Nielsen02Simple}
\begin{equation}
    \mathcal{F}(\mathcal{E})=\int d\psi\braket{\psi|\mathcal{E}(\ket{\psi}\!\bra{\psi})|\psi}
\end{equation}
for a quantum channel $\mathcal{E}$ acting on a discrete-variable Hilbert space $\mathcal{H}_{\vect{d}}$, where the integral is over the uniform measure $d\psi$ in state space. In practice this is most easily calculated via the entanglement fidelity
\begin{equation}\label{eq:entanglement_fid}
    \mathcal{F}_{e}(\mathcal{E})=\braket{ME|(\mathcal{I}\otimes\mathcal{E})(\ket{ME}\!\bra{ME})|ME},
\end{equation}
where $\ket{ME}\in\mathcal{H}_{\vect{d}}\otimes\mathcal{H}_{\vect{d}}$ is a maximally entangled state, and $\mathcal{I}$ is the identity quantum channel acting on $\mathcal{L}(\mathcal{H}_{\vect{d}})$. Then, we can use the relationship~\cite{Nielsen02Simple}
\begin{equation}\label{eq:Nielsen_simple}
    \mathcal{F}(\mathcal{E})=\frac{d\mathcal{F}_{e}(\mathcal{E})+1}{d+1},
\end{equation}
where $d=d_{1}\times d_{2}\times\cdots\times d_{n}$ is the dimension of the Hilbert space $\mathcal{H}_{\vect{d}}$. We note that in order to calculate the average gate fidelity, the logical noise map (and, equivalently, the noise map itself) must be CPTP.

Here we comment that the methods described above reflect the logical information one would obtain using ideal error correction, which is not possible in practice since it requires the use of ideal codestates. While we do not quantitatively model the more realistic case of error correction with approximate codestates, we briefly discuss the qualitative effect that this would have on our subsequent results. For small enough $\Delta$, one can approximately model the use of approximate codestates and inefficient homodyne detection in a teleportation-based error correction scheme as applying a Gaussian random displacement channel immediately before and after a round of ideal error correction, see for example Appendix F of Ref.~\cite{Rozpedek23}. To approximately model the effect of approximate error correction, we can simply post-compose the noise map $\mathcal{N}$ considered above with the an appropriate Gaussian random displacement channel. Such a modification would reduce the fidelity of the resulting logical channel and introduce more errors. However, the qualitative features described below, such as the presence of a global minimum in the infidelity, should not be affected since the stochastic displacement errors only add decoherently to the noise map $\mathcal{N}$.

To summarize, if one wishes to calculate the logical noise map corresponding to applying a noise map $\mathcal{N}$ on some non-ideal GKP codewords, one can follow the following steps:
\begin{enumerate}
    \item First, determine the map $\mathcal{C}$ that maps ideal codestates to non-ideal codestates.
    \item Then, calculate the characteristic function of the combined map $\mathcal{N}\circ\mathcal{C}$ [\cref{sec:Gaussian_chi_representation,eq:chi_composition}].
    \item Finally, substitute this into \cref{eq:logical_noise_channel} by truncating the infinite sums.
\end{enumerate}
These are the steps that we follow for the square GKP code in the remainder of this section. We start by looking at non-ideal codestates defined by the envelope operator in \cref{subsec:envelope}, and follow this by applying the loss (\cref{subsec:loss}), Gaussian random displacement (\cref{subsec:Gaussian}) and white-noise dephasing (\cref{subsec:dephasing}) noise channels to the codestates.

\subsection{Envelope Operator}\label{subsec:envelope}

We define approximate GKP codewords as
\begin{equation}\label{eq:approx}
    \ket{\bar{\psi}_{\Delta}}\propto e^{-\Delta^{2}\hat{a}^{\dag}\hat{a}}\ket{\bar{\psi}},
\end{equation}
where $\Delta$ quantifies the quality of the approximate codewords, and the constant of proportionality is defined such that $\ket{\bar{\psi}_{\Delta}}$ is normalized. $\Delta$ is also commonly quoted in decibels as $\Delta_{\mathrm{dB}}=-10\log_{10}(\Delta^{2})$. We call the non-unitary operator $e^{-\Delta^{2}\hat{a}^{\dag}\hat{a}}$ the \textit{envelope operator}, and we define the envelope map as $\mathcal{E}=\mathcal{J}[e^{-\Delta^{2}\hat{a}^{\dag}\hat{a}}]$, where $\mathcal{J}[\hat{O}](\hat{\rho})=\hat{O}\hat{\rho}\hat{O}^{\dag}$.

Since the envelope operator is non-unitary, the envelope map $\mathcal{E}$ is not trace-preserving. Moreover, the map that takes ideal codestates to normalized approximate codestates is not linear. To solve these problems, we apply an orthonormalization procedure (described in \cref{sec:orthonormalisation}) to the approximate codewords $\ket{\bar{\psi}_{\Delta}}\mapsto\ket{\bar{\psi}_{\Delta,\text{o}}}$ such that the orthonormalized encoding map $\mathcal{E}^{\Delta}_{\text{o}}$ is CPTP. The orthonormalization has a vanishingly small effect on the approximate codestates themselves as $\Delta\rightarrow0$. Conveniently, we can calculate the orthonormalized CPTP logical noise map $\mathcal{E}^{\Delta}_{\mathcal{L},\text{o}}$ from the non-trace-preserving \textit{logical} noise map $\mathcal{E}_{\mathcal{L}}^{\Delta}$ after we have applied the steps in \cref{subsec:method_overview}.

The envelope map has characteristic function (\cref{sec:Gaussian_chi_representation})
\begin{equation}\label{eq:envelope_chi}
    c_{\mathcal{E}}^{\Delta}(\vect{u},\vect{v})\propto \exp\bigg(\!{-}\frac{\pi}{2}\mathrm{coth}\big(\Delta^{2}/2\big)\big(|\vect{u}|^{2}+|\vect{v}|^{2}\big)\!\bigg),
\end{equation}
where we have omitted the constant of proportionality since the codewords are going to be orthonormalized anyway. Importantly, since the characteristic function \cref{eq:envelope_chi} decreases exponentially as $|\vect{u}|,|\vect{v}|\rightarrow\infty$, we can truncate the sum in \cref{eq:logical_noise_channel} to numerically obtain the logical envelope map $\mathcal{E}_{\mathcal{L}}^{\Delta}$. Moreover, each integral in \cref{eq:logical_noise_channel} can be evaluated analytically for the square GKP code (see \cref{sec:analytic_partial_trace}). Then, we apply the orthonormalization procedure in \cref{sec:orthonormalisation} to obtain a CPTP map $\mathcal{E}^{\Delta}_{\mathcal{L},\text{o}}$. Finally, we can calculate the average gate fidelity of $\mathcal{E}^{\Delta}_{\mathcal{L},\text{o}}$ using \cref{eq:entanglement_fid,eq:Nielsen_simple}. This average gate fidelity is plotted in the $\gamma=0$ curve of \cref{fig:loss_plots}(a). Since $c_{\mathcal{E}}^{\Delta}(\vect{u},\vect{v})$ becomes sharper around the origin as $\Delta\rightarrow0$, this method becomes more accurate and less computationally expensive as the approximation becomes more ideal. Indeed, our technique easily enables the simulation of highly squeezed states, and in principle there is no limit to the squeezing one could simulate (see \cref{sec:very_small_Delta}). Our result is similar in spirit to those of Ref.~\cite{Calcluth22}, where it is shown that circuits that are efficiently simulatable by classical computers with ideal $\ket{\bar{0}}$ states as input states are universal for quantum computing with approximate states as input, and the resourcefulness of the approximate states increases as $\Delta$ increases.

Let us briefly comment on the accuracy of the truncation of the infinite sum \cref{eq:logical_noise_channel}. To do this, we take the largest value of $\Delta$ that we used in our simulations ($\Delta\approx 0.94$, $\bar{n}\approx0.56$, $\Delta_{\text{dB}}\approx3.1$) and compare the average gate infidelity across different levels of truncation. In particular, we truncate the absolute value of each component of $\vect{s},\vect{t}$ to be less than or equal to some integer $s_{\text{max}}$. We find that the relative error
\begin{equation}\label{eq:rel_error_smax}
    \frac{\big|\mathcal{F}(s_{\text{max}}=2)-\mathcal{F}(s_{\text{max}}=1)\big|}{1-\mathcal{F}(s_{\text{max}}=2)}
\end{equation}
between the calculated average gate infidelities for $s_{\text{max}}=1$ and $2$ is only $7\times10^{-4}$, and the relative error between $s_{\text{max}}=2$ and $3$ is only $4\times10^{-9}$. Moreover, these relative errors should only decrease as $\Delta$ becomes smaller and the codestates become more ideal. These results show that the sum \cref{eq:logical_noise_channel} converges extremely rapidly even for large $\Delta$, allowing $s_{\text{max}}$ to be set to 1 for all our remaining calculations.

\subsection{Pure Loss}\label{subsec:loss}

\begin{figure*}
    \centering
    \includegraphics{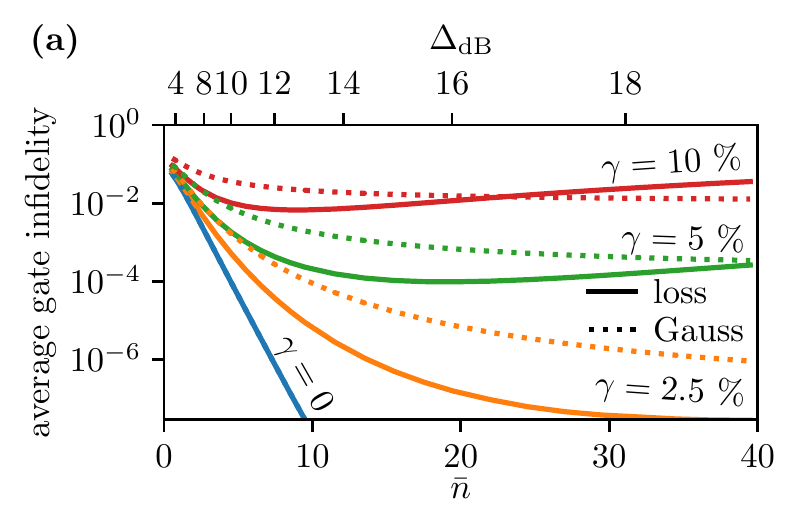}
    \hspace*{\fill}
    \includegraphics{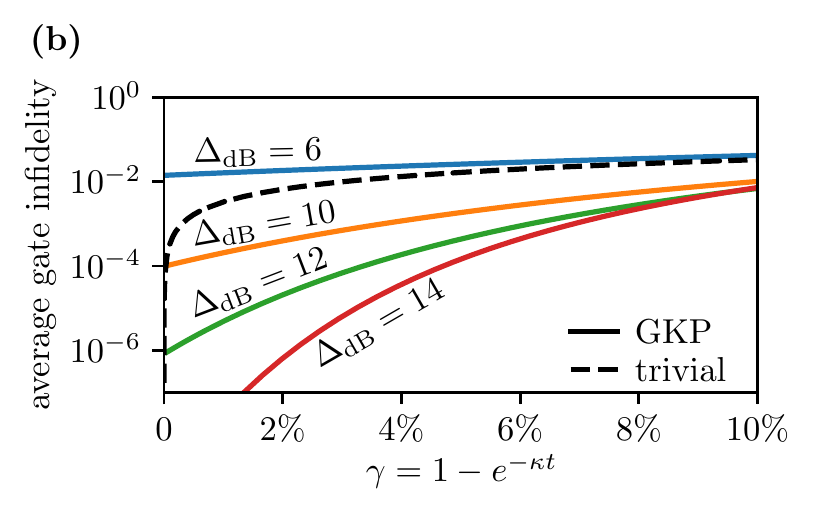}
    \caption{(Color) Average gate infidelities of the logical noise channels corresponding to loss acting on approximate single-mode square GKP qubit codestates $(\mathcal{L}^{\gamma}\circ\mathcal{E}^{\Delta})_{\mathcal{L},\text{o}}$, and random Gaussian displacements acting on approximate GKP codestates $(\mathcal{G}^{\sigma}\circ\mathcal{E}^{\Delta})_{\mathcal{L},\text{o}}$, where $\sigma^{2}=\gamma/(1-\gamma)$ is scaled such that the Gaussian displacement channel is equivalent to loss followed by a quantum-limited amplification via \cref{eq:gauss_loss_relationship}. (a) The average gate infidelities plotted as a function of $\Delta$, where $\bar{n}\approx1/(2\Delta^{2})-1/2$ is the average photon number of the approximate GKP encoded maximally mixed state. The $\gamma=0$ curve represents the errors solely resulting from the approximate GKP codestates. (b) Average gate fidelity of the loss channel as a function of $\gamma\approx\kappa t$, which represents the loss of logical information as approximate GKP codestates evolve in time under loss. This is compared to the loss of logical information stored in a trivial Fock encoding $\{\ket{0},\ket{1}\}$ under the same loss channel.}
    \label{fig:loss_plots}
\end{figure*}

We define loss from the Kraus operators
\begin{subequations}\label{eq:loss}
\begin{gather}
    \mathcal{L}^{\gamma}(\hat{\rho})=\sum_{j=0}^{\infty}\hat{L}_{j}^{\gamma}\hat{\rho}\hat{L}_{j}^{\gamma\dag},\\
    \hat{L}^{\gamma}_{j}=\Big(\frac{\gamma}{1-\gamma}\Big)^{j/2}\frac{\hat{a}^{j}}{\sqrt{j!}}(1-\gamma)^{\hat{n}/2},
\end{gather}
\end{subequations}
with characteristic function (\cref{sec:Gaussian_chi_representation})
\begin{equation}\label{eq:loss_chi}
    c_{\mathcal{L}}^{\gamma}(\vect{u},\vect{v})\propto\exp\bigg({-}\frac{\pi}{2}\frac{(1+\sqrt{1-\gamma})^{2}}{\gamma}\big(|\vect{u}-\vect{v}|^{2}+2i\vect{u}^{T}\Omega\vect{v}\big)\!\bigg).
\end{equation}
Here, $\gamma=1-e^{-\kappa t}$ represents the amount of loss applied to a system evolving under the master equation
\begin{equation}\label{eq:loss_me}
    \dot{\hat{\rho}}=\kappa\mathcal{D}[\hat{a}]\hat{\rho}
\end{equation}
for some time $t$, where $\mathcal{D}[\hat{a}]\hat{\rho}=\hat{a}\hat{\rho}\hat{a}^{\dag}-\frac{1}{2}\{\hat{a}^{\dag}\hat{a},\hat{\rho}\}$ is the Lindblad dissipator and $\{A,B\}=AB+BA$ is the anticommutator.

Simulating loss acting directly on ideal GKP codestates using our method is not possible since the modulus $|c_{\mathcal{L}}^{\gamma}(\vect{u},\vect{v})|$ is constant for $\vect{u}=\vect{v}$, even as $|\vect{u}|\rightarrow\infty$. Instead, we simulate loss acting on approximate codestates by considering a composition of maps $\mathcal{L}^{\gamma}\circ\mathcal{E}^{\Delta}$, whose characteristic function can be determined from \cref{eq:chi_composition}. Again, we calculate the logical map by truncating the infinite series \cref{eq:logical_noise_channel}, and we must orthonormalize the codewords using \cref{sec:orthonormalisation}, resulting in the orthonormalized logical map $(\mathcal{L}^{\gamma}\circ\mathcal{E}^{\Delta})_{\mathcal{L},\text{o}}$.

We plot the average gate fidelity of $(\mathcal{L}^{\gamma}\circ\mathcal{E}^{\Delta})_{\mathcal{L},\text{o}}$ for the square GKP code in \cref{fig:loss_plots}, and compare it to the average gate fidelity of loss acting on the trivial Fock encoding, $\{\ket{0},\ket{1}\}$, which represents the break-even point of the code. We find that for any fixed loss rate $\gamma$, there is an optimal $\Delta$ that maximizes the average gate fidelity. One can understand this intuitively by noting that loss has a larger effect on states with a large photon number. Since the average photon number of an approximate GKP codestate is given approximately by $\langle\hat{a}^{\dag}\hat{a}\rangle\approx1/(2\Delta^{2})-1/2$~\cite{Albert18}, as the GKP codestate becomes more ideal, the noise due to $\mathcal{E}^\Delta$ decreases while the noise due to $\mathcal{L}^{\gamma}$ increases, giving rise to a cross-over point which represents the optimal $\Delta$ that protects against the loss. As previously mentioned, we expect that these results would hold qualitatively (if not quantitatively) in the more realistic case of error-correction with approximate codestates.

Based on experimental parameters used in a recent paper that produced GKP Pauli eigenstates~\cite{Sivak22}, we can obtain a rough, order-of-magnitude estimate of the loss rates we might expect in near-term GKP experiments. Given a cavity lifetime $T_{1,\text{c}}\sim 600\ \mu$s, and an error-correction cycle time of $6$ $\mu$s, we can estimate the amount of loss we can expect during an error correction cycle to be on the order of $\gamma\sim 1\%$. At this loss rate, GKP codes perform very well, and the optimal $\Delta$ is far beyond what is experimentally feasible. Encouragingly, the square GKP code still outperforms the trivial encoding even for much larger amounts of loss.

\subsection{Gaussian Displacements}\label{subsec:random_walk}

Next, we consider a Gaussian random displacement noise model $\mathcal{G}^{\sigma}$ defined by its characteristic function
\begin{equation}\label{eq:chi_random_displacement}
    c_{\mathcal{G}}^{\sigma}(\vect{u},\vect{v})=\sigma^{-2}\exp\!\big({-}\pi|\vect{u}|^{2}/\sigma^{2}\big)\,\delta^{2}(\vect{u}-\vect{v}).
\end{equation}
Here, $\sigma^{2} = \kappa_{G} t$ represents the variance of random displacements applied to a system evolving under the master equation
\begin{equation}
    \dot{\hat{\rho}}=\kappa_{G}\big(\mathcal{D}[\hat{a}]+\mathcal{D}[\hat{a}^{\dag}]\big)\hat{\rho}
\end{equation}
for some time $t$. The Gaussian random-displacement channel $\mathcal{G}^{\sigma}$ is equivalent to a loss channel $\mathcal{L}^{\gamma}$ followed by a quantum-limited amplification channel~\cite{Noh18}
\begin{equation}\label{eq:gauss_loss_relationship}
    \mathcal{G}^{\sqrt{\gamma/(1-\gamma)}}=\mathcal{A}^{1/(1-\gamma)}\circ\mathcal{L}^{\gamma},
\end{equation}
where the quantum-limited amplification channel $\mathcal{A}^{g}(\hat{\rho})$ is equivalent to the state $\hat{\rho}$ evolving under the master equation
\begin{equation}\label{eq:amplification_me}
    \dot{\hat{\rho}}=\kappa_{A}\mathcal{D}[\hat{a}^{\dag}]\hat{\rho}
\end{equation}
for some time $t$, where $g=e^{\kappa_{A}t}$.

Using the same method as for loss, we calculate the orthonormalized logical noise map for the Gaussian random-displacement channel acting on approximate GKP codewords, and plot the average gate fidelity in \cref{fig:loss_plots}(a). To compare loss to Gaussian random-displacements, we use the relationship $\sigma^{2}=\gamma/(1-\gamma)$ from \cref{eq:gauss_loss_relationship}. For equivalent values of $\sigma$ and $\gamma$, $\mathcal{G}^{\sigma}$ introduces more noise into the system than $\mathcal{L}^{\gamma}$ in the region of small $\Delta_{\mathrm{dB}}$ and $\gamma$. However, the infidelity of $\mathcal{G}^{\sigma}$ becomes smaller than that of $\mathcal{L}^{\gamma}$ at large values of $\Delta_{\mathrm{dB}}$ or $\gamma$, reflecting the fact that $c_{\mathcal{G}}^{\sigma}(\vect{u},\vect{v})$ tends to $0$ as $|\vect{u}|,|\vect{v}|\rightarrow\infty$ even when acting on ideal codestates.

We again comment briefly on the accuracy of the truncation of \cref{eq:logical_noise_channel}. The data point with the highest infidelity in \cref{fig:loss_plots} corresponds to a Gaussian random displacement channel with variance $\sigma^{2}=0.1/(1-0.1)\approx0.11$ applied to a GKP codestate with very large $\Delta$ ($\Delta\approx0.94$, $\bar{n}\approx0.56$, $\Delta_{\text{dB}}\approx3.1$). Considering different truncations $s_{\text{max}}$ as in \cref{eq:rel_error_smax}, we find that the relative error in the average gate infidelities for this $\Delta$ and $\sigma^{2}$ between $s_{\text{max}}=1$ and $2$ is $2\times10^{-3}$, and between $s_{\text{max}}=2$ and $3$ is $2\times10^{-7}$, indicating that our truncation is still valid even for the channel with the largest infidelity.

The consequences of our results for correcting GKP codes against loss are interesting. In particular, in the regime of large $\gamma$ and large $\Delta_{\mathrm{dB}}$, it is better to apply a quantum-limited amplification channel before performing a standard round of GKP error-correction, as discussed in Ref.~\cite{Noh18}. The fact that this scheme outperforms pure loss followed by standard error-correction is not necessarily surprising, since the amplification is designed with knowledge of the noise model acting on the system, while standard error-correction is not. However, loss without amplification \textit{does} outperform loss followed by amplification in the small $\gamma$ and small $\Delta_{\mathrm{dB}}$ regime that is likely to be experimentally-relevant. This result is consistent with the results of Ref.~\cite{Hastrup22}.

\subsection{White-noise Dephasing}\label{subsec:dephasing}

\begin{figure*}
    \centering
    \includegraphics{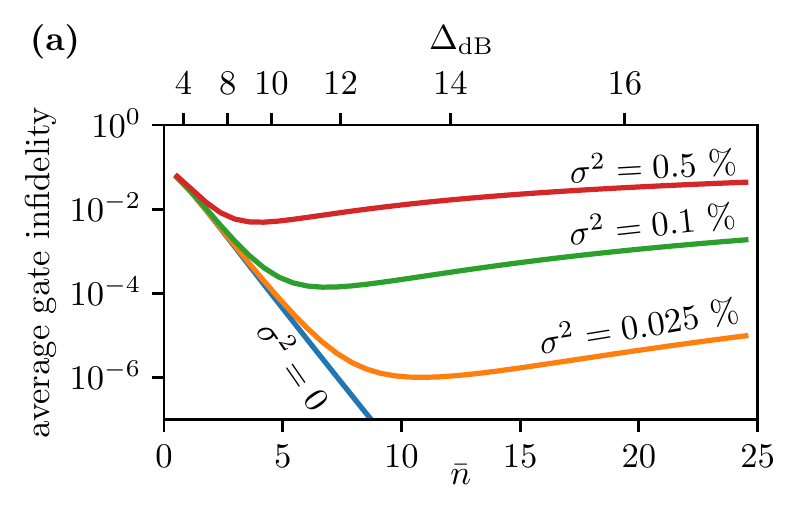}
    \hspace*{\fill}
    \includegraphics{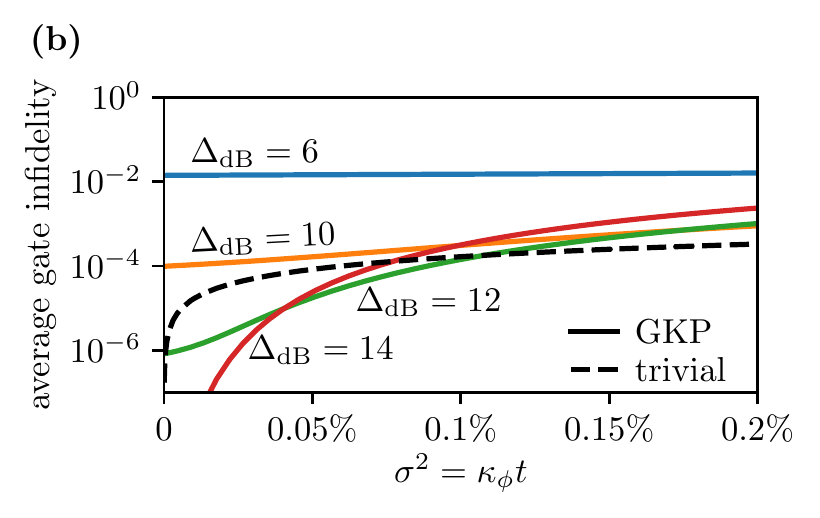}
    \caption{(Color) Average gate infidelities of the logical noise channels corresponding to dephasing acting on approximate single-mode square GKP qubit codestates $(\mathcal{D}^{\sigma}\circ\mathcal{E}^{\Delta})_{\mathcal{L},\text{o}}$. (a) The average gate infidelities plotted as a function of $\Delta$, where $\bar{n}\approx1/(2\Delta^{2})-1/2$ is the average photon number of the approximate GKP encoded maximally mixed state. (b) Average gate infidelities as a function of $\sigma^{2}=\kappa_{\phi} t$, which represents the loss of logical information as approximate GKP codestates evolve in time under dephasing. This is compared to the loss of logical information stored in a trivial Fock encoding $\{\ket{0},\ket{1}\}$ under the same pure dephasing channel.}
    \label{fig:dephasing_plots}
\end{figure*}

Finally, we conclude our numerical analysis by considering a white-noise dephasing channel, defined by its continuous Kraus decomposition
\begin{equation}\label{eq:dephasing}
    \mathcal{D}^{\sigma}(\hat{\rho})=\frac{1}{\sqrt{2\pi\sigma^{2}}}\int_{\mathbb{R}}\!d\phi\;e^{-\phi^{2}/(2\sigma^{2})}e^{i\phi \hat{a}^{\dag}\hat{a}}\hat{\rho} e^{-i\phi\hat{a}^{\dag}\hat{a}},
\end{equation}
and with characteristic function (\cref{sec:Gaussian_chi_representation})
\begin{equation}\label{eq:dephasing_chi}
    c_{\mathcal{D}}^{\sigma}(\vect{u},\vect{v})=\frac{1}{4\sqrt{2\pi\sigma^{2}}}\int_{\mathbb{R}}\!d\phi\;\frac{e^{-\phi^{2}/(2\sigma^{2})}}{\sin^{2}(\phi/2)}e^{-\frac{i\pi}{2}\mathrm{cot}\left(\!\frac{\phi}{2}\!\right)\left(|\vect{u}|^{2}-|\vect{v}|^{2}\right)}.
\end{equation}
This time, $\sigma^{2} = \kappa_{\phi}t$ represents the amount of dephasing applied to a system evolving under the master equation
\begin{equation}
    \dot{\hat{\rho}}=\kappa_{\phi}\mathcal{D}[\hat{a}^{\dag}\hat{a}]\hat{\rho}
\end{equation}
for some time $t$. The term \textquotedblleft white-noise\textquotedblright\ refers to the Gaussian distribution of the Kraus operators~\cref{eq:dephasing}, which is a simplification of general dephasing errors that can occur in an experiment. Although our results should give a general indication of the performance of GKP codes against dephasing, typical experiments observe dephasing with non-Gaussian, or colored, distributions.

Again, we calculate the orthonormalized logical noise channel $(\mathcal{D}^{\sigma}\circ\mathcal{E}^{\Delta})_{\mathcal{L},\text{o}}$ as a function of both $\Delta$ and $\sigma$. In the case of dephasing however, our result is not purely analytical as we perform the integral over $\phi$ arising from \cref{eq:dephasing_chi} numerically. The average gate fidelity of the logical noise channel is plotted in \cref{fig:dephasing_plots}, and compared to pure dephasing applied to the trivial Fock encoding $\{\ket{0},\ket{1}\}$. For any given $\sigma$, we again find an optimal $\Delta$ which corrects against the dephasing, which we also expect to hold qualitatively in the case of approximate error-correction. This can be understood intuitively by noting that dephasing applies random rotations to a state in phase space. As $\Delta\rightarrow0$, the approximate GKP codestate becomes more widely distributed in phase space and is thus more vulnerable to the effects of dephasing.

Comparing our results to the experimental data in Ref.~\cite{Sivak22}, we have $T_{\phi}\approx 1$ ms. At this rate, dephasing acting over a time of $6$ $\mu$s results in $\sigma^{2}\approx 0.6\%$, although this may be reduced if error correction can be performed faster in the future. In contrast to loss, this amount of dephasing severely affects GKP codestates, underperforming the trivial encoding for all values of $\Delta$. Indeed, for a GKP squeezing of $\Delta_{\mathrm{dB}}=10$, one needs to be performing error correction on a timescale of approximately $2$ $\mu$s to achieve an average gate infidelity of $10^{-3}$ even with state-of-the-art dephasing rates. Moreover, to break-even against a pure dephasing channel, one needs to be operating with a squeezing of greater than $\Delta_{\mathrm{dB}}=10$ \textit{and} a dephasing rate less than $0.1\%$. 

\section{Conclusion}\label{sec:conc}

We began the paper by constructing the stabilizer subsystem decomposition for the single-mode square GKP qubit code in \cref{sec:sq}, and discussing its key properties. In \cref{sec:notation} we introduced multi-mode GKP codes by describing the parameters $(\Sigma,\vect{d})$ that specify the lattice generators of a given GKP code. Then in \cref{sec:definition}, we introduced the stabilizer subsystem decomposition $\mathcal{H}=\mathcal{L}\otimes_{\mathcal{G}}\mathcal{S}$ of the CV Hilbert space into a logical and stabilizer subspace in the general case. We showed that the partial trace $\mathrm{tr}_{\mathcal{S}}$ corresponds to ideal decoding over the patch $\mathcal{P}$, justifying our use of the stabilizer subsystem decomposition over previously-developed alternatives. We explored the properties of the subsystem decomposition, and used three transformations (cell, Gaussian and dimension transformations) to connect states in the stabilizer subsystem decomposition to Zak states (\cref{sec:Zak}) and to describe logical Clifford gates in our formalism (\cref{sec:gates}). Finally, in \cref{sec:errors} we introduced a general method to calculate the logical effect of various noise operators on GKP codes, which we then used to analyze the envelope operator, pure loss channels, and Gaussian displacement channels in regimes that are unreachable by Fock space simulations.

In \cref{sec:gates,sec:errors} we provided a number of exciting applications of the stabilizer subsystem decomposition to implement logical gates and analyze the effects of noise. However, there is significantly more work that must be done to provide analysis that is useful for experiments. In particular, one could incorporate other techniques from fault-tolerant literature, such as gate teleportation, subsystem codes \cite{Poulin05} and gauge fixing \cite{Vuillot19-2}, into the GKP setting using our formalism. Furthermore, analysing more realistic noise sources such as colored dephasing, Kerr non-linearities, and approximate error correction is required to model the dominant sources of errors in current experiments. In particular, analysing the dissipative GKP error-correction schemes used in Ref.~\cite{Royer20,DeNeeve22} in terms of the subsystem decomposition may provide insights into how the schemes work. We add that our formalism may allow for more accurate simulations of concatenated GKP codes using the logical action of noise channels and logical gates defined by the stabilizer decomposition.

Finally, we note that our simulation methods rely on the integration of characteristic functions over a cell $\mathcal{P}$. In this work, we were able to produce our results by evaluating these integrals analytically over the single-mode square GKP Voronoi cell. However, higher-dimensional integrals over non-rectangular and/or multi-mode patches cannot be evaluated analytically and thus require numerical integration which may not scale favourably. As such, there is still work that can be done to optimize the numerical integration methods for multi-mode GKP codes.

\section{Acknowledgements}\label{sec:acknowledgements}
We acknowledge support from Australian Research Council via the Centre of Excellence in Engineered Quantum Systems (EQUS) project number CE170100009. MHS is also supported by an Australian Government Research Training Program (RTP) Scholarship. MHS would like to thank Giacomo Pantaleoni, Ben Baragiola, and Nicolas Menicucci for their discussions about subsystem decompositions and the Zak basis.
ALG is supported by the Australian Research Council,
through an Discovery Early Career Research Award project number
DE190100380.
We acknowledge the traditional owners of the land on which this work was undertaken at the University of Sydney, the Gadigal people of the Eora Nation.

\bibliography{my_bib}
\appendix
\section{Comparison to Ref.~\cite{Pantaleoni20}}\label{sec:comparison_to_giacomo}
In this appendix we briefly compare the stabilizer subsystem decomposition to the modular subsystem decomposition~\cite{Pantaleoni20}, for the single-mode square qubit GKP code $\mathcal{G}_{\text{sq}}=(I_{2},(2),\mathcal{V}_{\text{sq}})$ [\cref{eq:sq_G}]. We show that the stabilizer subsystem decomposition is not equivalent to the decomposition of Ref.~\cite{Pantaleoni20} by explicitly decomposing example states and operators into each subsystem decomposition. We also show explicitly the different choices of phase in the definitions of each decomposition in the Zak basis which lead to their different properties. Since the stabilizer subsystem decomposition is designed to describe ideal GKP codes, one way to view the decompositions of Ref.~\cite{Pantaleoni20} is that it defines a bosonic code that shares the same ideal codespace as the GKP code, but where the partial trace operation represents a non-ideal error correction procedure.

In order to define a modular subsystem decomposition following Ref.~\cite{Pantaleoni20}, one must choose a quadrature in which to decompose the Hilbert space $\mathcal{H}$. We choose the position basis for this purpose and refer to this decomposition as the modular-position subsystem decomposition ($\mathcal{Q}$).  Any real number $x$ can be decomposed into a sum 
\begin{equation}\label{eq:mod_x_decomposition}
    x=\sqrt{\pi}(2s+\mu)+r,
\end{equation}
where $s \in\mathbb{Z}$, $\mu\in\mathbb{Z}_{2}$, and $r\in(-\sqrt{\pi}/2,\sqrt{\pi}/2]$. Introducing the modular notation: $x=a\lfloor x\rceil_{a}+\{x\}_{a}$, where $\lfloor x\rceil_{a}\in\mathbb{Z}$ and $\{x\}_{a}\in(-a/2,a/2]$, we can write $r=\{x\}_{\sqrt{\pi}}$, $s=\big\lfloor\lfloor x\rceil_{\sqrt{\pi}}\big\rceil_{2}$ and $\mu=\big\{\lfloor x\rceil_{\sqrt{\pi}}\big\}_{2}$. Then, the modular-position subsystem decomposition is defined on the position eigenstates as
\begin{equation}\label{eq:mod_q_subsystem}
\ket{x}_{q}=\ket{\mu}\otimes_{\mathcal{Q}}\ket{\sqrt{\pi}s+r}_{q},
\end{equation}
where we use the subscript $q$ on the non-logical ``gauge’’ mode to signify that the state is a position eigenstate of the gauge mode. 

Due to this choice, the position and momentum quadratures are not treated symmetrically, resulting in a number of undesirable properties in the subsystem decomposition. To make our point explicit, let us give two concrete examples demonstrating the asymmetries of the modular-position subsystem decomposition. First, the modular-position subsystem decomposition treats the position and momentum quadratures asymmetrically, and as such the Fourier transform operator $e^{i\pi\hat{a}^{\dag}\hat{a}/2}$ does not have a neat decomposition over the subsystem. In contrast, we recall that the Fourier transform operator can be written in the stabilizer subsystem decomposition as a product of operators acting on $\mathcal{L}$ and $\mathcal{S}$ as
\begin{equation}
    e^{i\pi \hat{a}^{\dag}\hat{a}/2}=\hat{H}\otimes_{\text{sq}}\hat{R}(\pi/2)\tag{\ref{eq:sq_Had}},
\end{equation}
where $\hat{R}(\pi/2)\ket{k_{1},k_{2}}=\ket{k_{1},-k_{2}}$ rotates the vector $(k_{1},k_{2})$ by an angle $\pi/2$ anticlockwise.

Second, we take the partial trace of the two CV states $\ket{\phi_{\pm}}=\big(\!\ket{0}_{q}+\ket{\pm\sqrt{\pi}}_{q}\!\big)/\sqrt{2}$ (see \cref{tab:decoders}). These states are chosen to reveal an asymmetry between
left and right displacements in position in the modular-position subsystem decomposition, since the partial trace results in a pure final state $\ket{+}\!\bra{+}$ for $\ket{\phi_{+}}$ but the maximally mixed state $\hat{\rho}=\hat{I}/2$ for $\ket{\phi_{-}}$. In contrast, the stabilizer subsystem decomposition gives the same logical state for both CV states.

\begin{table}[]
    \centering
    \begin{tabular}{c|c|c}
    CV state & stabilizer partial trace & Mod-$q$ partial trace \\
    \hline
    $\displaystyle\ket{\phi_{+}}$ & $\displaystyle\frac{1}{2}\hat{I}+\frac{1}{\pi}\hat{X}$ & $\displaystyle\frac{1}{2}(\hat{I}+\hat{X})$ \rule{0pt}{1.8\normalbaselineskip}\\
    $\displaystyle\ket{\phi_{-}}$ & $\displaystyle\frac{1}{2}\hat{I}+\frac{1}{\pi}\hat{X}$ & $\displaystyle\frac{1}{2}\hat{I}$ \rule{0pt}{1.8\normalbaselineskip}\rule[-9pt]{0pt}{1em}
    \end{tabular}
    \caption{The partial trace of the states $\ket{\phi_{\pm}}=\frac{1}{\sqrt{2}}\big(\!\ket{0}_{q}+\ket{\pm\sqrt{\pi}}_{q}\!\big)$ in the square single-mode stabilizer and modular-position subsystem decompositions. The stabilizer subsystem decomposition gives the same result for both states, while the modular-position subsystem decomposition gives different results, revealing an asymmetry in superpositions of left and right displacements of the position eigenstates.}
    \label{tab:decoders}
\end{table}

Now we turn our attention to the Zak basis representation of each decomposition. Recall our definition of $a=\sqrt{2}$ Zak states
\begin{equation}\label{eq:Zak_state_2}
\ket{k_{1},k_{2}}_{\sqrt{2}}=\sqrt[4]{4\pi}e^{i\pi k_{1}k_{2}}\sum_{s\in\mathbb{Z}}e^{2\sqrt{2}i\pi k_{2}s}\ket{\sqrt{2\pi}k_{1}+2\sqrt{\pi}s}_{q}\tag{\ref{eq:Zak_states}$'$}
\end{equation}
for $k_{1},k_{2}\in\mathbb{R}$. The set of states $\ket{k_{1},k_{2}}_{\sqrt{2}}$ restricted to $k_{1}\in\big({-}2^{-3/2},3\times2^{-3/2}\big],k_{2}\in\big({-}2^{-3/2},2^{-3/2}\big]$
forms a basis. Next, recall the Zak representation of the $\mathcal{G}_{\text{sq}}$ decomposition:
\begin{equation}\label{eq:1_mode_defn_2}
    \ket{\mu}\otimes\ket{k_1,k_{2}}=e^{i\pi \mu k_{2}/\sqrt{2}}\ket{k_{1}+\mu/\sqrt{2},k_{2}}_{\sqrt{2}}\tag{\ref{eq:1_mode_defn}$'$}
\end{equation}
for $\mu=0,1$ and $k_{1},k_{2}\in\big({-}2^{-3/2},2^{-3/2}\big]$.

In recent work~\cite{Pantaleoni22}, a similar equation was developed for the modular-position subsystem decomposition. Here we provide a similar derivation using our notation in order to directly compare the two decompositions. One can see from the definition in \cref{eq:mod_q_subsystem} that for any gauge mode state $\ket{\phi}$ we have
\begin{equation}\label{eq:giacomo_translation_property}
\hat{W}\big(1/\sqrt{2},0\big)\ket{0}\otimes_{\mathcal{Q}}\ket{\phi}=\ket{1}\otimes_{\mathcal{Q}}\ket{\phi}.
\end{equation}
Next, consider the ``left half’’ of the Zak basis states $\ket{k_{1},k_{2}}_{\sqrt{2}}$ for which $k_{1}\leq2^{-3/2}$. These states have support only on position eigenstates $\ket{x}_{q}$ with $\mu=0$ in \cref{eq:mod_x_decomposition}, and thus can be decomposed into a state $\ket{0}\otimes_{\mathcal{Q}}\ket{\phi}$. We can then choose a Zak basis $\big\{\ket{k_{1},k_{2}}_{\zeta}\big|\;k_{1},k_{2}\in({-}2^{-3/2},2^{-3/2}]\big\}$ of the gauge mode such that
\begin{equation}
\ket{0}\otimes^{}_{\mathcal{Q}}\ket{k_{1},k_{2}}_{\zeta}=\ket{k_{1},k_{2}}_{\sqrt{2}}=\ket{0}\otimes\ket{k_{1},k_{2}},
\end{equation}
where the last $\otimes$ is across the stabilizer subsystem decomposition.
Now, we can apply \cref{eq:giacomo_translation_property} to find
\begin{equation}
\ket{1}\otimes_{\mathcal{Q}}\ket{k_{1},k_{2}}_{\zeta}=e^{-i\pi k_{2}/\sqrt{2}}\ket{k_{1}+1/\sqrt{2},k_{2}}_{\sqrt{2}},
\end{equation}
which differs from the equivalent equation for the stabilizer subsystem decomposition \cref{eq:1_mode_defn_2} only by a $e^{-\sqrt{2}i\pi k_{2}}$ phase. This in turn can be viewed as applying a $k_{2}$-dependent $Z$-axis rotation of the logical Bloch sphere, thus altering the properties of the partial trace operation. It is because of this phase that the stabilizer subsystem decomposition can be seen as a ``rephasing’’ of the modular-position subsystem decomposition which symmetrizes the treatment of the position and momentum quadratures. This result is consistent with recent work done in Ref.~\cite{Pantaleoni22}.

\section{Binned Quadrature Measurements and Logical State Tomography}\label{sec:logical_state_tomography}

In this appendix we discuss the subsystem decomposition of binned quadrature measurements (\cref{subsec:binned}), and the relationship between ideal decoding (as defined in \cref{subsec:latticeandcell}) and \textit{logical state tomography} using binned quadrature measurements (binned-LST), a procedure we define in \cref{subsec:binned-LST}. We show that binned quadrature measurements do \textit{not} correspond to measurements of the logical subsystem Pauli operators; and, moreover, that binned quadrature measurement operators do not decompose as tensor products in the stabilizer subsystem. Then, we show that binned-LST corresponds to a decoding procedure that is more prone to errors, and defines a map that is not CPTP. Throughout this appendix we will only consider single-mode qubit codes so that $\mathcal{G}=(\Sigma,(2),\mathcal{P})$ and the logical Pauli operators are given by
\begin{align}
    \bar{X}&=e^{-i\sqrt{\pi}\bar{p}},&\bar{Y}&=e^{i\sqrt{\pi}(\bar{q}-\bar{p})},&\bar{Z}&=e^{i\sqrt{\pi}\bar{q}},
\end{align}
in terms of the logical modes $\vect{\bar{\xi}}=\Sigma^{-1}\vect{\hat{\xi}}$ [see \cref{eq:logical_modes,eq:logical_mode_ops}]. 

\subsection{Binned Quadrature Measurements}\label{subsec:binned}

We begin by defining a binned operator $B(\hat{U})$ for any unitary operator $\hat{U}$ as follows. Given the spectral decomposition of the unitary operator
\begin{equation}
    \hat{U}=\sum_{\lambda\in L} e^{i\theta_{\lambda}}\ket{\lambda}\!\bra{\lambda}
\end{equation}
with $\theta_{\lambda}\in(-\pi/2,3\pi/2]$, we define the binned operator
\begin{equation}
    B(\hat{U})=\;\sum_{\mathclap{\lambda\in L_{+}}}\;\ket{\lambda}\!\bra{\lambda}-\;\sum_{\mathclap{\lambda\in L_{-}}}\;\ket{\lambda}\!\bra{\lambda},
\end{equation}
where where $L_{+}=\{\lambda\mid\theta_{\lambda}\in(-\pi/2,\pi/2]\}$ and $L_{-}=\{\lambda\mid\theta_{\lambda}\in(\pi/2,3\pi/2]\}$. In words, the binned operator $B(\hat{U})$ shares the same eigenstates $\{\ket{\lambda}\}$ as $\hat{U}$, but the complex eigenvalues of $\hat{U}$ are rounded to $+1$ if their real part is positive and $-1$ if their real part is negative. Note that $B(\hat{U})$ is both unitary and Hermitian for any unitary $\hat{U}$.

We define binned quadrature measurements as follows. For a binned Pauli $\bar{X}$ measurement:
\begin{enumerate}
    \item First, measure the logical quadrature $\bar{p}$.
    \item Then, round the result to the nearest multiple of $\sqrt{\pi}$.
    \item If the rounded result is even, assign a $+1$ measurement outcome to the $\bar{X}$ measurement; and if it is odd, assign a $-1$ measurement outcome.
\end{enumerate}
Binned $\bar{Y}$ and $\bar{Z}$ measurements are similarly defined by replacing $\bar{p}$ with $(\bar{q}-\bar{p})$ and $\bar{q}$, respectively. With this definition, a binned quadrature measurement for $\bar{X}$ is equivalent to a measurement of the Hermitian operator $B(\bar{X})$, and likewise for $\bar{Y}$ and $\bar{Z}$.

Now consider the action of $B(\bar{X})$ on a subsystem basis state $\ket{\psi}\otimes_{\mathcal{G}}\ket{\vect{k}}$. From \cref{eq:Paulis_in_subsystem}, $\bar{X}$ acts on the stabilizer states as
\begin{equation}
    \bar{X}\big(\!\ket{\psi}\otimes_{\mathcal{G}}\ket{\vect{k}}\!\big)=\hat{X}\ket{\psi}\otimes_{\mathcal{G}}e^{2i\pi\,\vect{k}^{T}\Omega\vect{\bar{m}}_{1}}\ket{\vect{k}}.
\end{equation}
Since $\hat{X}$ has eigenvalues $\pm1$, the action of $B(\bar{X})$ on the state is determined by whether the real part of $e^{2i\pi\,\vect{k}^{T}\Omega\vect{\bar{m}}_{1}}$ is positive or negative. In particular, we can write the action of $B(\bar{X})$ as
\begin{multline}\label{eq:binned_Paulis}
    B(\bar{X})\big(\!\ket{\psi}\otimes_{\mathcal{G}}\ket{\vect{k}}\!\big)=\\
    \left\{\!\begin{array}{r c}
    \big(\hat{X}\ket{\psi}\!\big)\otimes_{\mathcal{G}}\ket{\vect{k}},&\{\vect{k}^{T}\Omega\vect{\bar{m}}_{1}\}_{1}\in(-\frac{1}{4},\frac{1}{4}],\\
    \big({-}\hat{X}\ket{\psi}\!\big)\otimes_{\mathcal{G}}\ket{\vect{k}},&\text{else},\end{array}\right.
\end{multline}
where $\{x\}_{1}\in(-1/2,1/2]$ is the remainder of $x$ modulo 1. Similar results can be obtained for $B(\bar{Y})$ and $B(\bar{Z})$ by replacing $\vect{\bar{m}}_{1}$ with $(\vect{\bar{m}}_{1}+\vect{\bar{m}}_{2})$ and $\vect{\bar{m}}_{2}$, respectively.

\begin{figure}
    \centering
    \includegraphics{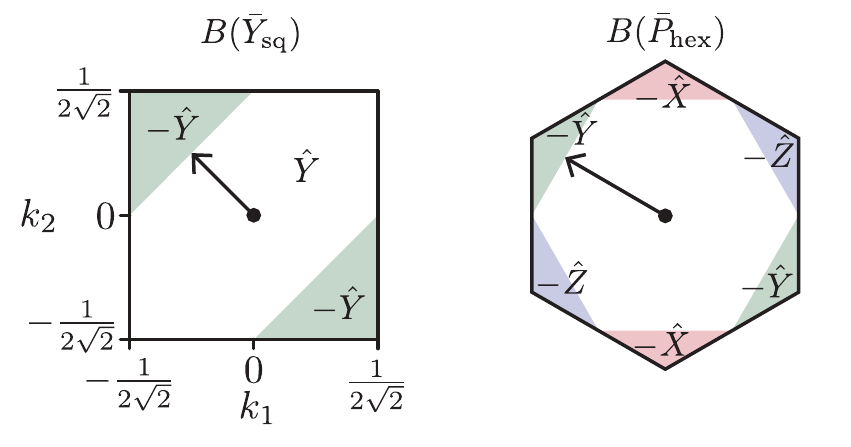}
    \caption{(Color) (a) Action of the binned logical $Y$ operator $B(\bar{Y}_{\text{sq}})$ on the square GKP qubit subsystem decomposition. Subsystem basis states in the unshaded subsystem satisfy $(k_{1}-k_{2})/\sqrt{2}\in(-1/4,1/4]$, and thus $B(\bar{Y}_{\text{sq}})$ acts as $\hat{Y}\otimes\hat{I}$ on these states. For subsystem basis states in the shaded regions $(k_{1}-k_{2})/\sqrt{2}$ lies outside $(-1/4,1/4]$ and $B(\bar{Y}_{\text{sq}})$ acts as $-\hat{Y}\otimes\hat{I}$ on these regions. The displacement $\hat{W}\left([-1,1]^{T}/(4\sqrt{2})\right)$ (depicted by the arrow) is the (equal) shortest displacement that causes a change in the logical measurement outcome. (b) Action of the binned logical Pauli operators $B(\bar{P}_{\text{hex}})$ on the hexagonal GKP qubit subsystem decomposition (for $P=X,Y,Z$). Each binned Pauli operator $B(\bar{P}_{\text{hex}})$ acts as $-\hat{P}\otimes\hat{I}$ in the corresponding shaded regions, and as $\hat{P}\otimes\hat{I}$ elsewhere.}
    \label{fig:logical_tomography_diagram}
\end{figure}

The significance of \cref{eq:binned_Paulis} is that the remainder $\{\vect{k}^{T}\Omega\vect{\bar{m}}_{1}\}_{1}$ [or the equivalent remainders for $B(\bar{Y})$ and $B(\bar{Z})$] is not always between $-1/4$ and $1/4$, even when $\vect{k}$ is in the Voronoi cell of the dual lattice. As an example, consider the square GKP code $\mathcal{G}_{\text{sq}}=(I_{2},(2),\mathcal{V}_{\text{sq}})$ [\cref{eq:sq_G}]. Then, the symplectic products that determine the binned logical operators simplify to
\begin{subequations}\label{eq:sq_bin}
\begin{align}
    \vect{k}^{T}\Omega\vect{\bar{m}}_{1}&=-k_{2}/\sqrt{2},\label{eq:sq_bin_X}\\
    \vect{k}^{T}\Omega(\vect{\bar{m}}_{1}{+}\vect{\bar{m}}_{2})&=(k_{1}-k_{2})/\sqrt{2},\label{eq:sq_bin_Y}\\
    \vect{k}^{T}\Omega\vect{\bar{m}}_{2}&=k_{1}/\sqrt{2}.\label{eq:sq_bin_Z}
\end{align}
\end{subequations}
Since the square Voronoi cell $\mathcal{V}_{\text{sq}}$ enforces $k_{1},k_{2}\in(-2^{-3/2},2^{-3/2}]$, \cref{eq:sq_bin_X,eq:sq_bin_Z} are guaranteed to be between $-1/4$ and $1/4$. As a result, the binned operators $B(\bar{X}_{\text{sq}})$ and $B(\bar{Z}_{\text{sq}})$ act as product operators $\hat{X}\otimes\hat{I}$ and $\hat{Z}\otimes\hat{I}$, since the second condition in the right-hand side of \cref{eq:binned_Paulis} is never satisfied. In contrast, there \emph{are} values of $k_{1}$ and $k_{2}$ inside the Voronoi cell for which \cref{eq:sq_bin_Y} is outside the range $(-1/4,1/4]$, for example $k_{1}=-1/(3\sqrt{2})$, $k_{2}=1/(3\sqrt{2})$. In this region, $B(\bar{Y}_{\text{sq}})$ acts as $-\hat{Y}\otimes\hat{I}$ on subsystem basis states as shown in \cref{fig:logical_tomography_diagram}(a).

Alternatively, one can understand this result as a consequence of the logical phase gate $\bar{S}_{\text{sq}}$ \textit{not} being a tensor product operator. Since $\bar{S}_{\text{sq}}$ maps $-\hat{p}\mapsto\hat{q}-\hat{p}$, we have $B(\bar{Y}_{\text{sq}})=\bar{S}_{\text{sq}}B(\bar{X}_{\text{sq}})\bar{S}_{\text{sq}}^{\dag}$. From \cref{eq:sq_phase} and using $B(\bar{X}_{\text{sq}})=\hat{X}\otimes\hat{I}$, one can quickly show that $B(\bar{Y}_{\text{sq}})$ is not a tensor product operator in the subsystem decomposition, in agreement with our direct calculation above.

Repeating the calculations in \cref{eq:binned_Paulis} for the hexagonal GKP code reveals that \textit{none} of the binned operators $B(\bar{X}_{\text{hex}})$, $B(\bar{Y}_{\text{hex}})$, $B(\bar{Z}_{\text{hex}})$ are product operators, and there are regions of the Voronoi cell in which each of them act as negative Pauli operators, as shown in \cref{fig:logical_tomography_diagram}(b).

\subsection{Logical State Tomography with Binned Quadrature Measurements}\label{subsec:binned-LST}

A natural way to define a decoder is to use the binned quadrature measurements to define a logical state tomography decoder that we label binned-LST. For a CV state $\hat{\rho}\in\mathcal{L}(\mathcal{H})$, the binned-LST decoder is defined by the map
\begin{equation}\label{eq:logical_state_tomography}
    \hat{\rho}\mapsto \frac{1}{2}\Big(\hat{I}+\mathrm{tr}\big(B(\bar{X})\hat{\rho}\big)\hat{X}+\mathrm{tr}\big(B(\bar{Y})\hat{\rho}\big)\hat{Y}+\mathrm{tr}\big(B(\bar{Z})\hat{\rho}\big)\hat{Z}\Big),
\end{equation}
which is identical to the logical state one would obtain by running a tomography experiment on the CV state using the binned Pauli measurements.

Since the binned measurement operators are not tensor product operators in the subsystem decomposition, \cref{eq:logical_state_tomography} is not equivalent to taking the partial trace over the stabilizer subsystem. Instead, the partial trace itself can be recovered by replacing the binned measurement operators $B(\bar{P})$ with ideal measurement operators $\bar{P}_{\text{m}}=\hat{P}\otimes_{\mathcal{G}}\hat{I}$ for $P=X,Y,Z$.

There are two consequences of this discrepancy between the binned-LST and subsystem decomposition decoders. First, for the square GKP code, if one starts in an ideal codestate $\ket{\psi}\otimes\ket{\vect{0}}$, the (non-unique) shortest displacement that alters the binned-LST decoded state is $\hat{W}_{\text{min}}=\hat{W}\big([-1,1]^{T}\!/(4\sqrt{2})\big)$ with length 1/4, see \cref{fig:logical_tomography_diagram}(a). In contrast the shortest displacement that causes an error in the subsystem decomposition has length $1/(2\sqrt{2})$. This reduction in distance is not unique for the square code and is also the case for the hexagonal code, as shown in \cref{fig:logical_tomography_diagram}(b).

Second, the logical error caused by $\hat{W}_{\text{min}}$ to the binned-LST decoded state is given by $\hat{X}\mapsto\hat{X}$, $\hat{Y}\mapsto-\hat{Y}$, $\hat{Z}\mapsto\hat{Z}$. In terms of density matrices, this is a transpose map, a well-known example of a positive but not \textit{completely} positive map. Recall that a map $\mathcal{N}$ is positive if it maps positive operators to positive operators, and is completely positive if $\mathcal{I}\otimes\mathcal{N}$ is positive where $\mathcal{I}$ is the identity map (acting on a Hilbert space of any dimension).

To see an example of a positive density operator being mapped to a negative operator under the binned-LST decoding map we must therefore consider an initial state in a composite Hilbert space, which we choose for simplicity to be between a qubit space and a CV space $\mathbb{C}^{2}\otimes\mathcal{H}$. We consider the initial Bell-like state
\begin{align}
    \ket{\phi}&=\frac{1}{\sqrt{2}}\Big(\!\ket{0}\otimes\big(\!\ket{0}\otimes_{\mathcal{G}}\ket{\vect{v}_{0}}\!\big)+\ket{1}\otimes\big(\!\ket{1}\otimes_{\mathcal{G}}\ket{\vect{v}_{0}}\!\big)\!\Big)\nonumber\\
    &\in \mathbb{C}^{2}\otimes\mathcal{H}=\mathbb{C}^{2}\otimes(\mathbb{C}^{2}\otimes_{\mathcal{G}}\mathcal{S}),
\end{align}
where $\vect{v}_{0}=[-1,1]^{T}/(3\sqrt{2})$ is in the shaded region of \cref{fig:logical_tomography_diagram}(a), and the unlabelled tensor product $\otimes$ is the tensor product between the $\mathbb{C}^{2}$ qubit space and the CV space $\mathcal{H}$ while the labelled tensor product $\otimes_{\mathcal{G}}$ is the tensor product between the square GKP logical subsystem $\mathbb{C}^{2}$ and stabilizer subsystem $\mathcal{S}$ of the CV space $\mathcal{H}$. The initial density operator can be equivalently written as
\begin{equation}\label{eq:initial_rho}
\begin{aligned}
    \ket{\phi}\!\bra{\phi}&=\frac{1}{2}\begin{bmatrix}1&0&0&1\\0&0&0&0\\0&0&0&0\\1&0&0&1\end{bmatrix}\otimes_{\mathcal{G}}\ket{\vect{v}_{0}}\!\bra{\vect{v}_{0}}\\
    &\in\mathcal{L}(\mathbb{C}^{4})\otimes_{\mathcal{G}}\mathcal{L}(\mathcal{S})\cong\mathcal{L}(\mathbb{C}^{2}\otimes\mathcal{H}),
\end{aligned}
\end{equation}
where we have represented the qubit state and the logical state of the CV mode together in the matrix.

Next, we perform binned-LST on the CV space while leaving the qubit state invariant. Since the binned-LST decoder maps the CV Hilbert space $\mathcal{H}$ to a qubit space $\mathbb{C}^{2}$, the initial density matrix $\ket{\phi}\!\bra{\phi}\in\mathcal{L}(\mathbb{C}^{2}\otimes\mathcal{H})$ is mapped to a 2-qubit operator $\hat{O}\in\mathcal{L}(\mathbb{C}^{4})$. Since $\vect{v}_{0}$ is in the shaded region of \cref{fig:logical_tomography_diagram}(a), performing logical tomography maps $\hat{Y}\mapsto-\hat{Y}$ on the logical subsystem of $\mathcal{H}$, while leaving the remaining logical subsystem Pauli operators and qubit Pauli operators invariant. This is equivalent to taking the transpose of each $2\times2$ block of the matrix in \cref{eq:initial_rho}. The final operator
\begin{equation}\renewcommand{\arraystretch}{1.1}
    \hat{O}=\frac{1}{2}\begin{bmatrix}\begin{pmatrix}1&0\\0&0\end{pmatrix}{\vphantom{\Big)}}^{\!\!T}&\begin{pmatrix}0&1\\0&0\end{pmatrix}{\vphantom{\Big)}}^{\!\!T}\\\begin{pmatrix}0&0\\1&0\end{pmatrix}{\vphantom{\Big)}}^{\!\!T}&\begin{pmatrix}0&0\\0&1\end{pmatrix}{\vphantom{\Big)}}^{\!\!T}\end{bmatrix}=\frac{1}{2}\begin{bmatrix}1&0&0&0\\0&0&1&0\\0&1&0&0\\0&0&0&1\end{bmatrix}
\end{equation}
has a $-1$ eigenvector $[0,1,-1,0]^{T}$, and thus $\hat{O}$ is not positive and cannot represent a physical state. Since the tensor product of binned-LST with the identity map has mapped a positive density operator $\ket{\phi}\!\bra{\phi}$ to a non-positive operator $\hat{O}$, binned-LST is not a completely-positive map. We note that this example is in essence the same as the well-known example that demonstrates the transpose map is not completely-positive.

To conclude this section, we note that the binned-LST map can be tweaked to be made CPTP by replacing the binned Pauli operators $B(\bar{P})$ with ideal Pauli measurement operators $\hat{P}_{\text{m}}=\hat{P}\otimes_{\mathcal{G}}\hat{I}$. These operators can be measured by applying a round of ideal error-correction (which guarantees $\vect{k}=\vect{0}$) followed by a standard binned Pauli measurement. Logical state tomography using these operators would, by definition, be equivalent to the partial trace operation over the stabilizer subsystem. However, we note that such ideal measurements are not possible in practice since they involve the preparation of an ideal GKP codestate to perform ideal error correction.

\section{Analytical formula for the partial trace of approximate square GKP state}\label{sec:analytic_partial_trace}

In this appendix we write the explicit analytical formula for the partial trace of a square approximate GKP codestate $\ket{\bar{\psi}_{\Delta}}\propto e^{-\Delta^{2}\hat{a}^{\dag}\hat{a}}\ket{\bar{\psi}}$, as discussed in \cref{sec:sq}. We start with the square subsystem decomposition of the approximate codestate
\begin{multline}
\ket{\bar{\psi}_{\Delta}}\propto\sum_{\vect{s}\in\mathbb{Z}^{2}}\hat{P}(\vect{s})\ket{\psi}\otimes\\
\int_{\mathrlap{\mathcal{V}}}\;\;d^{2}\vect{v}\,e^{-\frac{\pi}{2}\mathrm{coth}\big(\!\frac{\Delta^{2}}{2}\!\big)|\vect{v}+\vect{s}/\!\sqrt{2}|^{2}}e^{i\pi\vect{v}^{T}\Omega\vect{s}/\!\sqrt{2}}\ket{\vect{v}},\tag{\ref{eq:sq_approx}}
\end{multline}
where $\hat{P}(\vect{s})=e^{i\pi s_{1}s_{2}/2}\hat{X}^{s_{1}}\hat{Z}^{s_{2}}$, $\mathcal{V}=(-2^{-3/2},2^{-3/2}]^{2}$, and we've written $\ket{\vect{v}}=\ket{v_{1},v_{2}}$. We calculate the partial trace by first writing the density operator
\begin{multline}
    \ket{\bar{\psi}_{\Delta}}\!\bra{\bar{\psi}_{\Delta}}\propto\sum_{\vect{s},\vect{t}\in\mathbb{Z}^{2}}\hat{P}(\vect{s})\ket{\psi}\!\bra{\psi}\hat{P}(\vect{t})\,\otimes\\
    \iint_{\mathrlap{\mathcal{V}}}\;d^{2}\vect{v}d^{2}\vect{w}\,f_{\Delta}(\vect{v},\vect{s})f_{\Delta}(\vect{w},\vect{t})^{*}\ket{\vect{v}}\!\bra{\vect{w}},
\end{multline}
where we have written
\begin{equation}
    f_{\Delta}(\vect{v},\vect{s})=e^{-\frac{\pi}{2}\mathrm{coth}\big(\!\frac{\Delta^{2}}{2}\!\big)|\vect{v}+\vect{s}/\!\sqrt{2}|^{2}}e^{i\pi\vect{v}^{T}\Omega\vect{s}/\!\sqrt{2}}.
\end{equation}
Then, the partial trace is given by
\begin{equation}
    \mathrm{tr}_{\mathcal{S}}\big(\ket{\bar{\psi}_{\Delta}}\!\bra{\bar{\psi}_{\Delta}}\big)\propto\sum_{\vect{s},\vect{t}\in\mathbb{Z}^{2}}I^{\Delta}_{\vect{s},\vect{t}}\hat{P}(\vect{s})\ket{\psi}\!\bra{\psi}\hat{P}(\vect{t})\tag{\ref{eq:sq_approx_partial_trace}},
\end{equation}
where the coefficients $I_{\vect{s},\vect{t}}$ are
\begin{align}
    I_{\vect{s},\vect{t}}^{\Delta}&=\int_{\mathrlap{\mathcal{V}}}\;d^{2}\vect{v}\,f_{\Delta}(\vect{v},\vect{s})f_{\Delta}(\vect{v},\vect{t})^{*}\\
    &=\frac{1}{4}\mathrm{tanh}\bigg(\!\frac{\Delta^{2}}{2}\!\bigg)\exp\Big(\!{-}\frac{\pi}{4}\mathrm{coth}(\Delta^{2})|\vect{s}-\vect{t}|^{2}+i\frac{\pi}{2}\vect{s}^{T}\!\Omega\vect{t}\!\Big)\nonumber\\
    &\hspace{0.5 cm}\times\big(g_{\Delta}(s_{1}{+}t_{1}{-}1,t_{2}{-}s_{2})-g_{\Delta}(s_{1}{+}t_{1}{+}1,t_{2}{-}s_{2})\big)\nonumber\\
    &\hspace{0.5 cm}\times\big(g_{\Delta}(s_{2}{+}t_{2}{-}1,s_{1}{-}t_{1})-g_{\Delta}(s_{2}{+}t_{2}{+}1,s_{1}{-}t_{1})\big),
\end{align}
where
\begin{equation}
    g_{\Delta}(x,y)=\mathrm{erf}\bigg(\!\sqrt{\frac{\pi}{8}}\Big(x\sqrt{\coth(\Delta^{2}/2)}+iy\sqrt{\tanh(\Delta^{2}/2)}\Big)\!\bigg).
\end{equation}
We use this expression to produce \cref{fig:bloch_sphere}, which shows the location of the partial trace approximate codestate on the Bloch sphere, by truncating the sum in \cref{eq:sq_approx_partial_trace} to $s_{1},s_{2},t_{1},t_{2}=-2,-1,0,1,2$. We also note that similar analytical expressions can be obtained for the logical noise channels considered in \cref{subsec:envelope,subsec:loss,subsec:random_walk}, although these expressions are best obtained using a symbolic package such as Mathematica.

\section{Example: $n$-mode repetition code}\label{sec:repetition_code}

\begin{figure}
    \centering
    \includegraphics{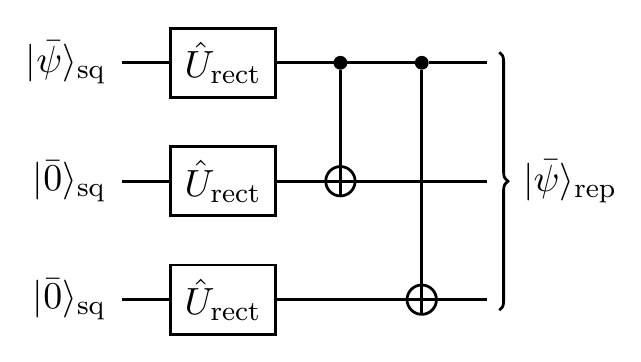}
    \caption{Encoding circuit for the 3-mode GKP-repetition code. The first mode is initialized in a square GKP qubit state $\ket{\bar{\psi}}_{\text{sq}}$, while the second and third modes are initialized in the square GKP 0 state $\ket{\bar{0}}_{\text{sq}}$. Then, each square GKP state is squeezed via $\hat{U}_{\text{rect}}$, resulting in a rectangular GKP state. Finally, the rectangular GKP codestates are encoded in a repetition code via two logical controlled-NOT gates, resulting in a 3-mode GKP-repetition codestate.}
    \label{fig:rep_encoding_circuit}
\end{figure}

For illustrative purposes, in this appendix we present the stabilizer subsystem decomposition of the $n$-mode rectangular GKP-repetition qubit code (which we simply refer to as the GKP-repetition code). We describe the structure of the GKP lattice $\Lambda_{\text{rep}}$ and dual lattice $\bar{\Lambda}_{\text{rep}}$, and their relationship to the parameters $(\Sigma_{\text{rep}},\vect{d}_{\text{rep}})$. We focus on the 3-mode GKP-repetition code since each of the lattices and primitive cells is a direct sum of two 3D lattices in the position and momentum quadratures, allowing for convenient visualization. We present the Voronoi cell of $\bar{\Lambda}_{\text{rep}}$, and describe the cell transformation between the Voronoi cell of the GKP-repetition code and the primitive cell that corresponds to a ``concatenated'' decoder, and find that the Voronoi cell has an improved distance compared to the concatenated decoding cell. Moreover, the GKP-repetition code can be tuned such that the distance in position and in momentum is the same, symmetrizing the effect of noise on the code. Although we are not suggesting that quantum computing with the GKP-repetition code is an optimal strategy (see Ref.~\cite{Harrington04,Royer22} for other multi-mode GKP codes), it does provide a neat illustration of how the bosonic nature of multi-mode GKP codes can be leveraged to enhance GKP-qubit code concatenations.

The GKP-repetition code is defined as follows. We start with the single-mode rectangular GKP qubit code $\mathcal{G}_{\text{rect}}$, which has stabilizer generators $\hat{S}_{1}=\hat{W}(\sqrt{2}\alpha,0)$ and $\hat{S}_{2}=\hat{W}(0,\sqrt{2}/\alpha)$, and parameters
\begin{subequations}
\begin{align}
    \mathcal{G}_{\text{rect}}&=\big(\mathrm{diag}(\alpha,1/\alpha),(2), \mathcal{V}_{\text{rect}}\big),\\
    \mathcal{V}_{\text{rect}}&=\big({-}\alpha\times 2^{-3/2},\,\alpha\times 2^{-3/2}\big]\nonumber\\
    &\hspace{1 cm}\times\big({-}2^{-3/2}/\alpha,\,2^{-3/2}/\alpha\big],
\end{align}
\end{subequations}
for $\alpha>0$. Then, we obtain the GKP-repetition code by taking $n$ copies of the $\mathcal{G}_{\text{rep},\alpha}$ and promoting the operators $\bar{Z}_{j}\bar{Z}_{j+1}$ to stabilizers, where we have written $\bar{Z}_{j}$ for the logical $\bar{Z}$ operator of the $j$-th mode. The GKP-repetition code logical operators are given in terms of the rectangular GKP logical operators by $\bar{X}_{\text{rep}}=\bar{X}_{1}\bar{X}_{2}\dots\bar{X}_{n}$ and $\bar{Z}_{\text{rep}}=\bar{Z}_{1}$.

There are many ways to obtain the parameters $(\Sigma_{\text{rep}},\vect{d}_{\text{rep}})$ of the GKP-repetition code, but one convenient way is via the encoding circuit in \cref{fig:rep_encoding_circuit}. Here, we have set $n=3$ for convenience, although our results generalize to arbitrary $n$. In this circuit, we start with a square GKP qubit codestate $\ket{\bar{\psi}}_{\text{sq}}$ in the first mode, and square GKP qubit zero states $\ket{\bar{0}}_{\text{sq}}$ on the second and third modes. The square GKP zero states can equivalently be seen as qunaught states of a rectangular $\alpha=\sqrt{2}$ GKP qunaught code. Next, we apply the Gaussian operator
\begin{align}\label{eq:rect_encoding_unitary}
    \hat{U}_{\text{rect}}&=\mathrm{exp}\bigg(\!{-}\frac{i\mathrm{ln}\alpha}{2}\big(\hat{q}\hat{p}+\hat{p}\hat{q}\big)\!\bigg),&S_{\text{rect}}&=\mathrm{diag}(\alpha,1/\alpha),
\end{align}
to each mode, where $S_{\text{rect}}$ is the symplectic matrix corresponding to $\hat{U}_{\text{rect}}$. After applying \cref{eq:rect_encoding_unitary}, each state is encoded in a rectangular GKP qubit codestate given by $\alpha$. Finally, we apply logical controlled-NOT gates between first and second and the first and third modes, where each logical controlled-NOT is a Gaussian unitary given by
\begin{align}\label{eq:cnot_unitary}
    \hat{U}_{C_{\text{NOT}}}&=\mathrm{exp}\big({-}i\hat{q}\otimes\hat{p}\big),& S_{C_{\text{NOT}}}=\begin{bmatrix}1&0&0&0\\1&1&0&0\\0&0&1&-1\\0&0&0&1\end{bmatrix}.
\end{align}
By comparison with \cref{fig:encoding_circuit}, we can take the product of the symplectic matrices in the encoding circuit from the initial square states, \cref{eq:rect_encoding_unitary} and \cref{eq:cnot_unitary} to obtain
\begin{subequations}
\begin{align}
    \Sigma_{\text{rep}}&=\begin{bmatrix}\Sigma_{q}&0\\0&\Sigma_{p}\end{bmatrix},\\
    \Sigma_{q}&=\alpha\begin{bmatrix}1&0&0\\1&\sqrt{2}&0\\1&0&\sqrt{2}\end{bmatrix},\\
    \Sigma_{p}&=\frac{1}{\alpha}\begin{bmatrix}1&-1/\sqrt{2}&-1/\sqrt{2}\\0&1/\sqrt{2}&0\\0&0&1/\sqrt{2}\end{bmatrix}.
\end{align}
\end{subequations}
The block-diagonal form of $\Sigma_{\text{rep}}$ represents the fact that the GKP-repetition code is ``CSS'' in the sense that each stabilizer generator translates either position quadratures or momentum quadratures, but not both.

\begin{figure}
    \centering
    \includegraphics{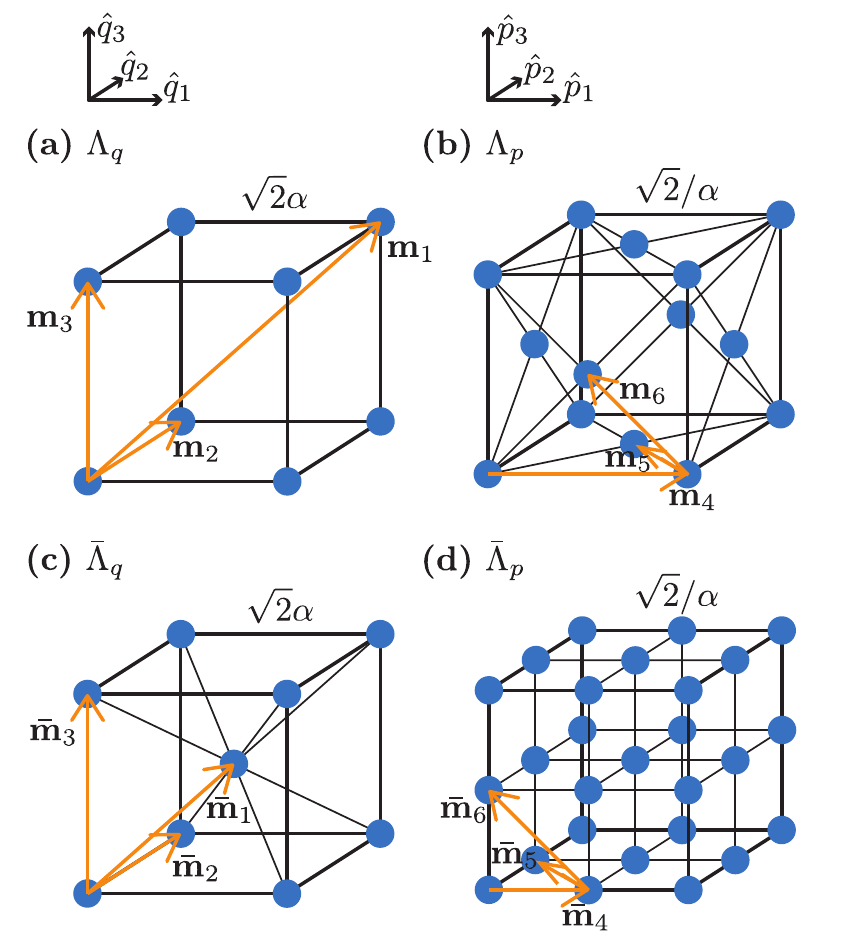}
    \caption{(Color) The GKP-repetition code lattice $\Lambda_{\text{rep}}=\Lambda_{q}\oplus\Lambda_{p}$ and dual lattice $\bar{\Lambda}_{\text{rep}}=\bar{\Lambda}_{q}\oplus\bar{\Lambda}_{p}$. In (a) and (c), each dimension represents a position operator on each mode, while in (b) and (d) each dimension represents a momentum operator. We have shown a cell of side-length $\sqrt{2}\alpha$ [for (a) and (c)] or $\sqrt{2}/\alpha$ [for (b) and (d)], since the lattices extend to infinity in all directions. The lattice generators $\vect{m}_{J}$ and dual lattice generators $\vect{\bar{m}}_{J}$ are shown in yellow; $\vect{m}_{5}$, $\vect{m}_{6}$, $\vect{\bar{m}}_{5}$ and $\vect{\bar{m}}_{6}$ have been displaced from the origin for display purposes. $\Lambda_{q}$ and $\bar{\Lambda}_{p}$ are cubic lattices, while $\Lambda_{p}$ is a face-centered cubic lattice and $\bar{\Lambda}_{q}$ is a body-centered cubic lattice.}
    \label{fig:rep_lattices}
\end{figure}

Since the GKP-repetition code encodes a single qubit, we have $\vect{d}_{\text{rep}}=(2,1,1)$, which coincides with the dimensions of the initial encoded states in \cref{fig:rep_encoding_circuit}. The GKP lattice and dual lattice generators are given by
\begin{align}
\vect{m}_{J}&=d_{J\;(\mathrm{mod}\,n)}^{1/2}\,(\Sigma_{\text{rep}})_{J},&\vect{\bar{m}}_{J}&=d_{J\;(\mathrm{mod}\,n)}^{-1/2}\,(\Sigma_{\text{rep}})_{J},
\end{align}
where $(\Sigma_{\text{rep}})_{J}$ is the $J$-th column of $\Sigma_{\text{rep}}$. Note that from \cref{eq:logical_Pauli} we have
\begin{align}
    \hat{W}(\vect{\bar{m}}_{1})&=\bar{X}_{\text{rep}},&\hat{W}(\vect{\bar{m}}_{4})&=\bar{Z}_{\text{rep}},
\end{align}
while the remaining displacements of dual lattice generators $\hat{W}(\vect{\bar{m}}_{2})=\hat{W}(\vect{\bar{m}}_{3})=\hat{W}(\vect{\bar{m}}_{5})=\hat{W}(\vect{\bar{m}}_{6})$ all represent logical identity gates (i.e.~stabilizers) since the qunaught Pauli operators trivially equal $X_{(1)}=Z_{(1)}=[1]$ by \cref{eq:PauliXZ}.

\begin{figure}
    \centering
    \includegraphics{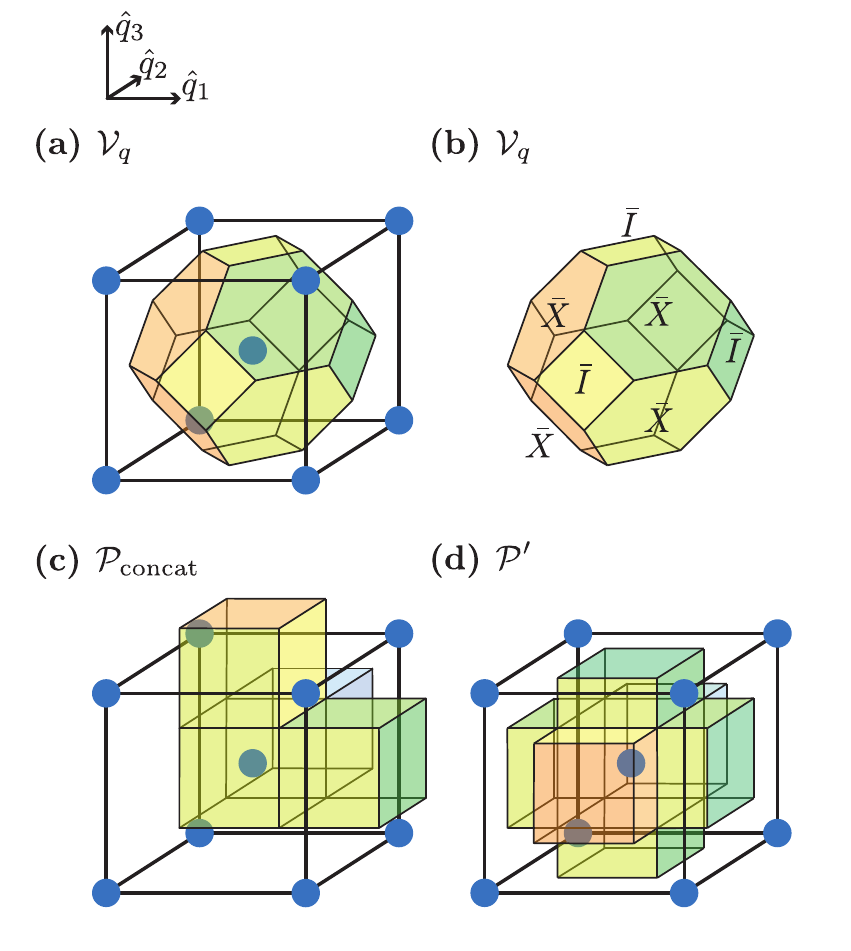}
    \caption{(Color) Primitive cells of the position sector of the dual lattice, $\bar{\Lambda}_{q}$. (a) The Voronoi cell $\mathcal{V}_{q}$ of the body-centered cubic lattice is a truncated octahedron. In the stabilizer subsystem decomposition, each face is associated with either an $\hat{I}$-boundary or a $\hat{X}$-boundary; as shown in (b) these correspond to the square and hexagonal faces respectively. (c) The primitive cell $\mathcal{P}_{\text{concat}}$ corresponding to a ``concatenated'' decoder. (d) After performing a cell transformation that \textit{only} acts along $\hat{I}$-boundaries, the symmetric primitive cell $\mathcal{P}'$ is obtained. The subsystem decompositions associated with $\mathcal{P}_{\text{concat}}$ and $\mathcal{P}$ contain identical logical information. The shortest displacement leading to a logical error is longer for the Voronoi cell $\mathcal{V}_{q}$ than $\mathcal{P}_{\text{concat}}$ or $\mathcal{P}'$.}
    \label{fig:rep_primitive_cells}
\end{figure}

Since the 3-mode repetition code is CSS, we can illustrate the GKP lattice $\Lambda_{\text{rep}}=\Lambda_{q}\oplus\Lambda_{p}$ and dual lattice $\bar{\Lambda}_{\text{rep}}=\bar{\Lambda}_{q}\oplus\bar{\Lambda}_{p}$ in the position and momentum sectors as 3D lattices as shown in \cref{fig:rep_lattices}. $\Lambda_{q}$ is a cubic lattice with spacing $\sqrt{2}\alpha$, coinciding with the single-mode rectangular GKP $\hat{S}_{1}$ stabilizers. $\Lambda_{p}$ is a face-centered cubic lattice with spacing $\sqrt{2}/\alpha$, where the cubic lattice points correspond to single-mode rectangular GKP $\hat{S}_{2}$ stabilizers while the face-centered lattice points correspond to stabilizers of the type $\bar{Z}_{j}\bar{Z}_{j'}$. For the dual lattice, $\bar{\Lambda}_{p}$ is a cubic lattice with spacing $1/(\sqrt{2}\alpha)$ since the GKP-repetition code logical $\bar{Z}_{\text{rep}}$ can be implemented with a single-mode rectangular $\bar{Z}_{j}$ on any of the modes. Finally, $\bar{\Lambda}_{q}$ is a body-centered cubic lattice with spacing $\sqrt{2}\alpha$ where the body-centered lattice points correspond to logical $\bar{X}$ operators.

The Voronoi cell $\mathcal{V}_{\text{rep}}$ of the dual lattice $\bar{\Lambda}_{\text{rep}}$ also has a CSS structure and can be split into Voronoi cells $\mathcal{V}_{q}$ of $\bar{\Lambda}_{q}$ and $\mathcal{V}_{p}$ of $\bar{\Lambda}_{p}$. Since $\bar{\Lambda}_{p}$ is a cubic lattice, $\mathcal{V}_{p}$ is a cube with side-length $1/(\sqrt{2}\alpha)$, and thus the shortest displacement that causes a logical error has length $1/(2\sqrt{2}\alpha)$. As a result, the GKP-repetition code offers no additional protection against shifts in momentum compared to the single-mode rectangular GKP code.

In contrast, the Voronoi cell $\mathcal{V}_{q}$ of the body-centered cubic lattice is a truncated octahedron, as illustrated in \cref{fig:rep_primitive_cells}(a) and (b). $\mathcal{V}_{q}$ has boundaries corresponding to both logical $\hat{X}$ errors (for each hexagonal face) and logical $\hat{I}$ operators (for each square face). The shortest displacement that causes a logical $\bar{X}$ error has length $\sqrt{3}\alpha/(2\sqrt{2})$, an improvement over the single-mode rectangular GKP code by a factor of $\sqrt{3}$. It is therefore natural to choose $\alpha=3^{-1/4}$ ($\alpha=n^{-1/4}$ in general) to equalize the logical distance in momentum and in position. This can be interpreted as choosing $\alpha$ to bias the noise of the single-mode rectangular GKP code to compensate for the fact that the repetition code provides additional protection only against shifts in position.

It is interesting to compare the Voronoi cell $\mathcal{V}_{q}$ to the primitive cell $\mathcal{P}_{\text{concat}}$ corresponding to a ``concatenated'' error-correction procedure which we define as follows:
\begin{enumerate}
    \item First, perform a round of error-correction on each single-mode rectangular GKP codestate.
    \item Then, measure the repetition code stabilizers $\bar{Z}_{1}\bar{Z}_{2}$ and $\bar{Z}_{1}\bar{Z}_{3}$, which take values $\pm1$ since each single-mode GKP state is in the codespace.
    \item Finally, perform the correction $\bar{X}_{1}^{\dag}$, $\bar{X}_{2}^{\dag}$ or $\bar{X}_{3}^{\dag}$ based on the syndrome from the repetition code stabilizers, returning the state to the GKP-repetition codespace.
\end{enumerate}
The corresponding primitive cell is identical to $\mathcal{V}_{p}$ in the momentum sector, while in the position sector it is given by $\mathcal{P}_{\text{concat}}$, see \cref{fig:rep_primitive_cells}(c). The shortest displacement outside of $\mathcal{P}_{\text{concat}}$ has length $\alpha/(2\sqrt{2})$; however, this displacement does not correspond to a logical error since it crosses a $\hat{I}$ boundary. Indeed, using \cref{eq:cell_transformation} we can perform a cell transformation $\mathcal{P}_{\text{concat}}\mapsto\mathcal{P}'$ [\cref{fig:rep_primitive_cells}(d)] along each of the $\hat{I}$-boundaries such that the cell transformation does not affect the logical subsystem. The shortest displacement outside of $\mathcal{P}'$ has length $\alpha/2$ and this time corresponds to a logical $\hat{X}$ error. This is a factor of $\sqrt{3/2}$ shorter than the corresponding distance of the Voronoi cell $\mathcal{V}_{q}$ (in general, this factor is $\sqrt{2n/(n+1)}$ for any odd $n$). One could perform a second cell transformation $\mathcal{P}'\mapsto\mathcal{V}_{q}$, but this would require translations along $\hat{X}$-boundaries, altering the information in the logical subsystem.

\section{Decomposing states into $\otimes_{\mathcal{G}}$}\label{sec:state_decompositions_eq}

In this appendix, we present the explicit formula for the decomposition of an arbitrary state $\ket{\phi}\in\mathcal{H}$ into an arbitrary subsystem decomposition $\otimes_{\mathcal{G}}$ in terms of the wavefunction $\vphantom{\ket{}}_{q}\!\braket{x|\hat{U}_{\Sigma}^{\dag}|\phi}$, as discussed in \cref{subsec:state_decompositions}. This is equivalent to calculating the overlap
\begin{equation}
    \big(\!\bra{\mu}\otimes_{\mathcal{G}}\bra{\vect{k}}\!\big)\ket{\phi}={\vphantom{\ket{}}}_{(\Sigma,\vect{d})}\!\braket{\mu,\vect{k}|\phi}\tag{\ref{eq:inner_product}}
\end{equation}
for $\vect{k}\in\mathcal{P}$. We continue the derivation from
\begin{equation}
\hat{U}_{\Sigma}\,e^{i\pi\vect{\bar{\ell}}(\mu)^{T}\Omega\vect{k}}\big|\Sigma^{-1}\big(\vect{k}+\vect{\bar{\ell}}(\mu)\big)\big\rangle_{\mathcal{Z}_{\sqrt{\vect{d}}}}.\tag{\ref{eq:stab_unfolding}}
\end{equation}
Next, we find the position representation of a multi-mode Zak state $\ket{\vect{k}}_{\mathcal{Z}_{\sqrt{\vect{d}}}}$ by applying \cref{eq:Zak_multi_mode} to \cref{eq:Zak_states}:
\begin{multline}
    \ket{\vect{k}}_{\mathcal{Z}_{\sqrt{\vect{d}}}}=\sqrt[4]{(2\pi)^{n}d}\,e^{i\pi\,\vect{k}_{q}\cdot\vect{k}_{p}}\\
    \times\sum_{\vect{s}\in\mathbb{Z}^{n}}e^{2i\pi \left(\!\sqrt{D}\vect{s}\right)\cdot\vect{k}_{p}}\big|\sqrt{2\pi}\big(\vect{k}_{q}+\sqrt{D}\vect{s}\big)\big\rangle_{q},
\end{multline}
where $d=d_{1}\times \cdots\times d_{n}$, $D=\mathrm{diag}(\vect{d})$, $\vect{k}_{q}=[k_{1},\dots,k_{n}]^{T}$, $\vect{k}_{p}=[k_{n+1},\dots,k_{2n}]^{T}$, `` $\cdot$ '' represents the dot product between two vectors in $\mathbb{R}^{n}$, and $\ket{\vect{x}}_{q}=\ket{x_{1}}_{q}\otimes\cdots\otimes\ket{x_{n}}_{q}$. Applying this to \cref{eq:stab_unfolding}, we obtain the equation, giving
\begin{multline}
    \big(\!\bra{\mu}\otimes_{\mathcal{G}}\bra{\vect{k}}\!\big)\ket{\phi}=\sqrt[4]{(2\pi)^{n}d}\,e^{-i\pi\vect{\bar{\ell}}(\mu)^{T}\Omega\vect{k}}\\
    \times e^{-i\pi\,\vect{\tilde{k}}_{q}\cdot\vect{\tilde{k}}_{p}}\sum_{\vect{s}\in\mathbb{Z}^{n}}\Big(e^{-2i\pi \left(\!\sqrt{D}\vect{s}\right)\cdot\vect{\tilde{k}}_{p}}\\
    \times{\vphantom{\big|}}_{q}\!\big\langle\sqrt{2\pi}\big(\vect{\tilde{k}}_{q}+\sqrt{D}\vect{s}\big)\big|\hat{U}_{\Sigma}^{-1}\big|\phi\big\rangle\Big),
\end{multline}
where we have written $\vect{\tilde{k}}=\Sigma^{-1}\big(\vect{k}+\vect{\bar{\ell}}(\mu)\big)$, concluding the derivation.

\section{Characteristic function of Gaussian channels}\label{sec:Gaussian_chi_representation}

Here, we present analytic expressions for the characteristic function of an arbitrary Gaussian unitary operator $\hat{U}_{S}$ [\cref{eq:Gaussian_unitary_chi}] and an arbitrary Gaussian channel $\mathcal{E}_{T,N}$ [\cref{eq:Gaussian_channel_chi}] that depend only on the matrices $S,T,N$ that define the Gaussian operator/channel (as described below). In doing so, we will also derive the characteristic function of the envelope operator, and the loss and dephasing noise channels discussed in \cref{sec:errors}.

To begin, we recall some basic properties of the characteristic function of operators and channels, written with our chosen scaling of the displacement operators $\hat{W}(\vect{v})$. The characteristic function of an arbitrary operator $\hat{O}$ is given by~\cite{Weedbrook12}
\begin{equation}\label{eq:chi_from_operator}
    c(\vect{v})=\mathrm{tr}\big(\hat{O}\hat{W}(\vect{v})^{\dag}\big),
\end{equation}
and satisfies
\begin{equation}\label{eq:operator_from_chi}
    \hat{O}=\int d^{2n}\vect{v}\,c(\vect{v})\hat{W}(\vect{v}),
\end{equation}
where all integrals in this section are over $\mathbb{R}^{2n}$. It follows from this definition that when an operator is conjugated by a Gaussian unitary via $\hat{O}\mapsto\hat{U}^{}_{S}\hat{O}\hat{U}_{S}^{\dag}$, its characteristic function transforms as
\begin{equation}\label{eq:chi_Gaussian_transformation}
    c(\vect{v})\mapsto c(S^{-1}\vect{v}).
\end{equation}
Moreover, when \cref{eq:chi_from_operator} is applied to a density operator $\hat{\rho}$, $\mathrm{tr}(\hat{\rho})=1$ implies $c(\vect{0})=1$ and $\hat{\rho}=\hat{\rho}^{\dag}$ implies $c(-\vect{v})=c(\vect{v})^{*}$. Alternatively, when \cref{eq:chi_from_operator} is applied to a unitary operator $\hat{U}$, $\hat{U}^{\dag}\hat{U}=\hat{I}$ implies
\begin{equation}
    \int d^{2n}\vect{v}\,c(\vect{u}+\vect{v})c(\vect{v})^{*}e^{-i\pi\vect{u}^{T}\!\Omega\vect{v}}=\delta^{2n}(\vect{u}).
\end{equation}

We define the characteristic function of a quantum channel $\mathcal{E}$ such that it satisfies the property~\cite{Conrad21}
\begin{equation}
    \mathcal{E}(\hat{\rho})=\iint d^{2n}\vect{u}\,d^{2n}\vect{v}\,c(\vect{u},\vect{v})\hat{W}(\vect{u})\hat{\rho}\hat{W}(\vect{v})^{\dag},\tag{\ref{eq:translation_operator_basis}}
\end{equation}
which is analogous to \cref{eq:operator_from_chi}. Then, the Hermitivity of $\mathcal{E}(\hat{\rho})$ implies $c(\vect{u},\vect{v})=c(\vect{v},\vect{u})^{*}$, and $\mathrm{tr}(\mathcal{E}(\hat{\rho}))=1$ implies
\begin{equation}
    \int d^{2n}\vect{v}\,c(\vect{u}+\vect{v},\vect{v})e^{-i\pi\vect{u}^{T}\!\Omega\vect{v}}=\delta^{2n}(\vect{u}).
\end{equation}

One way to obtain the characteristic function $c$ of an arbitrary quantum map is from its Kraus decomposition $\mathcal{E}(\hat{\rho})=\sum_{i}\hat{E}^{}_{i}\hat{\rho}\hat{E}_{i}^{\dag}$, in which case we have~\cite{Conrad21}
\begin{align}\label{eq:chi_from_Kraus}
    c(\vect{u},\vect{v})&=\sum_{i}c_{i}(\vect{u})c_{i}(\vect{v})^{*},&c_{i}(\vect{u})&=\mathrm{tr}\big(\hat{E}_{i}\hat{W}(\vect{u})^{\dag}\big).
\end{align}
Alternatively, we can use the characteristic function of the Liouville superoperator representation of the map $\hat{\mathcal{E}}=\sum_{i}\hat{E}^{}_{i}\otimes\hat{E}_{i}^{*}\in\mathcal{L}(\mathcal{H}\otimes\mathcal{H}^{*})$ (where $*$ here indicates complex conjugation in the Fock basis), in which case we have
\begin{equation}\label{eq:chi_from_Liouville}
    c\bigg(\!\begin{bmatrix}\vect{u}_{q}\\\vect{u}_{p}\end{bmatrix},\begin{bmatrix}\vect{v}_{q}\\\vect{v}_{p}\end{bmatrix}\!\bigg)=\mathrm{tr}\Bigg(\!\hat{\mathcal{E}}\,
    \hat{W}\bigg(\!\begin{bmatrix}\vect{u}_{q}\\\vect{u}_{p}\end{bmatrix}\!\bigg)^{\!\dag}\otimes
    \hat{W}\bigg(\!\begin{bmatrix}\vect{v}_{q}\\-\vect{v}_{p}\end{bmatrix}\!\bigg)^{\!\dag}\Bigg),
\end{equation}
where $\vect{u}_{q},\vect{u}_{p},\vect{v}_{q},\vect{v}_{p}\in\mathbb{R}^{n}$. Due to the uniqueness of the Liouville representation $\hat{\mathcal{E}}$, it is now clear that $c(\vect{u},\vect{v})$ does not depend on the Kraus decomposition of $\mathcal{E}$. We note that the minus sign $-\vect{v}_{p}$ in \cref{eq:chi_from_Liouville} arises from the fact that the complex conjugate of $\hat{p}$ in the Fock basis is $\hat{p}^{*}=-\hat{p}$. Although \cref{eq:chi_from_Kraus,eq:chi_from_Liouville} provide a general procedure to calculate the characteristic function of $\mathcal{E}$, in \cref{subsec:Gaussian_channel_chi} we will use a different strategy that leverages the Gaussianity of the channel.

Having reviewed these basic facts, we present and prove the characteristic function of Gaussian unitary operators and Gaussian channels in the following two subsections. Our proof strategy will be to verify the action of the characteristic function of the operator/channel on the characteristic function of states that they act upon. Then, we calculate the characteristic function of example operators and channels that were used in \cref{sec:errors}.

\subsection{Gaussian unitary operators}\label{subsec:Gaussian_unitary_chi}

First, we show the following:

\begin{center}
\setlength{\fboxsep}{1 em}
\fbox{\parbox{0.9\linewidth}{
The characteristic function of a Gaussian unitary $\hat{U}_{S}$ is given by
\begin{subequations}\label{eq:Gaussian_unitary_chi}
\begin{align}
    c_{S}(\vect{v})&=\mathrm{tr}\big(\hat{U}_{S}\hat{W}(\vect{v})^{\dag}\big)=\frac{\mathrm{exp}(i\pi\vect{v}^{T}M\vect{v})}{\sqrt{|\mathrm{det}(S-I)|}},\\
    M&=\frac{1}{2}\Omega(S+I)(S-I)^{-1},\label{eq:Gaussian_unitary_chi_M}
\end{align}
\end{subequations}
assuming $(S-I)$ is invertible.
}}
\end{center}
Note that it follows from $S$ being symplectic that $M$ is symmetric.

\textit{Proof:}
We show this by proving the condition \cref{eq:chi_Gaussian_transformation}. In particular, for any state $\hat{\rho}$ with characteristic function $c_{\rho}(\vect{v})$, we show that
\begin{gather}
    \iiint d^{2n}\vect{u} d^{2n}\vect{v} d^{2n}\vect{w}\,c_{\rho}(\vect{u})c_{S}(\vect{v})c_{S}(\vect{w})^{*}\hat{W}(\vect{v})\hat{W}(\vect{u})\hat{W}(\vect{w})^{\dag}\nonumber\\
    \qquad\qquad\qquad\qquad=\int d^{2n}\vect{u}\,c_{\rho}(S^{-1}\vect{u})\hat{W}(\vect{u}),\label{eq:chi_Gaussian_transformation_2}
\end{gather}
where we note that \cref{eq:chi_Gaussian_transformation_2} is equal to $\hat{U}^{}_{S}\hat{\rho}\hat{U}^{\dag}_{S}$.

Using the proposed characteristic function of $\hat{U}_{S}$ in \cref{eq:Gaussian_unitary_chi}, the left-hand side of \cref{eq:chi_Gaussian_transformation_2} becomes
\begin{multline}
    \iiint \frac{d^{2n}\vect{u}d^{2n}\vect{v}d^{2n}\vect{w}}{|\mathrm{det}(S-I)|}c_{\rho}(\vect{u})\mathrm{exp}\big(i\pi(\vect{v}^{T}M\vect{v}-\vect{w}^{T}M\vect{w})\big)\\
    \times\hat{W}(\vect{v})\hat{W}(\vect{u})\hat{W}(\vect{w})^{\dag}.
\end{multline}
Next, we combine the three displacements via
\begin{multline}\label{eq:combine_displacements}
    \hat{W}(\vect{v})\hat{W}(\vect{u})\hat{W}(\vect{w})^{\dag}\\
    =e^{i\pi(\vect{u}^{T}\Omega\vect{v}+\vect{u}^{T}\Omega\vect{w}+\vect{v}^{T}\Omega\vect{w})}\hat{W}(\vect{u}+\vect{v}-\vect{w})
\end{multline}
and perform a change of variables $\vect{u}\mapsto \vect{u}-\vect{v}$, $\vect{v}\mapsto\vect{v}+\vect{w}$, $\vect{w}\mapsto\vect{w}$, giving
\begin{multline}\label{eq:change_of_vars}
    \frac{1}{|\mathrm{det}(S-I)|}\int d^{2n}\vect{u}\,\hat{W}(\vect{u})\\
    \int d^{2n}\vect{v}\,c(\vect{u}-\vect{v})\mathrm{exp}\big(i\pi(\vect{v}^{T}M\vect{v}+\vect{u}^{T}\Omega\vect{v})\big)\\
    \int d^{2n}\vect{w}\,\mathrm{exp}\Big(2i\pi\big(\vect{v}^{T}(M-\Omega/2)+\vect{u}^{T}\Omega\big)\vect{w}\Big),
\end{multline}
where we have also used $M=M^{T}$. Next, we integrate over $\vect{w}$, giving
\begin{multline}\label{eq:w_delta_integral}
    \int d^{2n}\vect{w}\,\mathrm{exp}\Big(2i\pi\big(\vect{v}^{T}(M-\Omega/2)+\vect{u}^{T}\Omega\big)\vect{w}\Big)\\
    =\delta^{2n}\big((M+\Omega/2)\vect{v}-\Omega\vect{u}\big),
\end{multline}
where we have used $\Omega=-\Omega^{T}$. Applying the identity $\delta^{2n}(M\vect{v})=\delta^{2n}(\vect{v})/|\det M|$ allows us to simplify \cref{eq:w_delta_integral} to
\begin{equation}\label{eq:w_delta_simplified}
    \delta^{2n}\big(\vect{v}-(M+\Omega/2)^{-1}\Omega\vect{u}\big)/|\det(M+\Omega/2)|.
\end{equation}
It follows from \cref{eq:Gaussian_unitary_chi_M} that $M+\Omega/2=\Omega S(S-I)^{-1}$ and thus $\det(M+\Omega/2)=1/\det(S-I)$, since $\det S=1$ is guaranteed by $S$ being symplectic. Moreover, $(M+\Omega/2)^{-1}\Omega=I-S^{-1}$. Substituting \cref{eq:w_delta_simplified} into \cref{eq:change_of_vars} gives
\begin{gather}
\begin{multlined}
    \int d^{2n}\vect{u}\,\hat{W}(\vect{u})\bigg(\int d^{2n}\vect{v}\,c_{\rho}(\vect{u}-\vect{v})\\
    \times\mathrm{exp}\big(i\pi(\vect{v}^{T}M\vect{v}+\vect{u}^{T}\Omega\vect{v})\big)\delta^{2n}\big(\vect{v}-(I-S^{-1})\vect{u}\big)\bigg)\nonumber
\end{multlined}\\
\begin{multlined}
    =\int d^{2n}\vect{u}\,\hat{W}(\vect{u})c_{\rho}(S^{-1}\vect{u})\mathrm{exp}\Big(i\pi\vect{u}^{T}\big(\Omega(I-S^{-1})\\
    +(I-S^{-1})^{T}M(I-S^{-1})\big)\vect{u}\Big).
\end{multlined}
\end{gather}
The remaining exponent can be shown to be 0 again from the definition of \cref{eq:Gaussian_unitary_chi_M}. Thus, we have shown \cref{eq:chi_Gaussian_transformation_2}, and proven that the characteristic function of $\hat{U}_{S}$ is given by \cref{eq:Gaussian_unitary_chi}.\hspace{\fill}$\blacksquare$

It is worth noting at this point a subtlety regarding the phase of the unitary operator $\hat{U}_{S}$. The definition of $\hat{U}_{S}$ in \cref{eq:Gaussian_unitary} only defines $\hat{U}_{S}$ up to an overall phase $e^{i\theta}$. Consequently, our expression \cref{eq:Gaussian_unitary_chi} for the characteristic function of $\hat{U}_{S}$ only applies for the representative of $\hat{U}_{S}$ with $\mathrm{Arg}\big(\mathrm{tr}(\hat{U}_{S})\big)=0$. Nevertheless, this is typically of little consequence, since the overall phase of the unitary operator does not affect the state of the system.

As an example, consider the single mode rotation operator
\begin{subequations}
\begin{align}
    \hat{U}_{R(\theta)}&=-i\exp\Big(i\theta\big(\hat{a}^{\dag}\hat{a}+1/2\big)\Big),\\
    R(\theta)&=\begin{bmatrix}\cos\theta&-\sin\theta\\\sin\theta&\cos\theta\end{bmatrix},
\end{align}
\end{subequations}
with $\theta\in(0,2\pi)$. Note that the phase of $\hat{U}_{R(\theta)}$ is chosen such that $\mathrm{Arg}\big(\mathrm{tr}(\hat{U}_{R(\theta)})\big)=0$, which can be verified by inserting a coherent state resolution of the identity. Applying \cref{eq:Gaussian_unitary_chi} gives
\begin{subequations}
\begin{align}
    M&=\frac{1}{2}\Omega(R(\theta)+I)(R(\theta)-I)^{-1}\nonumber\\
    &=-\frac{1}{2}\cot(\theta/2)I,\\
    \mathrm{det}\big(R(\theta)-I\big)&=4\sin^{2}(\theta/2),
\end{align}
\end{subequations}
and therefore
\begin{equation}\label{eq:rot_chi_representation}
    \hat{U}_{R(\theta)}=\frac{1}{2\,\mathrm{sin}(\theta/2)}\int d^{2}\vect{v}\,\mathrm{exp}\Big({-}\frac{i\pi}{2}\mathrm{cot}(\theta/2)|\vect{v}|^{2}\Big)\hat{W}(\vect{v}).
\end{equation}
\Cref{eq:rot_chi_representation} is used directly to calculate the characteristic function of white-noise dephasing in \cref{eq:dephasing,eq:dephasing_chi}. One can also obtain the characteristic function of the operator $\mathrm{exp}(i\theta\hat{a}^{\dag}\hat{a})$ simply by multiplying \cref{eq:rot_chi_representation} by $i\exp(-i\theta/2)$, giving
\begin{equation}\label{eq:rot_chi_representation_2}
    e^{i\theta\hat{a}^{\dag}\hat{a}}=\frac{1}{1-e^{i\theta}}\int d^{2}\vect{v}\,\mathrm{exp}\Big({-}\frac{i\pi}{2}\mathrm{cot}(\theta/2)|\vect{v}|^{2}\Big)\hat{W}(\vect{v}).
\end{equation}
From \cref{eq:rot_chi_representation_2}, we can obtain the characteristic function of the envelope operator $e^{-\Delta^{2}\hat{a}^{\dag}\hat{a}}$ by substituting $\theta\mapsto i\Delta^{2}$, giving
\begin{multline}
    e^{-\Delta^{2}\hat{a}^{\dag}\hat{a}}=\frac{1}{1-e^{-\Delta^{2}}}\\
    \times\int d^{2}\vect{v}\,\mathrm{exp}\Big({-}\frac{\pi}{2}\mathrm{coth}(\Delta^{2}/2)|\vect{v}|^{2}\Big)\hat{W}(\vect{v}),
\end{multline}
as was used in \cref{eq:envelope_op,eq:envelope_chi}.

In the case where $(S-I)$ is not invertible, one can instead find a product of two symplectic matrices $S_{1}S_{2}=S$ for which $S_{1}-I$ and $S_{2}-I$ are both invertible and find the characteristic functions of the Gaussian unitary operators $\hat{U}_{S_{1}}$ and $\hat{U}_{S_{2}}$ separately. Then, one can use \cref{eq:chi_composition} to find the characteristic function of the overall unitary operator $\hat{U}_{S}$. However in these specific cases it may be easier to calculate the trace $\mathrm{tr}\big(\hat{U}_{S}\hat{W}(\vect{v})^{\dag}\big)$ directly. Intuitively, $0$-eigenvalues of $S-I$ correspond to $\delta$ functions in the corresponding characteristic function. For example, in the extreme case $S=I$ we trivially have $\mathrm{tr}\big(\hat{I}\hat{W}(\vect{v})^{\dag}\big)=\delta^{2n}(\vect{v})$.

\subsection{Gaussian Channels}\label{subsec:Gaussian_channel_chi}

The proof and result from \cref{subsec:Gaussian_unitary_chi} can be extended naturally to obtain the characteristic function of Gaussian channels, as we discuss below. Before doing so, we define a Gaussian channel $\mathcal{N}$ in the context of Gaussian quantum computing~\cite{Weedbrook12}.

First, a Gaussian state $\hat{\rho}$ is any state whose characteristic function takes the form
\begin{equation}
    c_{\vect{\mu},V}(\vect{v})=\mathrm{exp}\Big(\!-\pi\vect{v}^{T}\!\big(\Omega V\Omega^{T}\big)\vect{v}-i\pi(\Omega\vect{\mu})^{T}\vect{v}\Big),
\end{equation}
where $\vect{\mu}$ is the vector of expectation values $\mu_{i}=\langle\hat{\xi}_{i}\rangle$ and $V$ is the (symmetric) matrix of second-order expectation values with elements $V_{ij}=\langle\hat{\xi}_{i}\hat{\xi}_{j}+\hat{\xi}_{j}\hat{\xi}_{i}\rangle/2$. Then, a Gaussian channel is any quantum channel $\mathcal{E}$ that maps Gaussian states to Gaussian states. In particular, a Gaussian channel can be described by two matrices $T$ and $N$ and a vector $\vect{d}$, that transforms~\cite{Eisert05}
\begin{align}\label{eq:Gaussian_channel}
    V&\mapsto TVT^{T}+N,&\vect{\mu}&\mapsto T\vect{\mu}+\vect{d}.
\end{align}
In the remaining discussion we will notate $\mathcal{E}_{T,N}$ for a Gaussian channel with $\vect{d}=\vect{0}$. We have set $\vect{d}=\vect{0}$ here since any channel with non-zero $\vect{d}$ can be achieved by applying the channel $\mathcal{E}_{T,N}$ followed by a displacement $\hat{W}(\vect{d}/\sqrt{2\pi})$. In the special case where we also have $N=0$, the Gaussian channel corresponds to conjugation by a Gaussian unitary $\hat{U}_{S}$ where $T=S$. It is known that Gaussian channels transform the characteristic function of arbitrary (not just Gaussian) states via~\cite{Eisert05}
\begin{equation}\label{eq:chi_Gaussian_channel}
    c(\vect{v})\mapsto \exp\big(-\pi\vect{v}^{T}\Omega N\Omega^{T}\vect{v}\big)c(\Omega T^{T}\Omega^{T}\vect{v}),
\end{equation}
which provides a generalization of \cref{eq:chi_Gaussian_transformation} to Gaussian channels.

Now, we present our main result.
\begin{center}
\setlength{\fboxsep}{1 em}
\fbox{\parbox{0.9\linewidth}{
The characteristic function of a Gaussian channel $\mathcal{E}_{T,N}$ is given by
\begin{subequations}\label{eq:Gaussian_channel_chi}
\begin{align}
    c_{T,N}(\vect{u},\vect{v})&=\frac{1}{|\mathrm{det}(T-I)|}\mathrm{exp}\Big(2i\pi\vect{u}^{T}\!M_{\text{a}}\vect{v}\nonumber\\
    &\qquad\quad\!+i\pi(\vect{u}^{T}\!M_{\text{s}}\vect{u}-\vect{v}^{T}\!M_{\text{s}}\vect{v})\nonumber\\
    &\qquad\quad\;-\pi(\vect{u}-\vect{v})^{T}\!L(\vect{u}-\vect{v})\Big),\label{eq:Gaussian_channel_chi_main}\\
    L&=\Omega(T-I)^{-1}N(T-I)^{-T}\Omega^{T},\label{eq:Gaussian_channel_chi_L}\\
    M&=\frac{\Omega}{2}(T+I)(T-I)^{-1},\label{eq:Gaussian_channel_chi_M}
\end{align}
\end{subequations}
assuming $(T-I)$ is invertible, and where $M_{\text{s}}=(M+M^{T})/2$ is the symmetric part of $M$ and $M_{\text{a}}=(M-M^{T})/2$ the anti-symmetric part.
}}
\end{center}

\textit{Proof:}
The proof of \cref{eq:Gaussian_channel_chi} follows a similar strategy as in \cref{subsec:Gaussian_unitary_chi}, in that we wish to show that \cref{eq:Gaussian_channel_chi} satisfies \cref{eq:chi_Gaussian_channel}. In particular, we show that for any density matrix $\hat{\rho}$ with characteristic function $c_{\rho}(\vect{v})$, we have
\begin{gather}
    \iiint d^{2n}\vect{u}d^{2n}\vect{v}d^{2n}\vect{w}c_{\rho}(\vect{u})c_{T,N}(\vect{v},\vect{w})\hat{W}(\vect{v})\hat{W}(\vect{u})\hat{W}(\vect{w})^{\dag}\nonumber\\
    \qquad=\int d^{2n}\vect{u}\,e^{-\pi\vect{u}^{T}\!\Omega N\Omega^{T}\!\vect{u}}c_{\rho}(\Omega T^{T}\Omega^{T}\vect{u})\hat{W}(\vect{u}),\label{eq:chi_Gaussian_channel_2}
\end{gather}
which is equal to $\mathcal{E}_{T,N}(\hat{\rho})$.

Starting with the left-hand side of \cref{eq:chi_Gaussian_channel_2}, we substitute the proposed characteristic function of $\mathcal{E}_{T,N}$ in \cref{eq:Gaussian_channel_chi}, combine the three displacement operators using \cref{eq:combine_displacements}, and perform a change of variables $\vect{u}\mapsto\vect{u}-\vect{v}$, $\vect{v}\mapsto\vect{v}/2+\vect{w}$, $\vect{w}\mapsto\vect{w}-\vect{v}/2$ (which has unit Jacobean determinant), giving
\begin{multline}\label{eq:Gaussian_chi_long_equation}
    \frac{1}{|\mathrm{det}(T-I)|}\int d^{2n}\vect{u}\,\hat{W}(\vect{u})\\
    \int d^{2n}\vect{v}\,c_{\rho}(\vect{u}-\vect{v})\,\mathrm{exp}(-\pi\vect{v}^{T}\!L\vect{v})\\
    \int d^{2n}\vect{w}\,\mathrm{exp}\Big(2i\pi\big(\vect{v}^{T}(M-\Omega/2)+\vect{u}^{T}\Omega\big)\vect{w}\Big),
\end{multline}
where we have used $M=M_{\text{s}}+M_{\text{a}}$. Performing the integral over $\vect{w}$ gives
\begin{equation}\label{eq:w_delta_simplified_2}
    \delta^{2n}\big((M-\Omega/2)^{T}\!\vect{v}-\Omega\vect{u}\big)=\frac{\delta^{2n}\big(\vect{v}-(M-\Omega/2)^{-T}\Omega\vect{u}\big)}{|\mathrm{det}(M-\Omega/2)|}.
\end{equation}
Now, from \cref{eq:Gaussian_channel_chi_M}, we have that $M-\Omega/2=\Omega(T-I)^{-1}$ and thus $\mathrm{det}(M-\Omega/2)=1/\mathrm{det}(T-I)$. Substituting \cref{eq:w_delta_simplified_2} into \cref{eq:Gaussian_chi_long_equation} and integrating over $\vect{v}$ gives
\begin{align}
&\begin{multlined}
    \int d^{2n}\vect{u}\,\hat{W}(\vect{u})\bigg(\int d^{2n}\vect{v}c_{\rho}(\vect{u}-\vect{v})\mathrm{exp}(-\pi\vect{v}^{T}\!L\vect{v})\\
    \delta^{2n}\Big(\vect{v}-\big(I-\Omega T^{T}\!\Omega^{T}\big)\vect{u}\Big)\bigg)
\end{multlined}\nonumber\\
&\begin{multlined}[b]
    =\int d^{2n}\vect{u}\,\hat{W}(\vect{u})\,c_{\rho}\big(\Omega T^{T}\Omega^{T}\!\vect{u}\big)\\
    \times\exp\big({-}\pi\vect{u}^{T}\!\Omega^{T}\!(T-I)\Omega^{T}\!L\Omega(T-I)^{T}\Omega\vect{u}\big),
\end{multlined}
\end{align}
which yields the right-hand side of \cref{eq:chi_Gaussian_channel_2} upon substituting \cref{eq:Gaussian_channel_chi_L}.\hspace{\fill}$\blacksquare$

Finally, as an example, consider Gaussian channels where $T=\tau I$ and $N=\nu I$ are proportional to the identity matrix (with $\tau\neq1$). Then, \cref{eq:Gaussian_channel_chi} simplifies to
\begin{multline}
    c_{\tau I,\nu I}(\vect{u},\vect{v})=\frac{1}{(\tau-1)^{2n}}\mathrm{exp}\bigg(\!-\pi\frac{\nu}{(\tau-1)^{2}}|\vect{u}-\vect{v}|^{2}\\
    +i\pi\frac{\tau+1}{\tau-1}\vect{u}^{T}\Omega\vect{v}\bigg),
\end{multline}
where $n$ is the number of modes in the system. In fact, some of the most well-studied Gaussian channels take this form. For example, loss, defined in \cref{eq:loss} or (\ref{eq:loss_me}), is given by $\tau=\sqrt{1-\gamma}$ and $\nu=\gamma/2$, resulting in the characteristic function~\cref{eq:loss_chi}. Quantum-limited amplification, defined in \cref{eq:amplification_me}, is given by $\tau=\sqrt{g}$ and $\nu=(g-1)/2$. Composing the characteristic functions of quantum-limited amplification and loss via \cref{eq:chi_composition} and setting $g=1/(1-\gamma)$ as in \cref{eq:gauss_loss_relationship} results in the characteristic function of the Gaussian random displacement noise model \cref{eq:chi_random_displacement}, which has $\tau=1$ and $\nu=\sigma^{2}=\gamma/(1-\gamma)$ and thus cannot be directly calculated using \cref{eq:Gaussian_channel_chi}.

\section{Orthonormalization procedure}\label{sec:orthonormalisation}

\begin{figure}
    \centering
    \includegraphics{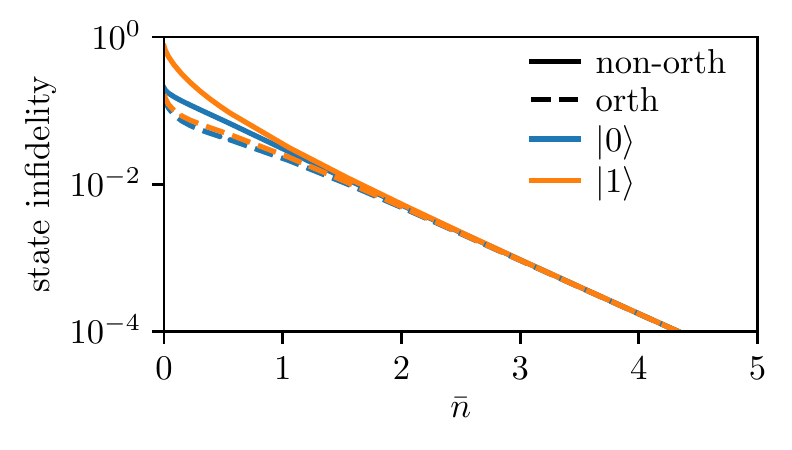}
    \caption{(Color) Comparison between the orthonormalized and non-orthonormalized logical noise maps acting on the computational basis states $\ket{\psi}=\ket{0},\ket{1}$. We plot the state infidelity $1-\bra{\psi}\mathcal{E}(\ket{\psi}\!\bra{\psi})\ket{\psi}$, where $\mathcal{E}$ is the logical noise map corresponding to the envelope operator $e^{-\Delta^{2}\hat{a}^{\dag}\hat{a}}$, against the average photon number $\bar{n}\approx 1/(2\Delta^{2})-1/2$ which is calculated numerically using $\bar{n}=\big(\!\bra{\bar{0}_{\Delta}}\hat{a}^{\dag}\hat{a}\ket{\bar{0}_{\Delta}}+\bra{\bar{1}_{\Delta}}\hat{a}^{\dag}\hat{a}\ket{\bar{1}_{\Delta}}\!\big)/2$. When we orthonormalize the codestates as described in \cref{sec:orthonormalisation}, the state infidelity is lower than the non-orthonormalized codestates, although this difference is only noticeable for very low average photon number $\bar{n}\leq1.5$, $\Delta_{\text{dB}}\leq6$. Note also that the infidelity for the $\ket{1}$ state is larger than that for the $\ket{0}$ state because the bosonic states $\ket{\bar{\psi}_{\Delta}}$ tend to the vacuum state as $\bar{n}\rightarrow0$, which, after the partial trace is applied, has a lower overlap with the qubit state $\ket{1}$ than $\ket{0}$.}
    \label{fig:orth}
\end{figure}

In this appendix, we present the derivation of the procedure to orthonormalize the codewords of a single-mode GKP qubit code. In particular, we are interested in a logical noise channel in which the approximate codestates are defined by the envelope operator $e^{-\Delta^{2}\hat{a}^{\dag}\hat{a}}$, i.e.
\begin{equation}\label{eq:logical_noise_map_orth}
\mathcal{N}_{\mathcal{L}}(\hat{\rho})=\mathrm{tr}_{S}\Big(\mathcal{N}_{2}\circ\mathcal{J}[e^{-\Delta^{2}\hat{a}^{\dag}\hat{a}}]\big(\hat{\rho}\otimes_{\mathcal{G}}\ket{\vect{0}}\!\bra{\vect{0}}\!\big)\Big),
\end{equation}
where $\mathcal{J}[\hat{O}]\hat{\rho}=\hat{O}\hat{\rho}\hat{O}^{\dag}$ and $\mathcal{N}_{2}$ is a (CPTP) noise channel such as loss or dephasing.

Since $e^{-\Delta^{2}\hat{a}^{\dag}\hat{a}}$ is non-unitary, the overall logical noise channel is completely positive (CP) but not trace-preserving (TP). However, in order to define a valid quantum channel, one can orthogonalize the approximate codewords $e^{-\Delta^{2}\hat{a}^{\dag}\hat{a}}\ket{\bar{0}}$ and $e^{-\Delta^{2}\hat{a}^{\dag}\hat{a}}\ket{\bar{1}}$ (where $\ket{\bar{\psi}}=\ket{\psi}\otimes_{\mathcal{G}}\ket{\vect{0}}$) via the equation
\begin{equation}
    \begin{bmatrix}\ket{\bar{0}_{\Delta,\text{o}}}\\\ket{\bar{1}_{\Delta,\text{o}}}\end{bmatrix}=C(\Delta)\begin{bmatrix}e^{-\Delta^{2}\hat{a}^{\dag}\hat{a}}\ket{\bar{0}}\\e^{-\Delta^{2}\hat{a}^{\dag}\hat{a}}\ket{\bar{1}}\end{bmatrix},
\end{equation}
where
\begin{subequations}\label{eq:C_Delta}
\begin{align}
 C(\Delta)&=\begin{bmatrix}R_{+}/(2N_{0})&e^{-i\phi}R_{-}/(2N_{1})\\e^{i\phi}R_{-}/(2N_{0})&R_{+}/(2N_{1})\end{bmatrix},\\
N_{\mu}&=\|e^{-\Delta^{2}\hat{a}^{\dag}\hat{a}}\ket{\bar{\mu}}\!\|,\\
R_{\pm}&=\frac{1}{\sqrt{1+R}}\pm\frac{1}{\sqrt{1-R}},\\
R&=\frac{\big|\!\bra{\bar{0}}e^{-2\Delta^{2}\hat{a}^{\dag}\hat{a}}\ket{\bar{1}}\!\big|}{N_{0}N_{1}},\\
\phi &= \mathrm{arg}\big(\!\bra{0}e^{-2\Delta^{2}\hat{a}^{\dag}\hat{a}}\ket{1}\!\big).
\end{align}
\end{subequations}
This orthogonalization procedure is equivalent to the L\"owdin orthogonalization~\cite{Lowdin50}, which orthogonalizes the codewords symmetrically.

Conveniently, the inner products $\bra{\bar{\mu}}e^{-2\Delta^{2}\hat{a}^{\dag}\hat{a}}\ket{\bar{\nu}}$ that define the orthonormalization can be obtained solely from the logical envelope channel $\mathcal{E}_{\mathcal{L}}^{\Delta}$ (i.e.~the map in \cref{eq:logical_noise_map_orth} with $\mathcal{N}_{2}$ set to the identity) via
\begin{equation}
\begin{aligned}
\mathrm{tr}(\mathcal{E}_{\mathcal{L}}^{\Delta}\ket{\nu}\!\bra{\mu})&=\mathrm{tr}_{\mathcal{L}}\big(\mathrm{tr}_{\mathcal{S}}\big(\mathcal{J}[e^{-\Delta^{2}\hat{a}^{\dag}\hat{a}}]\ket{\bar{\nu}}\bra{\bar{\mu}}\big)\big)\\
&=\bra{\bar{\mu}}e^{-2\Delta^{2}\hat{a}^{\dag}\hat{a}}\ket{\bar{\nu}},
\end{aligned}
\end{equation}
since tracing over the stabilizer subsystem followed by the logical subsystem is equivalent to the total trace over the entire mode.

To apply this to the original problem of defining a CPTP logical noise map, we use the orthonormalization matrix $C(\Delta)$ given in \cref{eq:C_Delta} such that $\ket{\bar{\psi}_{\Delta,\text{o}}}=e^{-\Delta^{2}\hat{a}^{\dag}\hat{a}}C(\Delta)\ket{\bar{\psi}}$. Thus, we can define a CPTP map $\mathcal{N}_{\mathcal{L},\text{o}}=\mathcal{N}_{\mathcal{L}}\circ\mathcal{J}[C(\Delta)]$. This resulting CPTP logical noise map $\mathcal{N}_{\mathcal{L},\text{o}}$ is identical to the non-trace-preserving map $\mathcal{N}_{\mathcal{L}}$ but with the codewords orthonormalized, as required.

Since we use this orthonormalization procedure to define the CPTP logical noise channel $\mathcal{N}_{\mathcal{L},\text{o}}$, it is important to consider the effect this has on our results. In \cref{fig:orth}, we present the state fidelity $\bra{\psi}\mathrm{tr}_{\mathcal{S}}\big(\ket{\bar{\psi}_{\Delta,\text{(o)}}}\!\bra{\bar{\psi}_{\Delta,\text{(o)}}}\big)\ket{\psi}$ between the decoded approximate codestates with the logical states they represent. The orthonormalized approximate codestates are orthonormalized via the above procedure, while the non-orthonormalized states are simply normalized by dividing by $\big|\!\braket{\bar{\psi}_{\Delta}|\bar{\psi}_{\Delta}}\!\big|^{2}$. Importantly, the effect of the orthonormalization procedure is negligible for any states with average photon number greater than roughly 1.5 (or $\Delta_{\text{dB}}>6$), demonstrating that our procedure does not significantly affect our results in the regime of interest. In \cref{fig:orth} we only show $\ket{\psi}=\ket{0},\ket{1}$, but similar results can also be obtained for other logical states.

\section{Envelope operator simulations as $\Delta\rightarrow0$}\label{sec:very_small_Delta}

\begin{figure}
    \centering
    \includegraphics{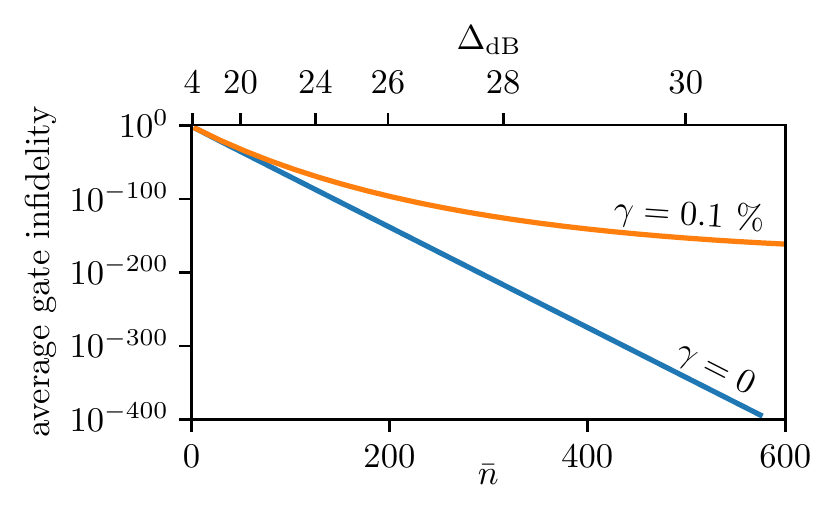}
    \caption{(Color) Average gate infidelity of the logical noise channel $(\mathcal{L}^{\gamma}\circ\mathcal{E}^{\Delta})_{\mathcal{L},\text{o}}$ corresponding to loss applied to approximate single-mode square GKP qubit codestates. We show two different values of loss given by $\gamma=0$, 0.1 \%, and show the plot as a function of $\Delta$, where $\bar{n}\approx\frac{1}{2\Delta^{2}}-\frac{1}{2}$ is the average photon number of the approximate GKP encoded maximally mixed state. This demonstrates that our methods can be applied easily to approximate GKP codestates with comically large average photon number.}
    \label{fig:very_large_plot}
\end{figure}

In this appendix we briefly discuss the utility of our simulations in the ideal limit $\Delta\rightarrow0$. In particular, our simulations of the envelope operator, loss, and Gaussian displacements for the square GKP code are all analytic. As long as the error rate is sufficiently low, one can truncate each of the sums in \cref{eq:logical_noise_channel} to $s_{i},t_{i}=-1,0,1$ to retain only the leading order sources of error. Then, the simulations can be run for arbitrarily squeezed states by simply evaluating each analytic expression for the integral and adding the relevant terms to the superoperator $\mathcal{E}^{\Delta}$. Following this, we again use \cref{sec:orthonormalisation} to calculate $\mathcal{E}^{\Delta}_{\text{o}}$ from $\mathcal{E}^{\Delta}$, and then we can directly extract the average gate infidelity. The run-time of this procedure is limited only by the analytic evaluation of the expressions for the superoperator with large enough precision to reach the extremely low infidelities.

As a proof of principle, in \cref{fig:very_large_plot} we present the average gate infidelities of an approximate GKP codestate with no other noise, and an approximate GKP codestate with a loss-rate of $\gamma=0.1$ \%, up to an average GKP photon number of $\sim600$ ($\Delta\approx0.029$, $\Delta_{\mathrm{dB}}\approx30.8$). At this level of squeezing, Fock space simulations would require a truncation dimension which excludes at most $\sim 10^{-400}$ of the total support of the state so that the leading source of error in the simulation is due to the approximate GKP codestate. Moreover, the variance in the photon number distribution of the approximate GKP codestate is also roughly 600, rendering Fock space simulations completely infeasible. While the $\gamma=0$ curve has a well-known approximate analytic expression which tends to being exact as $\Delta\rightarrow0$, it is less clear how one would determine a similar analytic expression for the infidelity associated with $\gamma=0.1~\%$, particularly in the region around $\bar{n}\approx600$ as the curve reaches the optimal $\Delta$. Although such photon numbers are unlikely to ever be experimentally realized, \cref{fig:very_large_plot} demonstrates the efficiency and numerical stability of our simulations when applied to square GKP codewords with arbitrary amounts of squeezing.
\end{document}